\documentclass[superscriptaddress,aps,showpacs,nofootinbib,11pt]{revtex4}
\usepackage{graphicx}
\usepackage{amsmath,amssymb}

\begin{document}

\title{Gravitino production in the Randall-Sundrum II braneworld cosmology}

\author{N. M. C. Santos}
\email{n.santos@thphys.uni-heidelberg.de}
\affiliation{%
Institut f\"{u}r Theoretische Physik, Universit\"{a}t Heidelberg\\
Philosophenweg 16, 69120 Heidelberg, Germany}

\begin{abstract}
Braneworld modifications to the Friedmann expansion law can have an
important effect on the cosmological evolution of the early
universe. In particular, the primordial particle abundances
crucially depend on the rate at which the universe expanded at early
times. In this article, we study the production of stable and
unstable gravitinos, both from thermal creation and from the decay
of a heavy scalar, in the Randall-Sundrum II braneworld context. We
conclude that, depending on the value of the 5D fundamental Planck
mass, some of the usual standard cosmology constraints on the
reheating temperature and on the mass of the heavy scalar can be
evaded.
\end{abstract}

\pacs{98.80.-k, 98.80.Cq, 04.50.+h, 04.65.+e}
\date{\today} \maketitle

\section{Introduction}

The (over)production of gravitinos can have striking impact on
cosmology~\cite{Khlopov:1984pf}. If gravitinos are unstable, their
overabundance would spoil the success of big-bang nucleosynthesis
(BBN). On the other hand, if they are stable particles, which is the
case if they are the lightest supersymmetric particle (LSP)
contributing to the total amount of dark matter, the present
abundance of gravitinos is constrained in order to avoid the
overclosure of the universe.

It is long known that gravitinos can be thermally produced through
scatterings in the plasma. However, it has been recently pointed
out~\cite{Endo:2006zj,Nakamura:2006uc}, in the context of the moduli
problem~\cite{Coughlan:1983ci}, that the decay rate of the scalar
field into a pair of gravitinos is much higher than previously
thought~\cite{Hashimoto:1998mu}. This ``new" source of gravitinos
coming from a scalar field decay might have important consequences
for the evolution of the universe. In fact, we expect that the
universe was once dominated by a scalar field, for example, the
inflaton (see, e.g., Ref.~\cite{Lyth:1998xn} for a review), or the
dilaton and moduli fields in superstring theories. The production of
gravitinos from inflaton/dilaton/scalar decays, and the problems
that might result, have been widely studied in standard cosmology
(SC)~\cite{Endo:2006zj,Nakamura:2006uc,Kawasaki:2006gs,Kawasaki:2006hm,
Asaka:2006bv,Endo:2006qk,Endo:2006tf,Dine:2006ii}.

In the last years, there has also been much interest in the
so-called braneworld cosmology (BC) scenarios. Whilst theories
formulated in extra dimensions have been around since the early
twentieth century, recent developments in string theory have opened
up the possibility that our universe could be a 1+3-surface, the
brane, embedded in a higher-dimensional space-time, called the bulk,
with all (minimal supersymmetric) standard model particles and
fields trapped on the brane, while gravity is free to access the
bulk (see Ref.~\cite{Maartens:2003tw} for a review). A remarkable
feature of BC is the modification of the expansion rate of the
universe before the BBN era\footnote{There are also brane
constructions in which the expansion rate is modified after the BBN
and gravity is modified at large distances, as in the
Dvali-Gabadadze-Porrati model~\cite{Dvali:2000hr}, which is a
possible explanation for the present accelerated expansion of the
universe.}. In the so-called Randall-Sundrum II (RSII) braneworld
construction~\cite{Randall:1999vf}, in which one has a single brane
with positive tension $\lambda$, the Friedmann equation receives an
additional term quadratic in the
density~\cite{Binetruy:1999ut,Csaki:1999jh,Cline:1999ts,Shiromizu:1999wj,
Binetruy:1999hy,Kraus:1999it,Ida:1999ui,Mukohyama:1999wi}. Setting
the 4D cosmological constant to zero and assuming that inflation
rapidly makes any dark radiation term negligible, one obtains
\begin{align}
H^2 = \frac{8\pi}{3 M_P^2} ~ \rho ~ \left(1 + \frac{\rho}{2\lambda}
\right)~, \quad\quad M_P = \sqrt{\frac{3}{4 \pi}}
\frac{M_5^3}{\sqrt{\lambda}}~,\label{eq:Fried}
\end{align}
where $M_P = 1.22 \times 10^{19}$~GeV is the 4D Planck mass and
$M_5$ the 5D fundamental mass. Notice that Eq.~(\ref{eq:Fried})
reduces to the usual Friedmann equation, $H\propto\sqrt{\rho}$, at
sufficiently low energies, \mbox{$\rho \ll \lambda$}, but at very
high energies one has $H\propto\rho$. Successful BBN requires that
the change in the expansion rate due to the new terms in the
Friedmann equation be sufficiently small at scales $\sim
\mathcal{O}$(MeV); this implies $M_5 \gtrsim 40~\mbox{TeV}$. A more
stringent bound, $M_5 \gtrsim 10^5~\mbox{TeV}$, is obtained by
requiring the theory to reduce to Newtonian gravity on scales larger
than 1~mm.

As is well known, the above modification in the Friedmann expansion
rate can have important implications for
inflation~\cite{Maartens:1999hf},
baryogenesis~\cite{Mazumdar:2000gj,Bento:2004pz}, dark
matter~\cite{Okada:2004nc} and thermal gravitino
production~\cite{Okada:2004mh}. In this paper we aim to investigate
its effect in the decay of a heavy scalar field into a pair of
gravitinos. After a brief discussion on the decay of heavy scalars
and the production of gravitinos in the RSII brane scenario, we
study the resulting constraints on the parameter space, both for the
stable and unstable gravitino cases.

\section{Heavy scalar decay and RSII cosmology}

We consider here the case in which the scalar particle $X$
dominates the energy of the universe when it decays, $\rho \simeq
\rho_X$. This scenario can be achieved if the scalar is, for
example, the inflaton. We will assume that $X$ is a singlet under
the standard model gauge group.

Of particular interest is the reheating temperature, $T_{rh}$, which
is defined here by assuming an instantaneous conversion of $X$
energy into radiation, when the decay width of $X$, $\Gamma_X$,
equals the expansion rate of the universe. If the total energy
density of the scalar $X$ is instantaneously converted into
radiation, then we can identify $\rho_X=\rho_R=(\pi^2/30)\,g_\ast\,
T^4\,$, where $g_\ast$ is the effective number of relativistic
degrees of freedom at the temperature $T$. In particular, $g_\ast =
915/4 = 228.75$ in the minimal supersymmetric standard model (MSSM),
for temperatures above the SUSY breaking scale $\sim
\mathcal{O}$(TeV). For a radiation dominated universe,
Eq.~(\ref{eq:Fried}) can then be written as
\begin{align}
H^2 = H_{SC}^2(T) \left[1+\left(\dfrac{T}{T_t}\right)^4\right]~,
\label{eq:Htemp}
\end{align}
where
\begin{align}
H_{SC}(T)= \dfrac{2 \pi}{3 }\, \sqrt{\dfrac{\pi\, g_*}{5}}\;
\dfrac{T^2}{M_P}~ \label{eq:HSCtemp}
\end{align}
is the usual SC expansion law, and we have defined
\begin{align}
T_t= \left(\dfrac{45 M_5^6}{\pi^3 g_* M_P^2}\right)^{1/4}=8.1
\times 10^4\,{\rm GeV}\,\left(\dfrac{M_5}{10^{10}\,{\rm
GeV}}\right)^{3/2}~, \label{eq:Tt}
\end{align}
as the transition temperature from the high energy brane regime
expansion to the standard expansion\footnote{To compute the
numerical value, we have used the value of $g_*$ as in the MSSM.
Hereafter, the same value will be used to obtain other numerical
estimates in the present work.}. From the condition
\begin{align}
H(T_{rh})=\Gamma_X \label{eq:rh}
\end{align}
one gets the reheating temperature $T_{rh}$, where $\Gamma_X$ is the
total decay rate of $X$. In the high-energy limit of BC, i.e. for
$T\gg T_t$, Eq.~(\ref{eq:Htemp}) reads
\begin{align}
H_{BC}\simeq \frac{2 \pi^3 g_\ast}{45} \frac{T^4}{M_5^3}~,
\label{eq:HrhBC}
\end{align}
and the condition in Eq.~(\ref{eq:rh}) leads then to the relation
\begin{align}
T_{rh}= \left(\frac{45}{2 \pi^3 g_\ast} \Gamma_X
M_5^3\right)^{1/4}~. \label{eq:TrhBC}
\end{align}
In the SC limit ($T \ll T_t$) one obtains
\begin{align}
T_{rh}=  \left(\frac{45}{4 \pi^3 g_\ast}\right)^{1/4} \sqrt{\Gamma_X
M_P}~. \label{eq:TrhSC}
\end{align}

The total decay rate of $X$ is determined from all the
interactions of $X$ with the other particles of the supersymmetric
model, and so it is model dependent. However we can in general
write
\begin{align}
\Gamma_X = \alpha_{t}^2 \,\dfrac{M_X^3}{M_P^2}~. \label{eq:GammaX}
\end{align}
In fact, when $X$ has only Planck suppressed interaction,
$\alpha_{t}$ is expected to be of order one. For example, if $X$ is
the heavy moduli field that couples to gauge supermultiplets in the
gauge kinetic function through a Planck suppressed interaction, the
decay rate of $X$ is of the form of Eq.~(\ref{eq:GammaX}) with
$\alpha_{t}={\cal O}(1)$~\cite{Endo:2006zj,Nakamura:2006uc}.

Using Eq.~(\ref{eq:GammaX}), the reheating temperature in the BC
regime can be written as
\begin{align}
T_{rh}=6.8 \times 10^4\,{\rm GeV} \,\alpha_{t}^{1/2}
\left(\dfrac{M_X}{10^{10}\,{\rm
GeV}}\right)^{3/4}\left(\dfrac{M_5}{10^{10}\,{\rm
GeV}}\right)^{3/4}~, \label{eq:TrhBC_M}
%= \left(\frac{45\, \alpha_t^2}{2 \pi^3 g_\ast}
%\dfrac{M_X^3\,M_5^3}{M_P^2}\right)^{1/4}
\end{align}
while in the SC limit one obtains
\begin{align}
T_{rh}= 5.7 \times 10^4\,{\rm GeV}\,\alpha_{t}
\left(\dfrac{M_X}{10^{10}\,{\rm GeV}}\right)^{3/2}~.
\label{eq:TrhSC_M}
%= \left(\frac{45}{4 \pi^3 g_\ast}\right)^{1/4}
%\sqrt{\dfrac{\alpha_{t}^2 M_X^3} {M_P}}
\end{align}

At this point one should notice that $T_{rh} > 7$~MeV in order to be
consistent with BBN theory. This implies that in BC regime the
relation  $M_X M_5 \gtrsim 4.8 \times 10^{10} \,\alpha_{t}^{-2/3}
\,{\rm GeV}^2$ should be satisfied, while in SC one should have $M_X
\gtrsim 2.5 \times 10^5\,\alpha_{t}^{-2/3} \,{\rm GeV} $. However,
in order to have the BC regime one should also impose $T_{rh}>T_t$,
i.e. to obtain such a low reheating temperature in the BC regime one
needs also a low $T_t$ which corresponds to a low value of $M_5$.
Such low values of $M_5$ are not compatible with the bound $M_5
\gtrsim 10^5~\mbox{TeV}$, obtained by requiring the theory to reduce
to Newtonian gravity on scales larger than 1~mm. This implies that
the only relevant bound is coming from the SC regime
\begin{align}
M_X \gtrsim 2.5 \times 10^5\,{\rm GeV}~
\dfrac{1}{\alpha_{t}^{2/3}}~.\label{eq:Mxlowbound}
\end{align}

One should also point out that in order to have $T_{rh}>T_t$, one
must impose
\begin{align}
M_X > \left(\dfrac{2}{\alpha_{t}^2}\right)^{1/3} M_5~.
\end{align}

Here we are mainly interested in the gravitino production, and
hence in the decay of a heavy scalar field $X$ into a pair of
gravitinos
\begin{align}
X \rightarrow \psi_{3/2} + \psi_{3/2}~.~\label{eq:decay}
\end{align}

Before we proceed with the discussion of the decay of the heavy
scalar into gravitinos, we should specify better our brane setup,
namely the localization of gravitinos on the brane. We are
considering here the supersymmetric extension of the RSII braneworld
model~\cite{Gherghetta:2000qt}. As already mentioned, in the
Randall-Sundrum construction under consideration, only gravity
propagates in the bulk: the zero-mode graviton is localized around
the ultraviolet (UV) brane (our brane), while the graviton
Kaluza-Klein (KK) modes are localized around the infrared brane
(which in the RSII is located at the infinite boundary). In the
supersymmetric generalization, all the fields in the same
supermultiplet will take the same configuration in the fifth
dimensional direction. Hence, the zero-mode gravitino will also be
localized around the UV brane. As an approximation we will consider
it as a field residing on the UV brane, as done previously in other
works~\cite{Okada:2004mh,Bento:2004pz}. The zero-mode gravitino will
obtain its mass through the super-Higgs mechanism, as in the usual
4D supergravity. Since supersymmetry is broken only locally on the
UV brane, the KK gravitinos will remain the same. We will not
consider here the production of these KK gravitinos and its possible
influence on the brane particle content. We will only consider the
zero-mode gravitino as a field residing on the brane and apply 4D
supersymmetry and the usual Boltzmann equation to describe its
production.

Let us now briefly discuss the decay channel in
Eq.~(\ref{eq:decay}). For simplicity, we will assume that $M_X >>
m_{3/2}$. In fact, the gravitino is likely to be much lighter than
either a heavy modulus\footnote{Many proposed solutions to solve the
moduli problem require the moduli to have large masses, e.g. $M_X
\gtrsim 100$~TeV~\cite{Randall:1994fr}.} or the inflaton, because a
too large gravitino mass requires a fine-tuning in the Higgs sector
due to the anomaly-mediated effects. With this assumption, the decay
rate for this process, as recently calculated in
Refs.~\cite{Endo:2006zj,Nakamura:2006uc}, can be written as
\begin{align}
\Gamma_{3/2} = \dfrac{\alpha_{3/2}^2}{36}\dfrac{M_X^3}{M_P^2}~.
\label{eq:Gamma32}
\end{align}
The coupling constant $\alpha_{3/2}$ is defined by the relation
\begin{align}
\alpha_{3/2}\,\, \dfrac{M_{3/2}^2}{M_X}= \left \langle
\left(G_{X\overline{X}}\right)^{-1/2}
 e^{G/2}\,G_X \right\rangle~,
\label{eq:alpha32}
\end{align}
with $G$ being the total K\"{a}hler potential, $G=K+\ln|W|^2$, where
$K$ and $W$ are the K\"{a}hler potential and the superpotential,
respectively. The subscript $X$ ($\overline{X}$) denotes
differentiation with respect to the $X$ ($\overline{X}$) field and
$\langle \, ... \; \rangle$ stands for the VEV. In general one
has~\cite{Kawasaki:2006gs,Kawasaki:2006hm,
Asaka:2006bv,Endo:2006zj,Nakamura:2006uc,Endo:2006tf,Endo:2006qk}
\begin{align}
 \alpha_{3/2} = c \; \dfrac{\left|\left\langle X
 \right\rangle\right|}{M_P}~,
\end{align}
where the value of the constant $c$ depends on the explicit model.
For example, in the scenario discussed in
Ref.~\cite{Kawasaki:2006gs}, one has $c=\sqrt{3}$. For a moduli
field, $\alpha_{3/2}$ is expected to be order
unity~\cite{Asaka:2006bv}, and hence the decay of the moduli field
into a pair of gravitinos can be significant, thus leading to
cosmological disasters. If $X$ is the inflaton, then $\alpha_{3/2}$
can take values much smaller than unity, as in some new and hybrid
inflationary models. If $\alpha_{3/2}$ is zero then no gravitinos
are produced in the decay of $X$. This is what happens in the
simplest chaotic inflationary models, for which $G_X =0$ in the
vacuum (see Ref.~\cite{Asaka:2006bv,Kawasaki:2006hm,Endo:2006qk}).
In the following we will leave $\alpha_{3/2}$ as a free parameter.

One should note that the expression in Eq.~(\ref{eq:Gamma32}) for
the decay rate of the gravitino cannot be applicable for
$H>m_{3/2}$. As pointed out in \cite{Kawasaki:2006hm}, the decay
proceeds only if the Hubble parameter is smaller than the gravitino
mass, since the chirality flip of the gravitino forbids the decay to
proceed otherwise. This imposes a bound on the maximum value that
$M_X$ can take, and also on the maximum reheating temperature. The
condition $H(T_{rh})<m_{3/2}$ is, by the definition of $T_{rh}$
[cf.~Eq.~(\ref{eq:rh})], the same as $\Gamma_X<m_{3/2}$. Hence
\begin{align}
2.5 \times 10^5 \,{\rm GeV}\,\alpha_{t}^{-2/3} \lesssim M_X
\lesssim 5.3 \times 10^{13}\,{\rm GeV}\,\alpha_t^{-2/3} \,
\left(\dfrac{m_{3/2}}{10^{3}~{\rm
GeV}}\right)^{1/3}~,\label{eq:Mxlowupbound}
\end{align}
where the first inequality is the one already presented in
Eq.~(\ref{eq:Mxlowbound}). The upper bound on the reheating
temperature turns out to be
\begin{align}
T_{rh} \lesssim 4.2 \times 10^{7}\,{\rm GeV}
\left(\dfrac{m_{3/2}}{10^{3}~{\rm GeV}}\right)^{1/4}
\left(\dfrac{M_5}{10^{10}~{\rm
GeV}}\right)^{3/4}\label{eq:TrhboundHlargem32BC}
\end{align}
in the high-energy brane regime, and
\begin{align}
T_{rh} \lesssim 2.2 \times 10^{10}\,{\rm GeV}
\left(\dfrac{m_{3/2}}{10^{3}~{\rm
GeV}}\right)^{1/2}\label{eq:TrhboundHlargem32SC}
\end{align}
in the SC limit.

Summarizing, one would expect that the brane effects will change
the gravitino production for reheating temperatures in the range
\begin{align}
8.1 \times 10^4\,{\rm GeV}\,\left(\dfrac{M_5}{10^{10}\,{\rm
GeV}}\right)^{3/2} \lesssim T_{rh} \lesssim  4.2 \times
10^{7}\,{\rm GeV} \left(\dfrac{m_{3/2}}{10^{3}~{\rm
GeV}}\right)^{1/4} \left(\dfrac{M_5}{10^{10}~{\rm
GeV}}\right)^{3/4} \label{eq:TrhBCbounds}
\end{align}
or, equivalently, heavy scalar masses in the interval
\begin{align}
\left(\dfrac{2}{\alpha_{t}^2}\right)^{1/3} M_5\, \lesssim \,M_X\,
\lesssim \,5.3 \times 10^{13}\,{\rm GeV}\,\alpha_t^{-2/3} \,
\left(\dfrac{m_{3/2}}{10^{3}~{\rm
GeV}}\right)^{1/3}~.\label{eq:MxBCbounds}
\end{align}

\section{Gravitino Abundance}

As referred in the Introduction, a viable cosmological
supergravity-inspired scenario has to avoid the so-called gravitino
problem. For unstable gravitinos, their decay products should not
alter the BBN predictions for the abundance of light elements in the
universe. On the other hand, if they are stable particles they will
contribute to the dark matter content, and thus one has to avoid the
overclosure of the universe.

In the following we will estimate the contributions to the gravitino
abundance, both from thermal production and from the decay of a
heavy scalar. Then we will discuss the constraints coming for stable
or unstable gravitinos.

\subsection{Brane cosmology vs standard cosmology}

Gravitinos can be produced by the decay of the heavy scalar particle
$X$, and, during the reheating phase, they are also thermally
produced through scatterings in the plasma. The total abundance is
then given by both contributions
\begin{align}
Y_{3/2} \equiv \frac{n_{3/2}}{s} = Y_{3/2}^{Th} +
Y_{3/2}^X~,\label{eq:grav_total}
\end{align}
where $s(T)=2 \pi^2 g_* T^3/45$ is the entropy density, and
$Y_{3/2}^{Th (X)}$ stands for the contribution from thermal
scatterings ($X$ decay) to the total gravitino abundance\footnote{In
fact, there are other potential sources of gravitinos in addition to
these two~\cite{Asaka:2006bv}. The gravitinos can be produced in the
decay $X \rightarrow \widetilde{X} + \psi_{3/2}$, where
$\widetilde{X}$ is the fermionic partner of $X$, however for this to
happen is required a large mass hierarchy between $X$ and
$\widetilde{X}$~\cite{Kohri:2004qu}. Also, the decay of $X$ usually
produces superparticles, followed by cascade decays into gravitinos.
Finally, gravitinos can be produced before the decay of $X$ in such
a way that their abundance remains sizable after the dilution by the
reheating. In this work, we will not consider any of these
contributions.}.

Let us estimate the contribution of thermal scatterings at
reheating. The gravitino abundance at a given temperature $T <
T_{rh}$ is obtained from the Boltzmann equation
\begin{align}
\frac{d n_{3/2}}{dt} + 3 H n_{3/2} = C_{3/2} (T)~, \label{eq:bolt}
\end{align}
where $C_{3/2}(T)$ is the collision term given by \cite{Bolz:2000fu}
\begin{align}
C_{3/2}(T)\simeq \alpha (T)\left( 1+ \beta (T) \frac{m_{{\tilde
g}}^2}{m_{3/2}^2}\right) \frac{T^6}{M_P^2}~, \label{eq:c32}
\end{align}
where $m_{{\tilde g}}$ is the low-energy gluino mass; $\alpha (T)$
and $\beta (T)$ are slowly-varying functions of the temperature.
Assuming constant entropy, the integration of Eq.~(\ref{eq:bolt})
yields
\begin{align}
Y_{3/2}^{Th}\simeq {\cal A}(T;m_{3/2},m_{{\tilde
g}})\left(2\,T_t-\dfrac{T_t^2}{T_{rh}}-T\right)~,
\label{eq:gravabundTh}
\end{align}
with
\begin{align}
{\cal A}(T;m_{3/2},m_{{\tilde g}})=\left(\dfrac{45}{\pi
g_*}\right)^{3/2}  \dfrac{\alpha (T)}{4 \pi^2 M_P}\left( 1+ \beta
(T) \frac{m_{{\tilde g}}^2}{m_{3/2}^2}\right)~. \label{eq:AT}
\end{align}
This is an approximate solution, obtained by integrating the
Boltzmann equation in the BC regime for $T_{rh}>T>T_t$ and the SC
limit for $T<T_t$, considering a small dependence of ${\cal
A}(T;m_{3/2},m_{{\tilde g}})$ on the temperature. In
Ref.~\cite{Okada:2004mh}, an analytic solution in terms of a Gauss
hypergeometric function was found. However the result presented in
Eqs.~(\ref{eq:gravabundTh}) and (\ref{eq:AT}) is, for our purposes,
a very good approximation.

In the high energy limit of BC, Eq.~(\ref{eq:gravabundTh}) leads to
\begin{align}
Y_{3/2}^{Th} \simeq 2 \,{\cal A}(T_t;m_{3/2},m_{{\tilde g}})
\,T_t~,\label{eq:ythbc}
\end{align}
while in the SC limit one obtains the usual result
\begin{align}
Y_{3/2}^{Th} \simeq {\cal A}(T_{rh};m_{3/2},m_{{\tilde g}})
\,T_{rh}~.\label{eq:ythsc}
\end{align}
In the BC regime, the abundance of gravitinos produced by thermal
scatterings is proportional to the transition temperature instead of
the reheating temperature. This has been proposed as a possible
solution for the thermal gravitino problem in the context of
supersymmetric thermal leptogenesis~\cite{Okada:2004mh}: in SC, the
gravitino bound imposed on the reheating temperature,
$T_{rh}\lesssim 10^6-10^{10}$~GeV (depending on the gravitino mass
and the hadronic branching
ratio~\cite{Kohri:2005wn,Kawasaki:2006hm,Endo:2006qk}), is barely
compatible with the simplest models of supersymmetric thermal
leptogenesis which need $T_{rh}\gtrsim 10^{7}-10^{9}$~GeV.

Let us now look at the contribution of the decay
process~(\ref{eq:decay}) to the total gravitino abundance [cf.
Eq.~(\ref{eq:grav_total})]. The amount of gravitinos produced by
this process can be written as
\begin{align}
Y_{3/2}^{X} = \dfrac{3}{2}\, B_{3/2} \,\dfrac{T_{rh}}{M_X}~,
\label{eq:gravabundX1}
\end{align}
where
\begin{align}
B_{3/2} \equiv \dfrac{\Gamma_{3/2}}{\Gamma_X}=
\dfrac{\alpha_{3/2}^2}{36 \,\alpha_{t}^2}\label{eq:B32_1}
\end{align}
is the branching ratio of the decay channel (\ref{eq:decay}). If
$\alpha_{3/2} \ll 1$ (or/and $\alpha_{t} \gg 1$), $B_{3/2}$ can be
much smaller than unity.

The abundance of gravitinos produced from the decay of $X$ can
then be written as
\begin{align}
Y_{3/2}^{X} = \dfrac{\alpha_{3/2}^2}{24\,\alpha_{t}^2}
\,\dfrac{T_{rh}}{M_X} = \dfrac{\alpha_{3/2}^2\, M_X^2\,T_{rh} }
{24 \, H(T_{rh})\,M_P^2}~.\label{eq:gravabundX2}
\end{align}
Notice that, in the brane scenario, $H(T_{rh})$ is in fact a
function of $T_{rh}$ and $T_t$ (or $M_5$). In the BC and SC limits,
this equation leads, respectively, to
\begin{align}
Y_{3/2}^{X} &= 8.9 \times 10^{-11}\,\alpha_{3/2}^2
\left(\dfrac{M_X}{10^{10}~{\rm GeV}}\right)^2
\left(\dfrac{M_5}{10^{10}~{\rm GeV}}\right)^3
\left(\dfrac{T_{rh}}{10^{6}~{\rm GeV}}\right)^{-3} \nonumber\\
&=1.0 \times 10^{-5}\,B_{3/2}\,\alpha_t^{1/2}
\left(\dfrac{M_5}{10^{10}~{\rm GeV}}\right)^{3/4}
\left(\dfrac{M_X}{10^{10}~{\rm
GeV}}\right)^{-1/4}~,\label{eq:YxtrBC}
\end{align}
and
\begin{align}
Y_{3/2}^{X} &= 1.4 \times 10^{-8}\,\alpha_{3/2}^2
\left(\dfrac{M_X}{10^{10}~{\rm GeV}}\right)^2
\left(\dfrac{T_{rh}}{10^{6}~{\rm GeV}}\right)^{-1} \nonumber\\
&=8.6 \times 10^{-6}\,B_{3/2}\,\alpha_t
\left(\dfrac{M_X}{10^{10}~{\rm GeV}}\right)^{1/2}~.\label{eq:YxtrSC}
\end{align}
Note that in both regimes the gravitino abundance is inversely
proportional to $T_{rh}$, in opposition to the thermal abundance
in SC.

%-------------------------------------------------------------
\begin{figure*}[t]
\includegraphics[width=7cm]{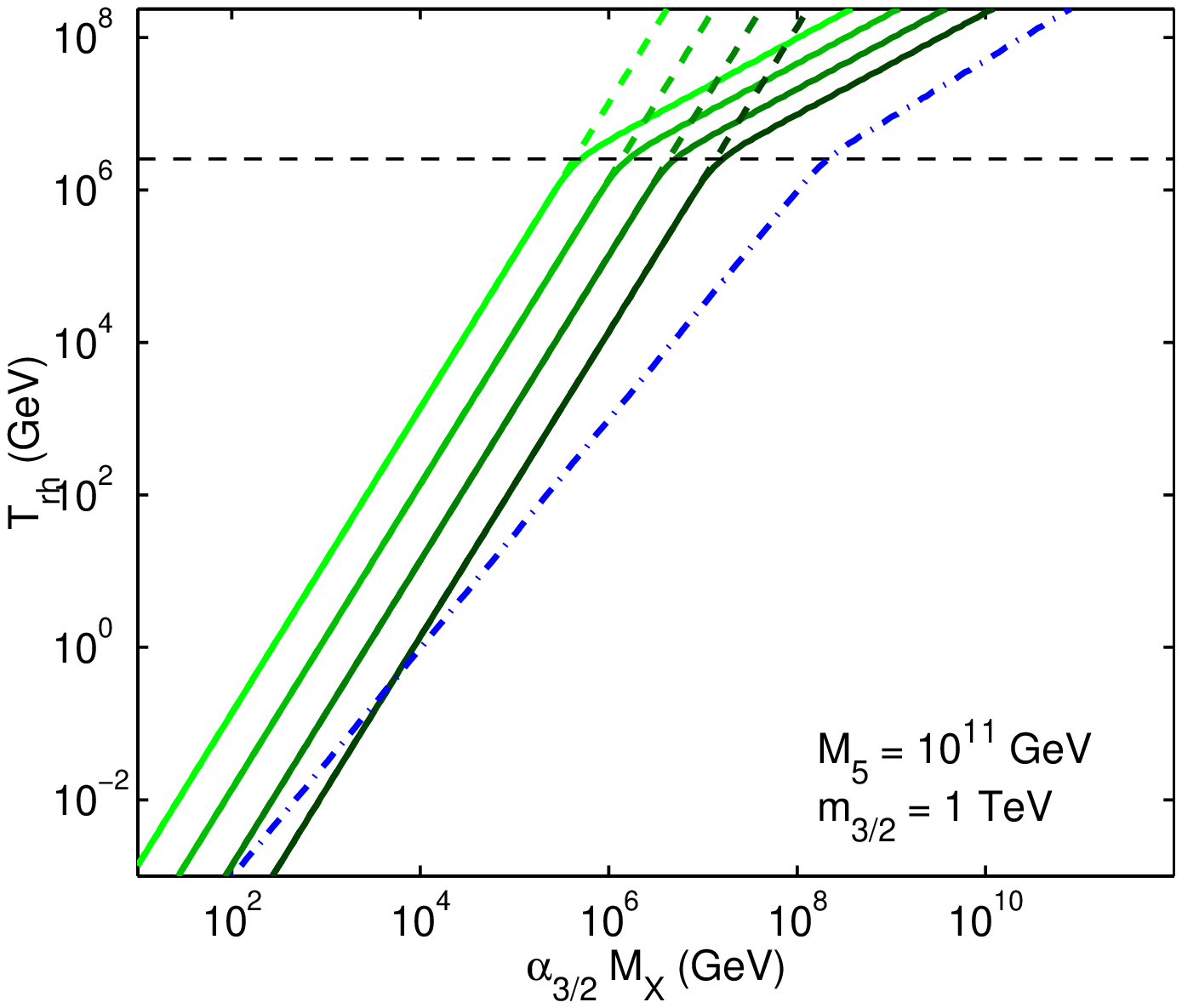}
\includegraphics[width=7cm]{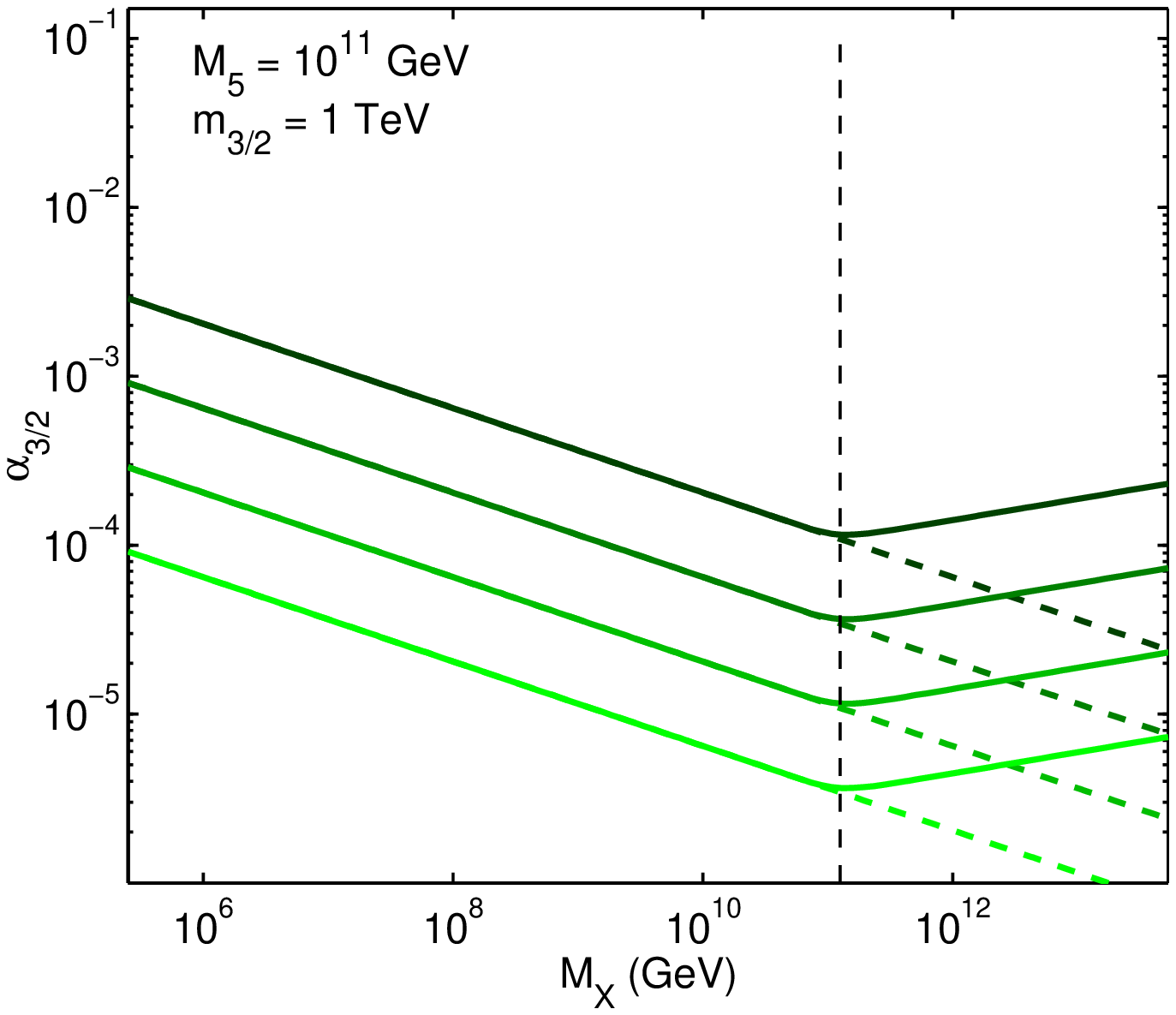}
\caption{(Color online) Contours of constant
$Y_{3/2}^{X}=10^{-17},\,10^{-16},\,10^{-15},\,10^{-14}$ (from
lighter to darker green) in the $T_{rh}- \alpha_{3/2}M_X$ and
$\alpha_{3/2}-M_X$ planes, for BC with $M_5=10^{11}$~GeV (full
lines) and SC (dashed lines). The (black) horizontal dashed line
in the left panel represents the value of $T_t$; the (black)
vertical line in the right panel shows the value of $M_X =
\left(2/\alpha_{t}^2\right)^{1/3} M_5$. The (blue) dash-dotted
line represents the lower bound on $T_{rh}$ from $B_{3/2}\leq 1$
for $\alpha_{3/2}=10^{-10}$. The gravitino mass is taken to be
$m_{3/2}=1$~TeV and, in the right panel, we fixed $\alpha_{t}=1$.
The maximum values for $T_{rh}$ and $M_X$ shown in the plots are
the ones given by the upper bounds of Eqs.~(\ref{eq:TrhBCbounds})
and (\ref{eq:MxBCbounds}), respectively.} \label{fig:Yxcontours}
\end{figure*}
%-----------------------------------------------------------

%-----------------------------------------------------------
\begin{figure*}[t]
\includegraphics[width=7cm]{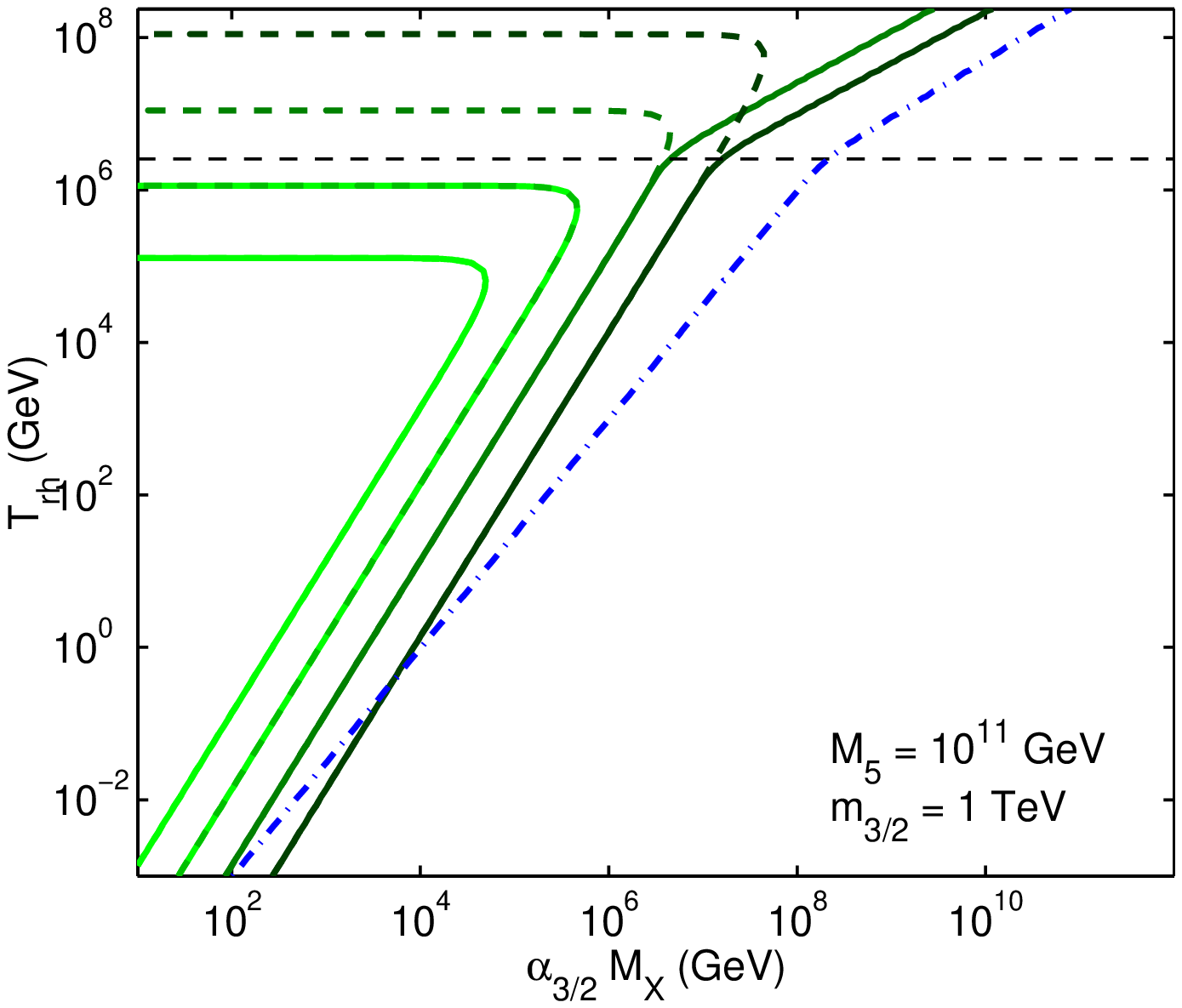}
\includegraphics[width=7cm]{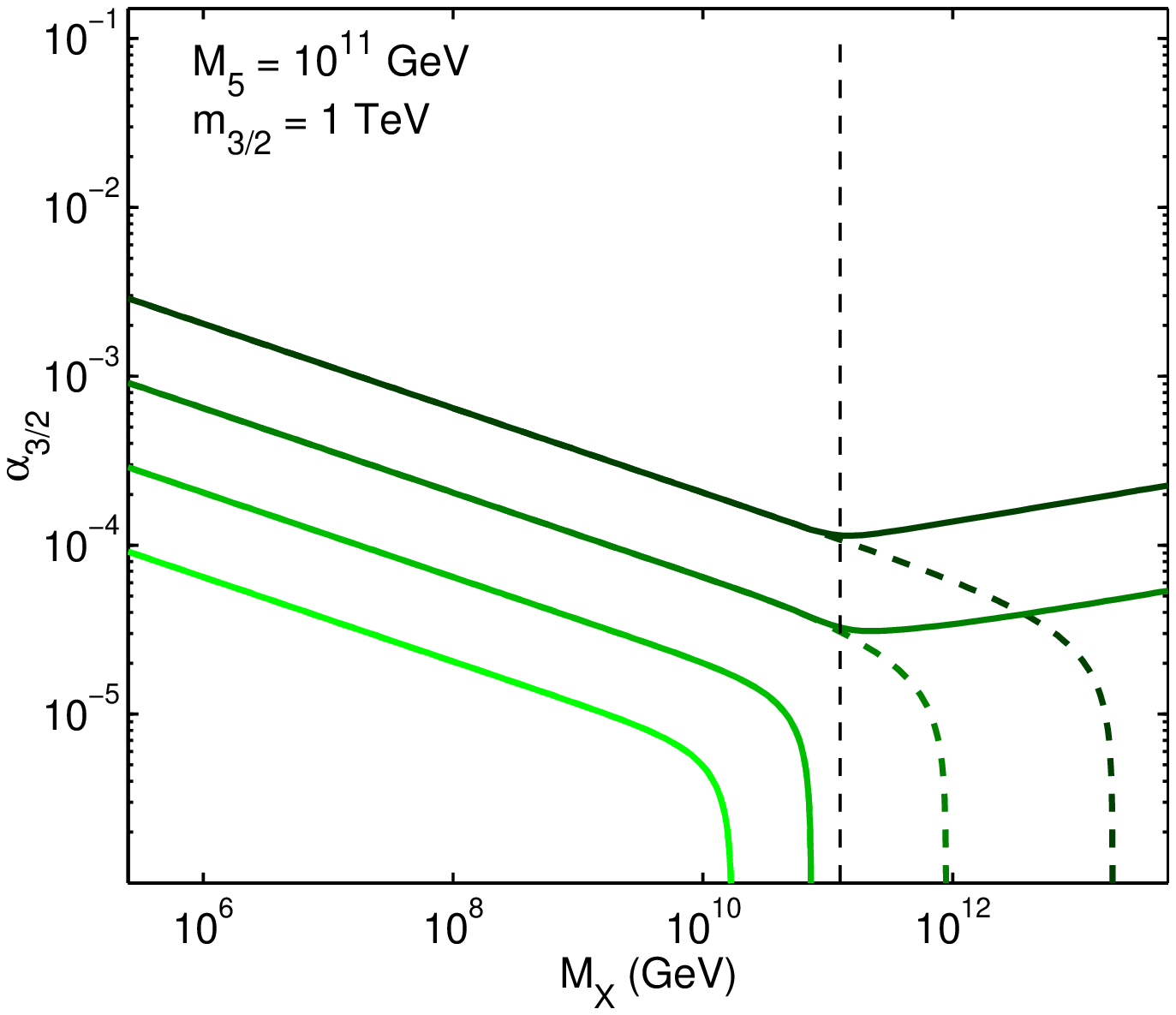}
\caption{(Color online) Same as in Fig.~\ref{fig:Yxcontours}, but
for the total abundance of gravitinos:
$Y_{3/2}=10^{-17},\,10^{-16},\,10^{-15},\,10^{-14}$ (from lighter
to darker green). In the computation of the thermal abundance, the
gluino mass was taken to be $m_{{\tilde g}}=1$~TeV.}
\label{fig:Ytotcontours}
\end{figure*}
%---------------------------------------------------------------

In Fig.~\ref{fig:Yxcontours} we present the contours of constant
$Y_{3/2}^X$ in the $T_{rh}- \alpha_{3/2}M_X$ and $\alpha_{3/2}-M_X$
planes. We show the SC case (dashed lines) and the BC one (full
lines), for a fixed value of $M_5$ ($M_5=10^{11}$~GeV). We have
chosen the gravitino mass to be $m_{3/2}=1$~TeV. In the right panel,
we see the change of behavior in the contours for $M_X >
\left(2/\alpha_{t}^2\right)^{1/3} M_5$, for which the high-energy
regime of BC is valid. This regime occurs for $T_{rh}$ larger than
$T_t$, as we can see on the left panel. For the contour plot in the
$\alpha_{3/2}-M_X$ plane we have chosen $\alpha_t=1$. One should
notice that if we fix $\alpha_t$ at a smaller value, then the lines
of constant $Y_{3/2}^X$ would move downwards, i.e. only smaller
values of $\alpha_{3/2}$ would be allowed.

In the left panel plot, the lower bound on $T_{rh}$ from
$B_{3/2}\leq 1$ is also shown by choosing $\alpha_{3/2}=10^{-10}$
(dash-dotted line). This bound can be derived from
Eq.~(\ref{eq:B32_1}), which can also be rewritten as
\begin{align}
B_{3/2} = \dfrac{\alpha_{3/2}^2\,M_X^3}{36
\,H(T_{rh})\,M_P^2}~,\label{eq:B32_2}
\end{align}
and that in the high-energy limit of BC reads
\begin{align}
B_{3/2} = 5.9 \times 10^{-7} \, \alpha_{3/2}^2
\left(\dfrac{M_5}{10^{10}~{\rm GeV}}\right)^3
\left(\dfrac{M_X}{10^{10}~{\rm GeV}}\right)^3
\left(\dfrac{T_{rh}}{10^{6}~{\rm GeV}}\right)^{-4}~.
\end{align}
This leads to
\begin{align}
T_{rh} \geq 4.9 \,
\alpha_{3/2}^{-1/4}\,\left(\dfrac{M_5}{10^{10}~{\rm
GeV}}\right)^{3/4}\,\left(\dfrac{\alpha_{3/2}\,M_X}{10^{5}~{\rm
GeV}}\right)^{3/4}~.\label{eq:TrhboundB32BC}
\end{align}
In SC one has
\begin{align}
B_{3/2} = 9.1 \times 10^{-5} \, \alpha_{3/2}^2
\left(\dfrac{M_X}{10^{10}~{\rm GeV}}\right)^3
\left(\dfrac{T_{rh}}{10^{6}~{\rm GeV}}\right)^{-2}~,
\end{align}
and hence
\begin{align}
T_{rh} \geq 3.0 \times 10^{-4}\,
\alpha_{3/2}^{-1/2}\,\left(\dfrac{\alpha_{3/2}\,M_X}{10^{5}~{\rm
GeV}}\right)^{3/2}~.\label{eq:TrhboundB32SC}
\end{align}
Of course, if one requires a smaller value for $B_{3/2}$, a larger
value of the reheating temperature is needed. And if $\alpha_{3/2}$
is larger, the bound will be less stringent, in both cases, for a
fixed $\alpha_{3/2}\,M_X$.

In Fig.~\ref{fig:Ytotcontours} we show the contours of constant
$Y_{3/2} = Y_{3/2}^{Th} + Y_{3/2}^X$ in the same plane as in
Fig.~\ref{fig:Yxcontours}, assuming $m_{{\tilde g}}=1$~TeV in the
computation of the thermal contribution to the gravitino abundance
(hereafter we will consider this mass for the gluino). In the right
panel, we see that in the SC regime, i.e. for $M_X <
\left(2/\alpha_{t}^2\right)^{1/3} M_5$, the heavy scalar mass $M_X$
has an upper bound (coming from the thermal production
contribution), which corresponds to the upper bound on the reheating
temperature (see left panel). These bounds can be easily derived
from Eq.~(\ref{eq:ythsc}), which can be rewritten as \footnote{For
$m_{{\tilde g}}=1$~TeV, $m_{3/2}=1-10^2$~TeV and
$T=10^4-10^{12}$~GeV one has ${\cal A}(T;m_{3/2},m_{{\tilde g}})\sim
10^{-22}$~GeV$^{-1}$. }
\begin{align}
Y_{3/2}^{th} &\simeq 10^{-16} \left[\dfrac{{\cal A}(T_{rh};m_{3/2}
,m_{{\tilde g}})}{10^{-22}~{\rm GeV}^{-1}}\right]
\left(\dfrac{T_{rh}}{10^{6}~{\rm GeV}}\right) \nonumber\\
&\simeq 5.7 \times 10^{-18} \,\alpha_t \,\left[\dfrac{{\cal
A}(T_{rh};m_{3/2} ,m_{{\tilde g}})}{10^{-22}~{\rm GeV}^{-1}}\right]
\left(\dfrac{M_X}{10^{10}~{\rm GeV}}\right)^{3/2} ~,
\end{align}
from which one gets
\begin{align}
T_{rh} \lesssim 10^6~{\rm GeV} \left[\dfrac{{\cal A}(T_{rh};m_{3/2}
,m_{{\tilde g}})}{10^{-22}~{\rm GeV}^{-1}}\right]^{-1}
\left(\dfrac{Y_{3/2}^{BBN}}{10^{-16}}\right) ~,
\label{eq:TrhboundYbbnSC}
\end{align}
and
\begin{align}
M_X \lesssim 6.7 \times 10^{10}~{\rm GeV}\,\alpha_t^{-2/3}\,
\left[\dfrac{{\cal A}(T_{rh};m_{3/2} ,m_{{\tilde g}})}{10^{-22}~{\rm
GeV}^{-1}}\right]^{-2/3}
\left(\dfrac{Y_{3/2}^{BBN}}{10^{-16}}\right)^{2/3}~.\label{eq:MxboundYbbnSC}
\end{align}

On the other hand, in order to avoid the overproduction of
gravitinos by $X$ decays, the coupling constant $\alpha_{3/2}$ has
to satisfy
\begin{align}
\alpha_{3/2} \lesssim 2.1 \times 10^{-5}\,\alpha_t^{1/2}
\left(\dfrac{M_X}{10^{10}~{\rm GeV}}\right)^{-1/4}
\left(\dfrac{Y_{3/2}^{BBN}}{10^{-16}}\right)^{1/2}~,\label{eq:alpha32boundYbbnSC}
\end{align}
or, in terms of the branching ratio,
\begin{align}
B_{3/2} \lesssim 1.2 \times 10^{-11}\,\alpha_t^{-1}
\left(\dfrac{M_X}{10^{10}~{\rm GeV}}\right)^{-1/2}
\left(\dfrac{Y_{3/2}^{BBN}}{10^{-16}}\right)
~,\label{eq:B32boundYbbnSC}
\end{align}
as already derived in \cite{Asaka:2006bv}.

For $M_X \gtrsim \left(2/\alpha_{t}^2\right)^{1/3} M_5$, or
equivalently $T_{rh}>T_t$, the high-energy regime of BC holds. In
this regime, the upper bound on the $T_{rh}$ disappears. Indeed,
this bound is due to the thermal production of gravitinos, and, as
can be seen from Eq.~(\ref{eq:ythbc}), the abundance of gravitinos
produced thermally is proportional to $T_t$,
\begin{align}
Y_{3/2}^{th} &\simeq 2\times10^{-16} \left[\dfrac{{\cal
A}(T_{t};m_{3/2} ,m_{{\tilde g}})}{10^{-22}~{\rm GeV}^{-1}}\right]
\left(\dfrac{T_{t}}{10^{6}~{\rm GeV}}\right) \nonumber \\
&\simeq 1.6 \times 10^{-17} \,\left[\dfrac{{\cal A}(T_{t};m_{3/2}
,m_{{\tilde g}})}{10^{-22}~{\rm GeV}^{-1}}\right]
\left(\dfrac{M_5}{10^{10}~{\rm GeV}}\right)^{3/2}~.
\end{align}
Hence, in the high-energy regime the bound on the reheating
temperature is replaced by a bound on the transition temperature,
\begin{align}
T_{t} \lesssim 5 \times 10^5~{\rm GeV} \left[\dfrac{{\cal
A}(T_{t};m_{3/2} ,m_{{\tilde g}})}{10^{-22}~{\rm
GeV}^{-1}}\right]^{-1}
\left(\dfrac{Y_{3/2}^{BBN}}{10^{-16}}\right) ~,
\end{align}
or, equivalently, on the 5D fundamental mass $M_5$
\begin{align}
M_5 \lesssim 3.4 \times 10^{10}\,{\rm GeV}\, \left[\dfrac{{\cal
A}(T_{t};m_{3/2} ,m_{{\tilde g}})}{10^{-22}~{\rm
GeV}^{-1}}\right]^{-2/3}
\left(\dfrac{Y_{3/2}^{BBN}}{10^{-16}}\right)^{2/3}~.\label{eq:M5boundYbbn}
\end{align}

This is also reflected in the disappearance of the upper bound on
the mass of the heavy scalar imposed by the limits on the gravitino
abundance, and the only bound that remains on this quantity is the
one in Eq.~(\ref{eq:Mxlowupbound}), which comes from requiring
$m_{3/2}> H$. In fact, in this regime, in order to avoid
overproduction of gravitinos in the $X$ decay, one must have
\begin{align}
T_{rh} \gtrsim 4.5 \times 10^{4}
\left(\dfrac{\alpha_{3/2}\,M_X}{10^{5}~{\rm GeV}}\right)^{2/3}
\left(\dfrac{M_5}{10^{10}~{\rm GeV}}\right)
\left(\dfrac{Y_{3/2}^{BBN}}{10^{-16}}\right)^{-1/3}~,\label{eq:TrhboundMXBC}
\end{align}
easily obtained from Eq.~(\ref{eq:YxtrBC}). This means that, like
in SC case, the $X$ decay into gravitinos imposes a lower bound on
the reheating temperature. In SC one must have [cf.
Eq.~(\ref{eq:YxtrSC})]
\begin{align}
T_{rh} \gtrsim  1.4 \times 10^{4}
\left(\dfrac{\alpha_{3/2}\,M_X}{10^{5}~{\rm GeV}}\right)^{2}
\left(\dfrac{Y_{3/2}^{BBN}}{10^{-16}}\right)^{-1}~.\label{eq:TrhboundMXSC}
\end{align}
The lower bound coming from $B_{3/2}$, presented in
Eqs.~(\ref{eq:TrhboundB32BC}) and (\ref{eq:TrhboundB32SC}), can be
stronger than those in Eqs.~(\ref{eq:TrhboundMXBC}) and
(\ref{eq:TrhboundMXSC}), depending on the value of $B_{3/2}$ and/or
$\alpha_{3/2}$. One finds that the bounds from $B_{3/2}$ can be
disregarded, if one considers sufficiently large $\alpha_{3/2}$.

We should also notice that in the high energy regime the upper bound
on the coupling constant $\alpha_{3/2}$ becomes
\begin{align}
\alpha_{3/2} \lesssim  1.9 \times 10^{-5}\,\alpha_t^{3/4}
\left(\dfrac{M_5}{10^{10}~{\rm GeV}}\right)^{-3/8}
\left(\dfrac{M_X}{10^{10}~{\rm GeV}}\right)^{1/8}
\left(\dfrac{Y_{3/2}^{BBN}}{10^{-16}}\right)^{1/2}~,\label{eq:alpha32boundYbbnBC}
\end{align}
and the one on the branching ratio
\begin{align}
B_{3/2} \lesssim 9.8 \times 10^{-12}\,\alpha_t^{-1/2}
\left(\dfrac{M_5}{10^{10}~{\rm GeV}}\right)^{-3/4}
\left(\dfrac{M_X}{10^{10}~{\rm GeV}}\right)^{1/4}
\left(\dfrac{Y_{3/2}^{BBN}}{10^{-16}}\right)
~.\label{eq:B32boundYbbnBC}
\end{align}
As in SC [cf. Eqs.~(\ref{eq:alpha32boundYbbnSC}) and
(\ref{eq:B32boundYbbnSC})], a very low value for $\alpha_{3/2}$ (or
equivalently for $B_{3/2}$) is also required in the high-energy
regime. This implies that the decay of the heavy scalar into
gravitinos is very constrained. We will comment more on this in the
conclusion.

In the following we will discuss in more detail the additional
cosmological constraints on the gravitino amount produced at the
reheating by the $X$ decay. Namely, we will use BBN constraints on
the amount of unstable gravitinos to impose bounds on the relevant
quantities. Afterwards, we will discuss the case when the gravitino
is stable and use the WMAP limits on the amount of cold dark matter
to derive constraints.

\subsection{Unstable gravitino and BBN constraints}

Let us start with the unstable gravitino case. In conventional
scenarios, the gravitino mass is expected to be comparable to the
masses of the supersymmetric partners of the standard model
particles and, therefore, $m_{3/2}\lesssim$ a few TeV in order to
solve the gauge hierarchy problem. Since the gravitino coupling to
matter is suppressed by the Planck mass $M_{P}$ , its lifetime is
$\tau_{3/2} \sim M_{P}^{2}/m_{3/2}^{3} \sim 10^{8}(100$~GeV
$/m_{3/2})^{3}$~s. For a gravitino mass $\gtrsim 100$~GeV, it decays
during or soon after the BBN epoch. Therefore, if gravitinos are
overabundant, the high energetic photons and hadrons emitted in
their decays may destroy the light elements and hence spoil the
success of BBN. However, it has been recently pointed
out~\cite{Kohri:2005wn} that the hadronic decay gives a more
stringent constraint on the abundance of gravitinos, because mesons
and nucleons produced in the decay and subsequent hadronization
processes significantly affect BBN. For $m_{3/2}=1$~TeV, the bounds
that come from BBN are~\cite{Kawasaki:2006hm,Endo:2006qk}
\begin{align}
Y_{3/2}^{BBN} \lesssim \left\{
\begin{array}{l}
  4 \times 10^{-17} \quad {\rm for} \quad B_h=1\quad {\rm (Case~I)}~, \\ % from Endo:2006qk
  5 \times 10^{-14} \quad {\rm for} \quad B_h=10^{-3}~, \\ % from Kawasaki:2006hm
\end{array}\right.
\end{align}
where $B_h$ denotes the hadronic branching ratio of the gravitino.
While, for $m_{3/2}=10$~TeV one
has~\cite{Kawasaki:2006hm,Endo:2006qk}
\begin{align}
Y_{3/2}^{BBN} \lesssim \left\{
\begin{array}{l}
  2 \times 10^{-14} \quad {\rm for} \quad B_h=1~, \\ % from Kawasaki:2006hm
  2 \times 10^{-13} \quad {\rm for} \quad B_h=10^{-3}\quad {\rm (Case~II)}~. \\ % from Kawasaki:2006hm
\end{array}\right.
\end{align}
The region $m_{3/2} \gtrsim 10^3$~TeV is disfavored since large
gravitinos masses do not solve the gauge hierarchy problem, as
mentioned before. The BBN bound disappears for $m_{3/2} \sim
10^2-10^3$~TeV and, for this mass range, the constraint comes from
the abundance of the relic LSP produced by the gravitino
decay~\cite{Nakamura:2006uc,Kawasaki:2006hm}.

%--------------------------------------------------
\begin{figure}[t]
\includegraphics[width=7cm]{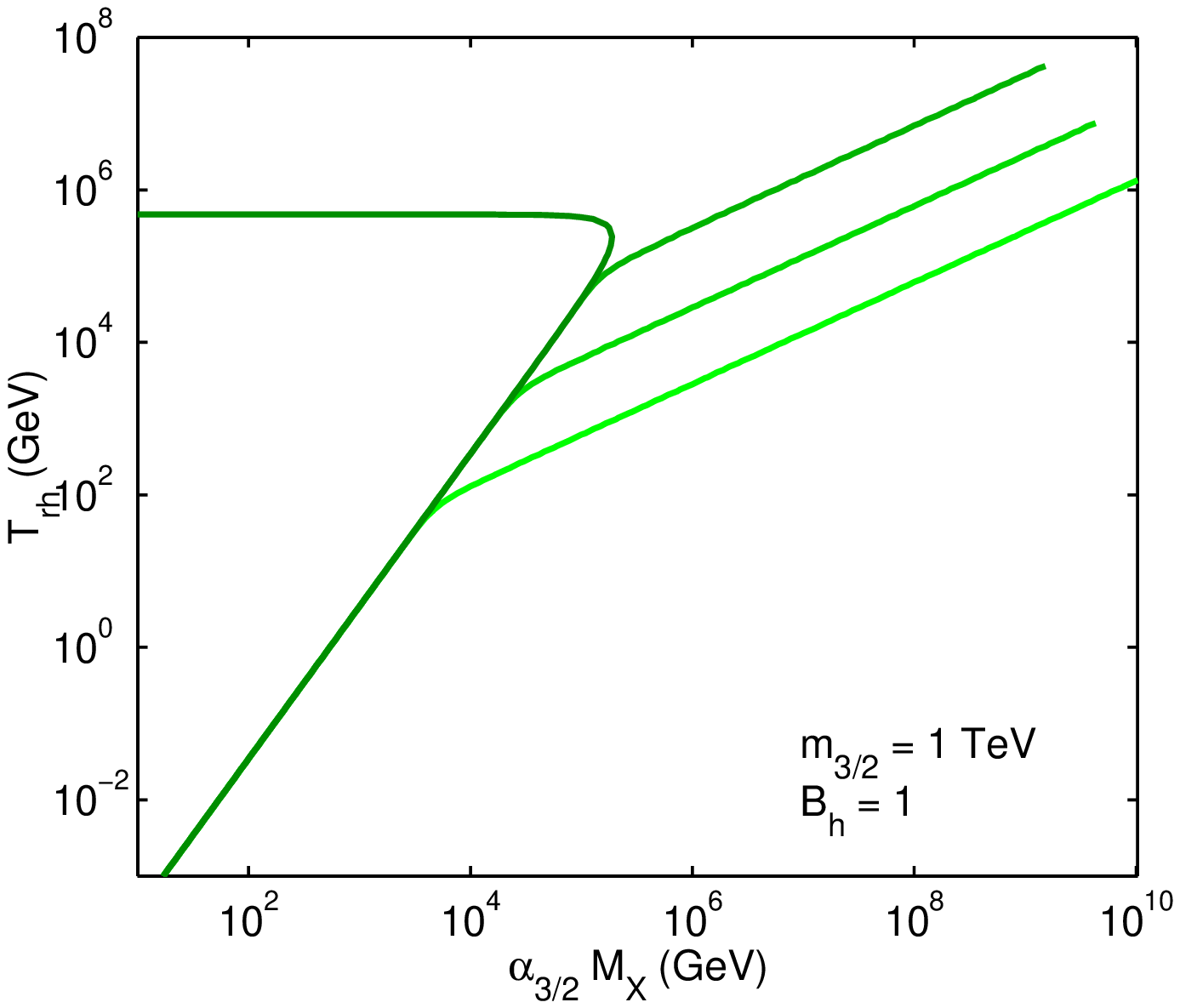}
\includegraphics[width=7cm]{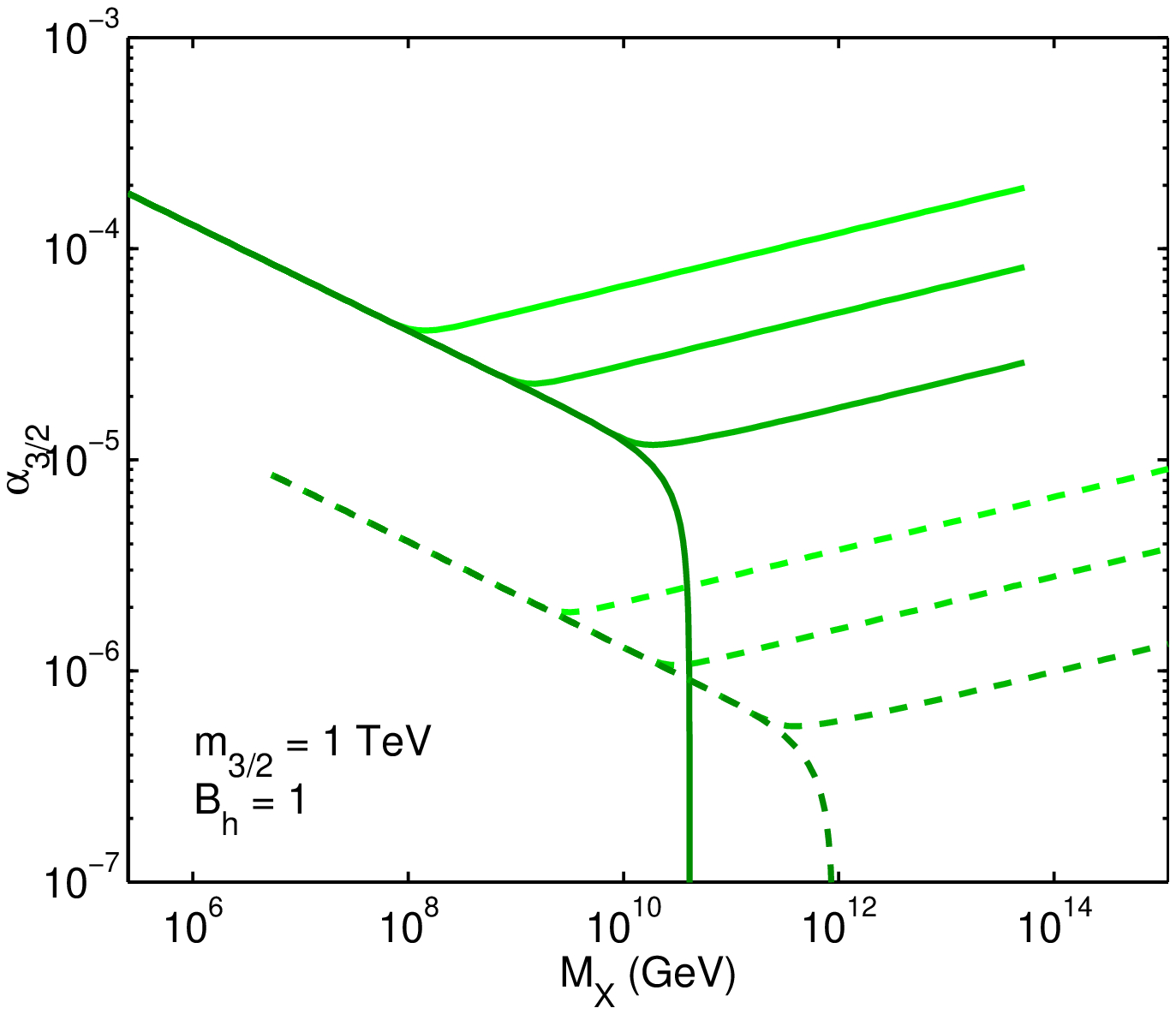}
\includegraphics[width=7cm]{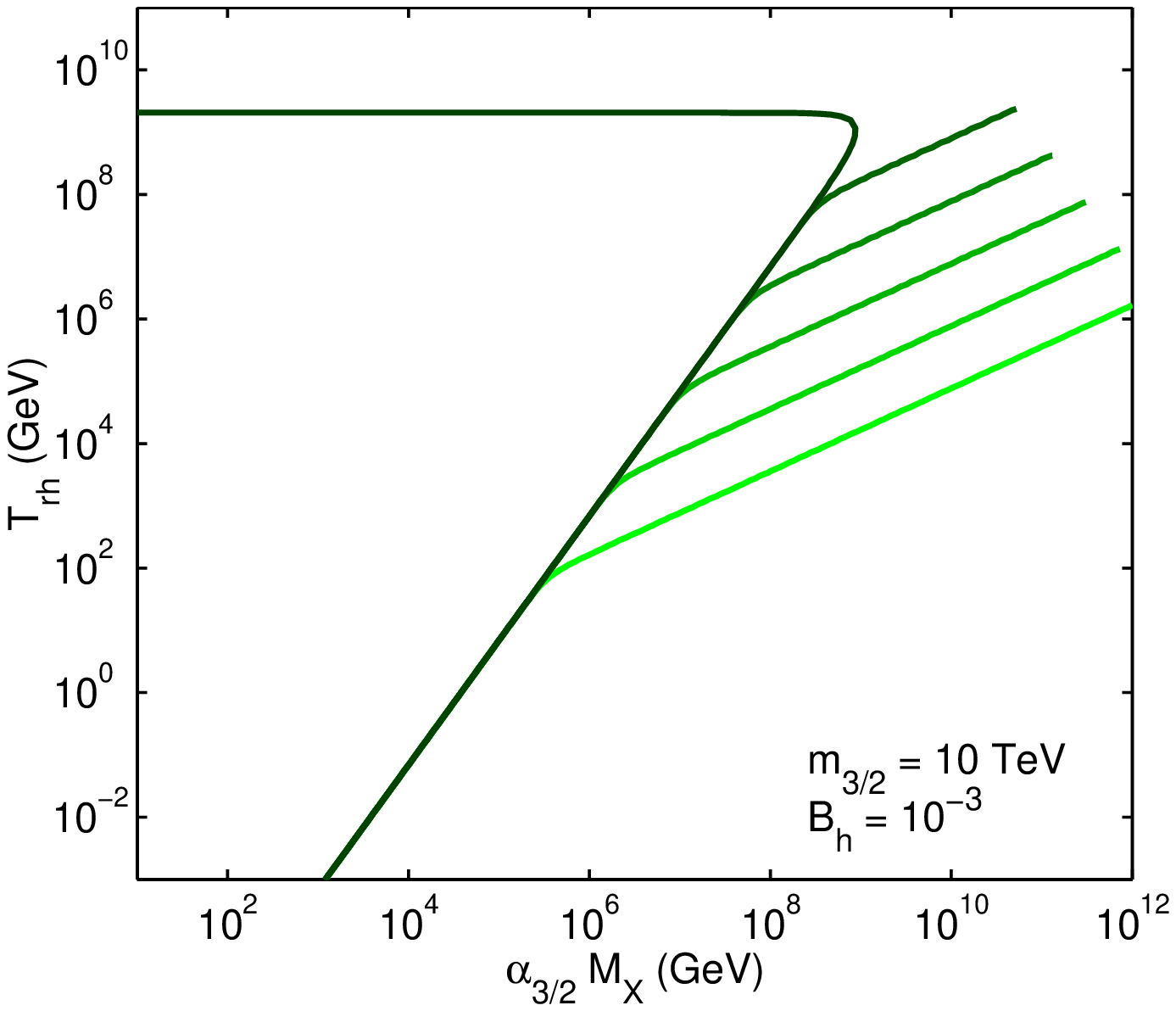}
\includegraphics[width=7cm]{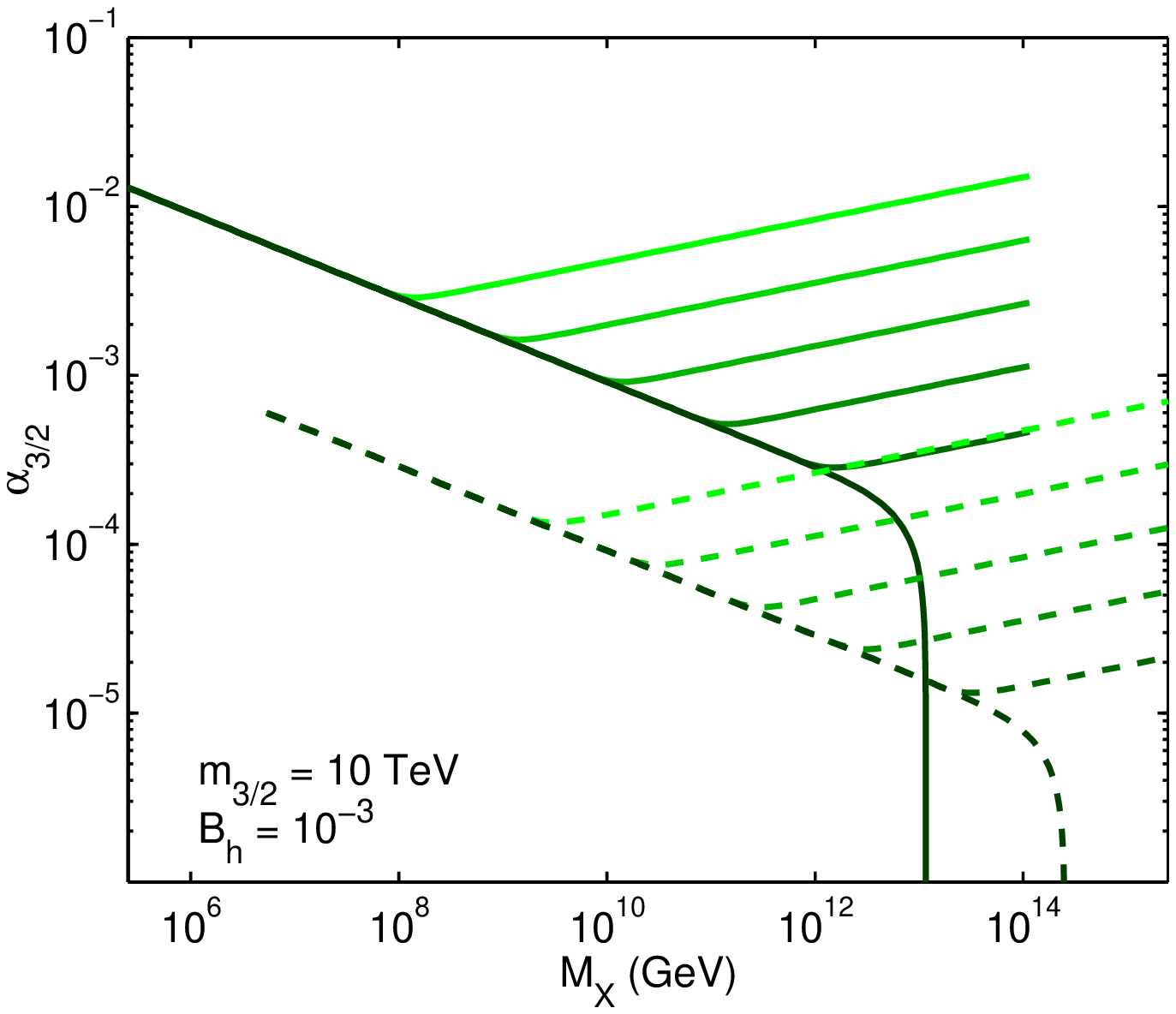}
\caption{(Color online) Contours of constant $Y_{3/2}=Y_{3/2}^{BBN}$
for $M_5= 10^8,\, 10^9,\, 10^{10},\, 10^{11},\, 10^{12},\,
10^{13}$~GeV (from lighter to darker green). In the top panels we
have $Y_{3/2}^{BBN}=4 \times 10^{-17}$ ($m_{3/2}=1$~TeV and
$B_h=1$), and in the bottom ones $Y_{3/2}^{BBN}=2 \times 10^{-13}$
($m_{3/2}=10$~TeV and $B_h=10^{-3}$). In the right panels we show
two sets of curves for which we fixed $\alpha_{t}=1$ (full lines)
and $\alpha_{t}=10^{-2}$ (dashed lines). For $m_{3/2}= 1$~TeV, the
$M_5\sim 10^{11}$~GeV curves already reproduce the results from SC,
while for $m_{3/2}= 10$~TeV we obtain the SC results only for
$M_5\sim 10^{13}$~GeV. \label{fig:unstable1}}
\end{figure}
%-----------------------------------------------------

In Fig.~\ref{fig:unstable1} we present the contours of constant
$Y_{3/2}=Y_{3/2}^{BBN}$ for several values of $M_5$. We present the
contours for $m_{3/2}=1$~TeV and $B_h=1$ (case I), for which
$Y_{3/2}^{BBN} \lesssim 4 \times 10^{-17}$, and $m_{3/2}=10$~TeV and
$B_h=10^{-3}$ (case II), for which $Y_{3/2}^{BBN} \lesssim 2 \times
10^{-13}$. In the right panels we show two sets of curves, one for
$\alpha_{t}=1$ and the other for $\alpha_{t}=10^{-2}$. Once more we
have fixed $m_{{\tilde g}}=1$~TeV in the computation of the thermal
contribution to the gravitino abundance. Some of the curves are
broken due to the upper limits on $T_{rh}$ and $M_X$ [cf.
Eqs.~(\ref{eq:TrhBCbounds}) and (\ref{eq:MxBCbounds}), respectively]
which come from $m_{3/2}> H$.

One should notice that for $m_{3/2}= 1$~TeV, the curves
corresponding to $M_5 \gtrsim 10^{11}$~GeV already reproduce the
results from SC, while for $m_{3/2}= 10$~TeV we obtain the SC
results only for $M_5 \gtrsim 10^{13}$~GeV. The value of $M_5$ below
which the BC effects affect the production of gravitinos can be
easily obtained from Eq.~(\ref{eq:M5boundYbbn}). One has
\begin{align}
M_5 \lesssim 3.3 \times 10^{10}\,{\rm GeV}~,\quad {\rm for}~m_{3/2}=
1~{\rm TeV}~,
\end{align}
and
\begin{align}
M_5 \lesssim 9.7 \times 10^{12}\,{\rm GeV}~,\quad {\rm for}~m_{3/2}=
10~{\rm TeV}~.
\end{align}

For 5D Planck masses above these values, the upper bounds on
$T_{rh}$, $M_X$ and $\alpha_{3/2}$ derived for
SC~\cite{Asaka:2006bv} are valid. For the reheating temperature one
has [cf. Fig.~\ref{fig:unstable1} (left) and
Eq.~(\ref{eq:TrhboundYbbnSC})]
\begin{align}
T_{rh} \lesssim \left\{
\begin{array}{l}
  4.7 \times 10^{5} \,{\rm GeV} \quad  {\rm (I)}~, \\
  2.1 \times 10^{9} \,{\rm GeV}  \quad  {\rm (II)}~, \\
\end{array}\right.
\end{align}
while for the heavy scalar mass one finds [cf.
Fig.~\ref{fig:unstable1} (right) and Eq.~(\ref{eq:MxboundYbbnSC})]
\begin{align}
M_{X} \lesssim \left\{
\begin{array}{l}
  4.1 \times 10^{10} \alpha_t^{-2/3}\,{\rm GeV} \quad  {\rm (I)}~, \\
  1.2 \times 10^{13} \alpha_t^{-2/3}\,{\rm GeV}  \quad  {\rm (II)}~, \\
\end{array}\right.
\end{align}
and for $\alpha_{3/2}$ [cf. Fig.~\ref{fig:unstable1} (right) and
Eq.~(\ref{eq:alpha32boundYbbnSC})]
\begin{align}
\alpha_{3/2} \lesssim \left\{
\begin{array}{l}
  1.3 \times 10^{-5} \,\alpha_t^{1/2}
\left(\dfrac{M_X}{10^{10}~{\rm GeV}}\right)^{-1/4} \quad  {\rm (I)}~, \\
  9.4 \times 10^{-4} \,\alpha_t^{1/2}
\left(\dfrac{M_X}{10^{10}~{\rm GeV}}\right)^{-1/4}  \quad  {\rm (II)}~. \\
\end{array}\right.
\end{align}
The lower bound on the reheating temperature presented in
Eq.~(\ref{eq:TrhboundMXSC}) reads in this case
\begin{align}
T_{rh} \gtrsim \left\{
\begin{array}{l}
  3.4 \times 10^{4} \,{\rm GeV}\left(\dfrac{\alpha_{3/2}\,M_X}{10^{5}~{\rm GeV}}\right)^{2} \quad  {\rm (I)}~, \\
  6.8 \,{\rm GeV}\left(\dfrac{\alpha_{3/2}\,M_X}{10^{5}~{\rm GeV}}\right)^{2}  \quad  {\rm (II)}~. \\
\end{array}\right.
\end{align}

For lower values of the 5D Planck mass, the disappearance of the
upper bounds on $T_{rh}$ and $M_X$ coming from BBN limits on the
amount of gravitinos will allow the Universe to reheat at a larger
temperature and, at the same time, the scalar to have a larger mass
and a larger decay rate into gravitinos than the ones allowed in SC.
The upper bound in $M_X$ is simply the one given in
Eq.~(\ref{eq:MxBCbounds}), while the reheating temperature, in
addition to $T_{rh}>T_t$, should satisfy
\begin{align}
\left.
\begin{array}{l}
  6.1 \times 10^{4} \,{\rm GeV}  \left(\dfrac{\alpha_{3/2}\,M_X}{10^{5}~{\rm GeV}}\right)^{2/3}
\left(\dfrac{M_5}{10^{10}~{\rm GeV}}\right)\\
  3.5 \times 10^{3} \,{\rm GeV} \left(\dfrac{\alpha_{3/2}\,M_X}{10^{5}~{\rm GeV}}\right)^{2/3}
\left(\dfrac{M_5}{10^{10}~{\rm GeV}}\right) \\
\end{array}\right\}
\lesssim  T_{rh} \lesssim \left\{
\begin{array}{l}
  4.2 \times 10^{7} \,{\rm GeV}\left(\dfrac{M_5}{10^{10}~{\rm
GeV}}\right)^{3/4} \quad  {\rm (I)}~, \\
  7.5 \times 10^{7} \,{\rm GeV}\left(\dfrac{M_5}{10^{10}~{\rm
GeV}}\right)^{3/4}  \quad  {\rm (II)}~, \\
\end{array}\right.
\end{align}
obtained from Eqs.~(\ref{eq:TrhboundMXBC}) and
(\ref{eq:TrhboundHlargem32BC}), respectively. In this regime, the
upper bound on the coupling constant $\alpha_{3/2}$ will grow with
$M_X$, in contrast with the SC scenario. From
Eq.~(\ref{eq:alpha32boundYbbnBC}) we obtain
\begin{align}
\alpha_{3/2} \lesssim \left\{
\begin{array}{l}
  1.2 \times 10^{-5} \,\alpha_t^{3/4}
\left(\dfrac{M_X}{10^{10}~{\rm GeV}}\right)^{1/8}
\left(\dfrac{M_5}{10^{10}~{\rm GeV}}\right)^{-3/8} \quad  {\rm (I)}~, \\
  8.4 \times 10^{-4} \,\alpha_t^{3/4}
\left(\dfrac{M_X}{10^{10}~{\rm GeV}}\right)^{1/8}
\left(\dfrac{M_5}{10^{10}~{\rm GeV}}\right)^{-3/8}  \quad  {\rm (II)}~. \\
\end{array}\right.
\end{align}

Before concluding this section, it is worth commenting on the
possible additional constraints coming from the LSP abundance, which
is stable if R-parity is conserved. In this case one must impose
that the abundance of LSP does not overclose the universe, as it
contributes to the dark matter content. The LSP can be produced both
from gravitino and heavy scalar decays. The LSP abundance from
gravitino decays does not give additional bounds provided that
$m_{3/2}= 100~{\rm GeV}-10~{\rm TeV}$. The constraints coming from
the production of the LSP from heavy scalar decays will be model
dependent. In Ref.~\cite{Asaka:2006bv} the bounds arising from
considering the neutral wino $\tilde{W}$ as the LSP, which maximizes
the annihilation cross section and hence leads to the most
conservative bounds, were derived in SC. In order to avoid the
overclosure of the universe, a lower bound in $M_X$ is found, namely
$M_X\gtrsim 2 \times 10^{6}$~GeV for $M_{\tilde{W}}=100$~GeV, which
becomes more stringent for larger wino masses. Since the brane
effects are only present for heavy scalar masses $M_X \gtrsim
\alpha_t^{-2/3}\,M_5$ [cf. Eq.~(\ref{eq:MxBCbounds})], one should
expect that for sufficiently high values of $M_5$ these bounds will
not change.

\subsection{Stable gravitino and dark matter}

In the case of a stable gravitino, when the gravitino is itself the
LSP with exact R-parity, one should check whether the contribution
of gravitinos to the energy density of the universe does not exceed
the observed matter density limit. From the gravitino abundance, we
can estimate their contribution to the closure density
\begin{align}
\Omega_{3/2} h^2 = m_{3/2}\,Y_{3/2}\, s_{0}\, h^2 \rho_c^{-1}\,.
\label{eq:Omega32}
\end{align}
Here $\rho_c=3 H_0^2 M_P^2/8\pi = 8.07 \times 10^{-47} h^2$~GeV$^4$
is the critical density and $s_{ 0}=2.22 \times 10^{-38}$~GeV$^3$ is
the present entropy density. The total fraction of gravitinos
contributing to dark matter, $\Omega_{3/2}$, is the sum of the two
contributions
\begin{align}
\Omega_{3/2} = \Omega_{3/2}^{X} + \Omega_{3/2}^{th}\,,
\label{eq:Omega32_sum}
\end{align}
the first coming from the abundance of gravitinos due to the decay
of the heavy scalar, Eq.~(\ref{eq:gravabundX2}), and the second from
thermal production, Eq.~(\ref{eq:gravabundTh}). To avoid overclosure
one must require
\begin{align}
\Omega_{3/2} h^2 \leq \Omega_{DM} h^2~,
\end{align}
where we will use the WMAP bound on the matter density of the
universe~\cite{Bennett:2003bz},
\begin{align}\Omega_{DM} h^2 = 0.143~.\label{eq:Omega_dm_WMAP}
\end{align}
If $\Omega_{3/2} = \Omega_{DM}$, the gravitinos constitute (all)
the dark matter of the universe.

Let us now consider the two contributions to the gravitino dark
matter. From the decay of the heavy scalar one has
\begin{align}
\Omega_{3/2}^X h^2 & = 2.8 \times 10^{-2}
\left(\dfrac{m_{3/2}}{10^{-1}\,{\rm
GeV}}\right)\,\left(\dfrac{Y^X_{3/2}}{10^{-9}}\right) \nonumber\\
& =4.1 \times 10^{-1}\,\left(
\dfrac{B_{3/2}}{10^{-4}}\right)\,\left(\dfrac{m_{3/2}}{10^{-1}\,{\rm
GeV}}\right)\,\left(\dfrac{M_{X}}{10^{10}\,{\rm
GeV}}\right)^{-1}\,\left(\dfrac{T_{rh}}{10^{6}\,{\rm GeV}}\right)
\,,
\end{align}
and from the thermal production
\begin{align}
\Omega_{3/2}^{th} h^2 = 2.8 \times 10^{-2}
\left(\dfrac{m_{3/2}}{10^{-1}\,{\rm
GeV}}\right)\,\left(\dfrac{Y^{th}_{3/2}}{10^{-9}}\right)\,.
\end{align}
In the SC case one can write
\begin{align}
\Omega_{3/2}^X h^2 &= 2.3 \times 10^{-2}\,\alpha_t\,\left(
\dfrac{B_{3/2}}{10^{-4}}\right)\,\left(\dfrac{m_{3/2}}{10^{-1}\,{\rm
GeV}}\right)\,\left(\dfrac{M_{X}}{10^{10}\,{\rm
GeV}}\right)^{1/2} \,,\\
\Omega_{3/2}^{th} h^2 &= 1.6 \times 10^{-2}\, \alpha_t\,
\left[\dfrac{{\cal A}(T_{rh};m_{3/2} ,m_{{\tilde g}})}{10^{-14}~{\rm
GeV}^{-1}}\right]\, \left(\dfrac{m_{3/2}}{10^{-1}\,{\rm
GeV}}\right)\, \left(\dfrac{M_{X}}{10^{10}\,{\rm GeV}}\right)^{3/2}
\,,
\end{align}
while in the BC regime one has
\begin{align}
\Omega_{3/2}^X h^2 &= 2.8 \times 10^{-2}\,\alpha_t^{1/2}\,\left(
\dfrac{B_{3/2}}{10^{-4}}\right)\,\left(\dfrac{m_{3/2}}{10^{-1}\,{\rm
GeV}}\right)\,\left(\dfrac{M_{X}}{10^{10}\,{\rm
GeV}}\right)^{-1/4}\,\left(\dfrac{M_{5}}{10^{10}\,{\rm
GeV}}\right)^{3/4} \,,\\
\Omega_{3/2}^{th} h^2 &=4.4 \times 10^{-2}\, \left[\dfrac{{\cal
A}(T_{t};m_{3/2} ,m_{{\tilde g}})}{10^{-14}~{\rm
GeV}^{-1}}\right]\, \left(\dfrac{m_{3/2}}{10^{-1}\,{\rm
GeV}}\right)\, \left(\dfrac{M_{5}}{10^{10}\,{\rm
GeV}}\right)^{3/2} \,.
\end{align}
Notice that, in both BC and SC, $\Omega_{3/2}^X h^2$ is proportional
to $m_{3/2}$, while $\Omega_{3/2}^{th} h^2$ is inversely
proportional to $m_{3/2}$, since ${\cal A}(T_{t};m_{3/2} ,m_{{\tilde
g}}) \sim m_{3/2}^{-2}$. This means that for large gravitino masses
the $X$ decay contribution will dominate, while for small values of
$m_{3/2}$, the main contribution comes from thermal production. On
the other hand, the dependence of $\Omega_{3/2}^X h^2$ on $M_X$ is
different in BC from that of SC. Once more, this implies that in the
brane high-energy regime, there is no upper limit on the heavy
scalar mass coming from the gravitino dark matter fraction.

One should notice that the momentum of the gravitinos produced by
$X$ decay can be much larger than its mass since $p_{3/2} \simeq
M_X/2$. Therefore, they can behave as warm dark matter and influence
the structure formation: with a non negligible velocity, and before
the time of matter-radiation equality, they can freely stream out of
overdense regions and into underdense regions, and this can erase
the small structures that are observed today. This leads to an upper
bound on the present dispersion velocity of the gravitino
$v_{3/2}^0$~\cite{Borgani:1996ag,Jedamzik:2005sx,Steffen:2006hw}.
The present velocity of dispersion of the gravitinos produced by the
$X$ decay is estimated to be~\cite{Jedamzik:2005sx}
\begin{align}
v_{3/2}^0 \simeq \dfrac{1}{2} \dfrac{M_X}{m_{3/2}}
\dfrac{T_0}{T_{rh}}\left(\dfrac{g_*(T_0)}{g_*(T_{rh})}\right)^{1/3}~,
\end{align}
where $T_0=2.35 \times 10^{-13}$~GeV is the present photon
temperature and $g_*(T_0)=43/11$. Bounds on this velocity are
obtained by means of the power spectrum inferred from
Lyman-$\alpha$ together with cosmic microwave background radiation
and galaxy clustering
constraints~\cite{Viel:2005qj,Seljak:2006qw}. From
Ref.~\cite{Seljak:2006qw} it was found~\cite{Asaka:2006bv} that
$v_{3/2}^0 \lesssim 4 \times 10^{-8}$.

From the limit on the present velocity of dispersion one can
estimate an approximate upper bound on the branching ratio of the
decay channel into gravitinos $B_{3/2}$. Considering the
contribution from the heavy scalar decay to the amount of
gravitinos we can write
\begin{align}
\Omega_{3/2}^X= \dfrac{3}{2}\,B_{3/2}\,\dfrac{m_{3/2}
T_{rh}}{M_X}\,s_0\,\rho_c^{-1} \,,
\end{align}
from which
\begin{align}
v_{3/2}^0 = 1.26 \times 10^{-5} \dfrac{B_{3/2}}{\Omega_{3/2}^X
h^2}~.
\end{align}
Assuming $\Omega_{3/2}^X h^2 = \Omega_{DM}^X h^2$, one gets the
approximate bound
\begin{align}
B_{3/2} \lesssim 4.5 \times 10^{-4} \,. \label{eq:B32_warm}
\end{align}
Notice that this bound is independent of the background cosmology.
One can also compute the velocity of dispersion for gravitinos
produced thermally, however one finds that for $m_{3/2}\gtrsim
100$~keV they have a negligible free-streaming
behavior~\cite{Steffen:2006hw}. Here we will consider masses well
above this bound. The above constraint in $B_{3/2}$ was determined
assuming that all dark matter is constituted by gravitinos,
$\Omega_{3/2}^X= \Omega_{DM}$. However, if gravitinos are only a
fraction of dark matter, then the bound becomes weaker, and can even
be evaded if
\begin{align}
\Omega^X_{3/2} \lesssim 0.12 \,\Omega_{DM} \,,
\end{align}
as derived in Ref.~\cite{Viel:2005qj}.

%---------------------------------------------------
\begin{figure}[t]
\includegraphics[width=5.4cm]{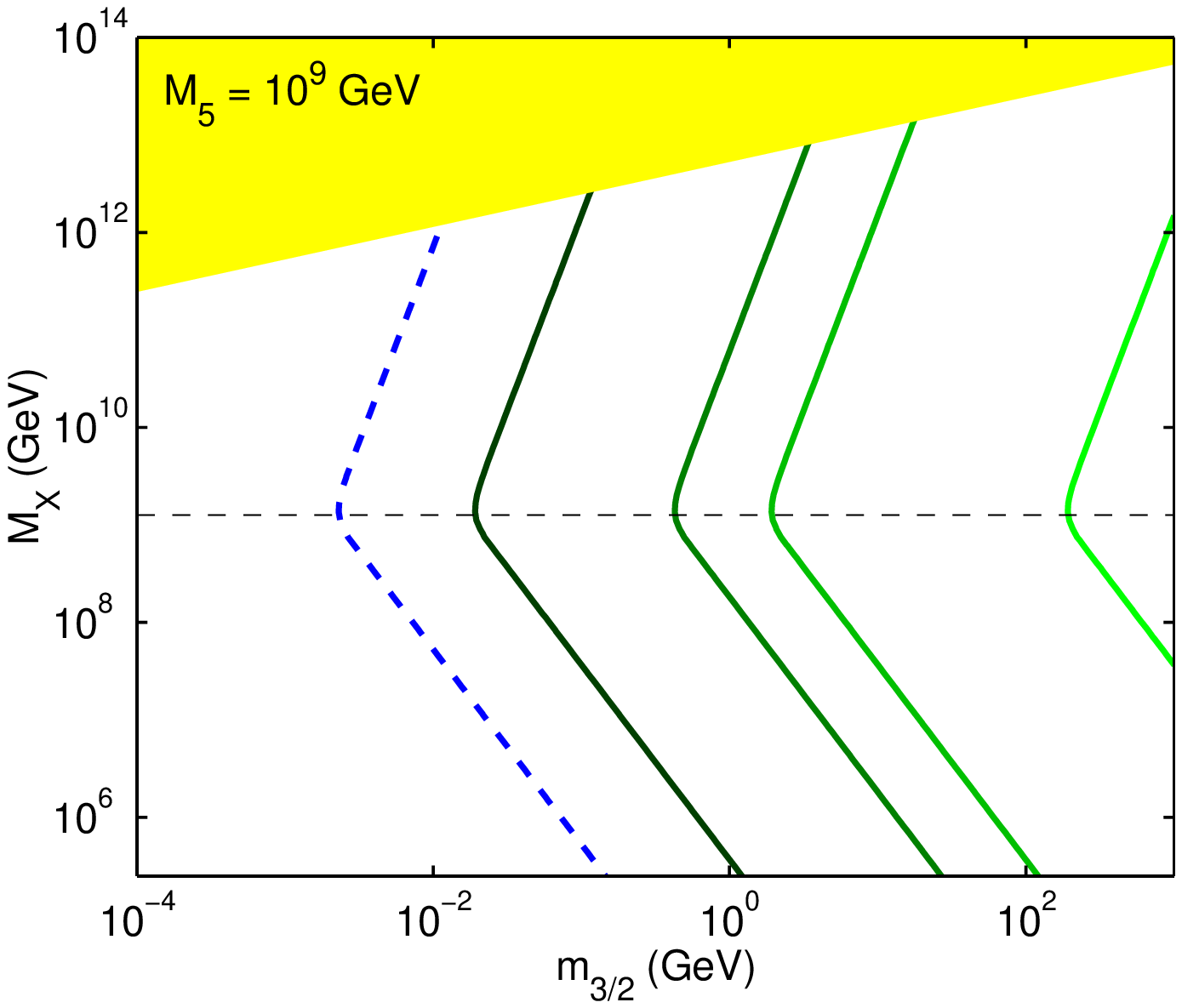}
\includegraphics[width=5.4cm]{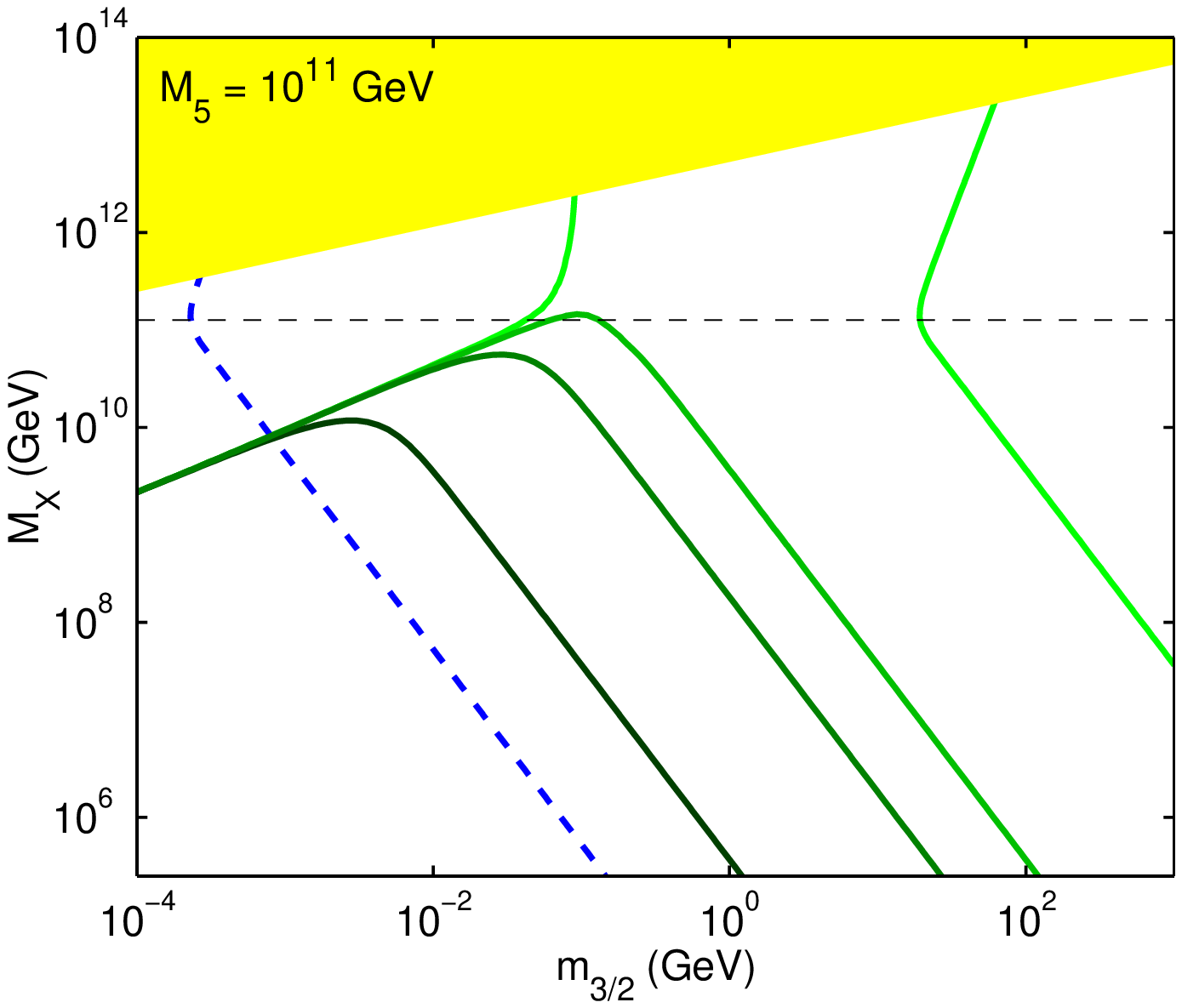}
\includegraphics[width=5.4cm]{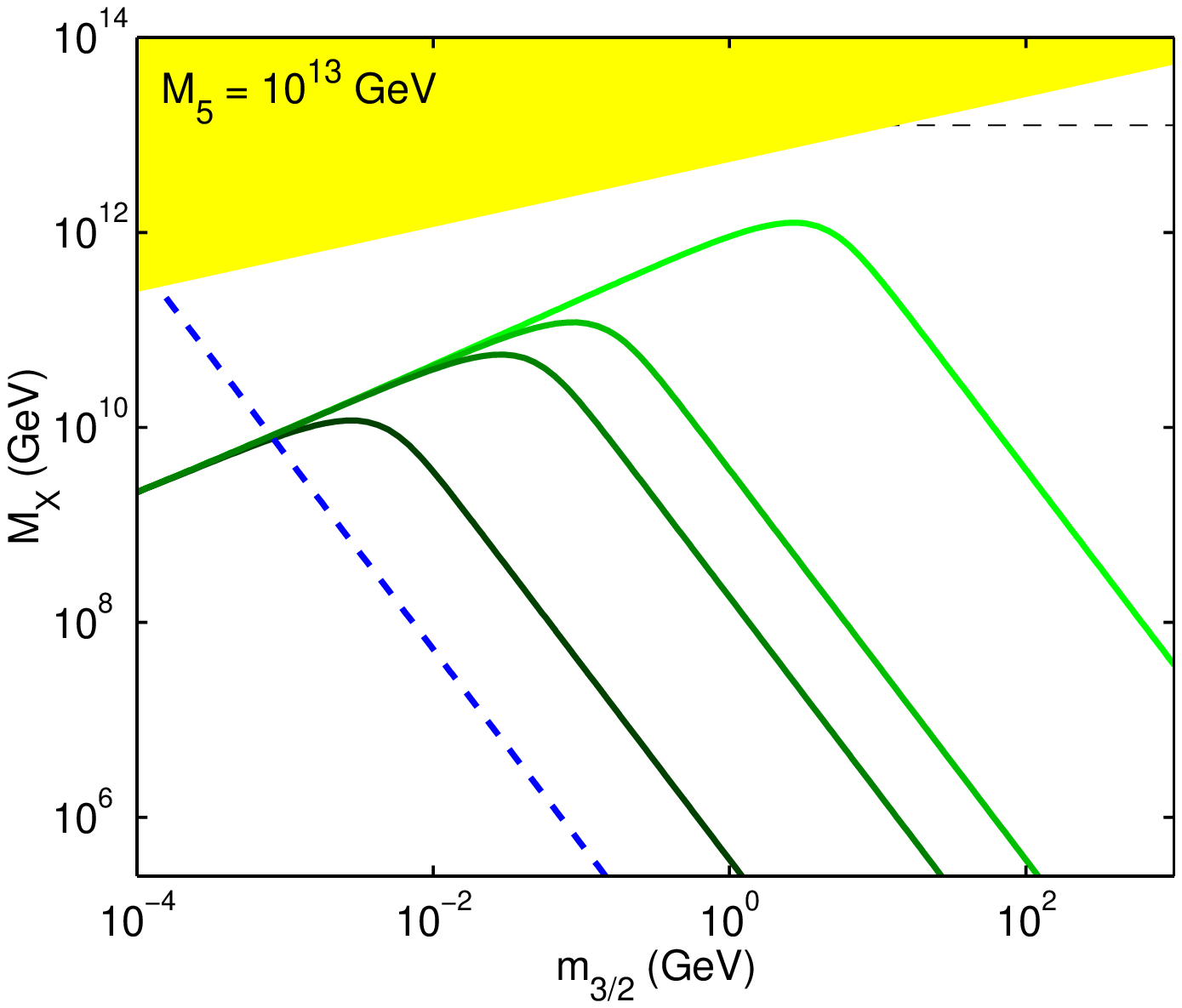}
\caption{(Color online) Contour of constant
$\Omega_{3/2}h^2=\Omega_{DM}h^2=0.143$ for $B_{3/2}= 10^{-2},\, 4.5
\times 10^{-4},\, 10^{-4},\, 10^{-6}$ (darker to lighter, or left to
right, green full lines) in the $M_X-m_{3/2}$ plane. The different
panels correspond to various values of $M_5$. We have chosen
$m_{{\tilde g}}=1$~TeV and $\alpha_{t}=1$. Also shown is the contour
for $\Omega_{3/2}h^2=0.12\Omega_{DM}h^2$ (blue dashed line), when
$B_{3/2}= 10^{-2}$. The yellow shaded area is excluded because of
the condition $H(T_{rh})<m_{3/2}$ [cf. Eq.~(\ref{eq:MxBCbounds})].
The horizontal line represents the value of $M_X$ for which the
transition from the high to the low energy regime occurs.
\label{fig:stable1}}
\end{figure}
%---------------------------------------------------

In Fig.~\ref{fig:stable1} we present the contours of constant
$\Omega_{3/2}h^2=\Omega_{DM}h^2,$ in the $M_X-m_{3/2}$ plane, with
$\Omega_{DM}h^2$ given by the WMAP bound in
Eq.~(\ref{eq:Omega_dm_WMAP}). The different (green) full lines
correspond to various values of $B_{3/2}$, where we fixed
$\alpha_{t}=1$ and $m_{{\tilde g}}=1$~TeV. The allowed range of
masses lies to the left and/or below the contour lines. Moreover the
gravitino dark matter is only viable to the right/above the line
which corresponds to the upper bound on the branching ratio shown in
Eq.~(\ref{eq:B32_warm}), $B_{3/2}= 4.5 \times 10^{-4}$. The yellow
shaded area is excluded because of the condition $H(T_{rh})<m_{3/2}$
[cf. Eq.~(\ref{eq:MxBCbounds})]. The horizontal line represents the
value of $M_X = \left(2/\alpha_{t}^2\right)^{1/3} M_5$, for which
the transition from the high to the low energy regimes occurs. The
contours are shown for three different values of the 5D Planck mass,
namely, $M_5= 10^9,\,10^{11},\,10^{13}$~GeV.

In SC there is an upper limit on the mass of the heavy scalar $M_X$
coming from the bounds on the gravitino abundance. The maximum value
of $M_X$ is obtained when both contributions to the gravitino
production, from the heavy scalar decay and thermal scatterings, are
equal; hence, at that point, one can write that the total abundance
is
$\Omega_{3/2}=2\sqrt{\Omega_{3/2}^X\,\Omega_{3/2}^{th}}<\Omega_{DM}$
and obtain
\begin{align}
M_X \lesssim 7 \times 10^9 \, {\rm GeV} \, \alpha_{3/2}^{-1} \, .
\end{align}
We have taken into consideration that, for the gravitino mass in the
range $10^{-4}-10^2$~GeV , ${\cal A}(T_{rh};m_{3/2} ,m_{{\tilde g}})
\simeq 10^{-17} m_{3/2}^{-2}$ is a good approximation. As said
before, in the high energy regime there is no such a bound, and the
allowed values for $M_X$ are only constrained by the condition
$H(T_{rh})<m_{3/2}$ [cf. Eq.~(\ref{eq:MxBCbounds})]. We see that,
for low values of $M_5$ and a fixed $B_{3/2}$, it is possible to
have larger values of both $M_X$ and $m_{3/2}$ and yet satisfy the
bounds from dark matter. For example, for $M_5 \sim 10^9$~GeV and
sufficiently low branching ratio, it is possible to have $M_X \sim
10^{12}$~GeV while having $m_{3/2} \sim 100$~GeV.

%---------------------------------------------------
\begin{figure}[t]
\includegraphics[width=5.4cm]{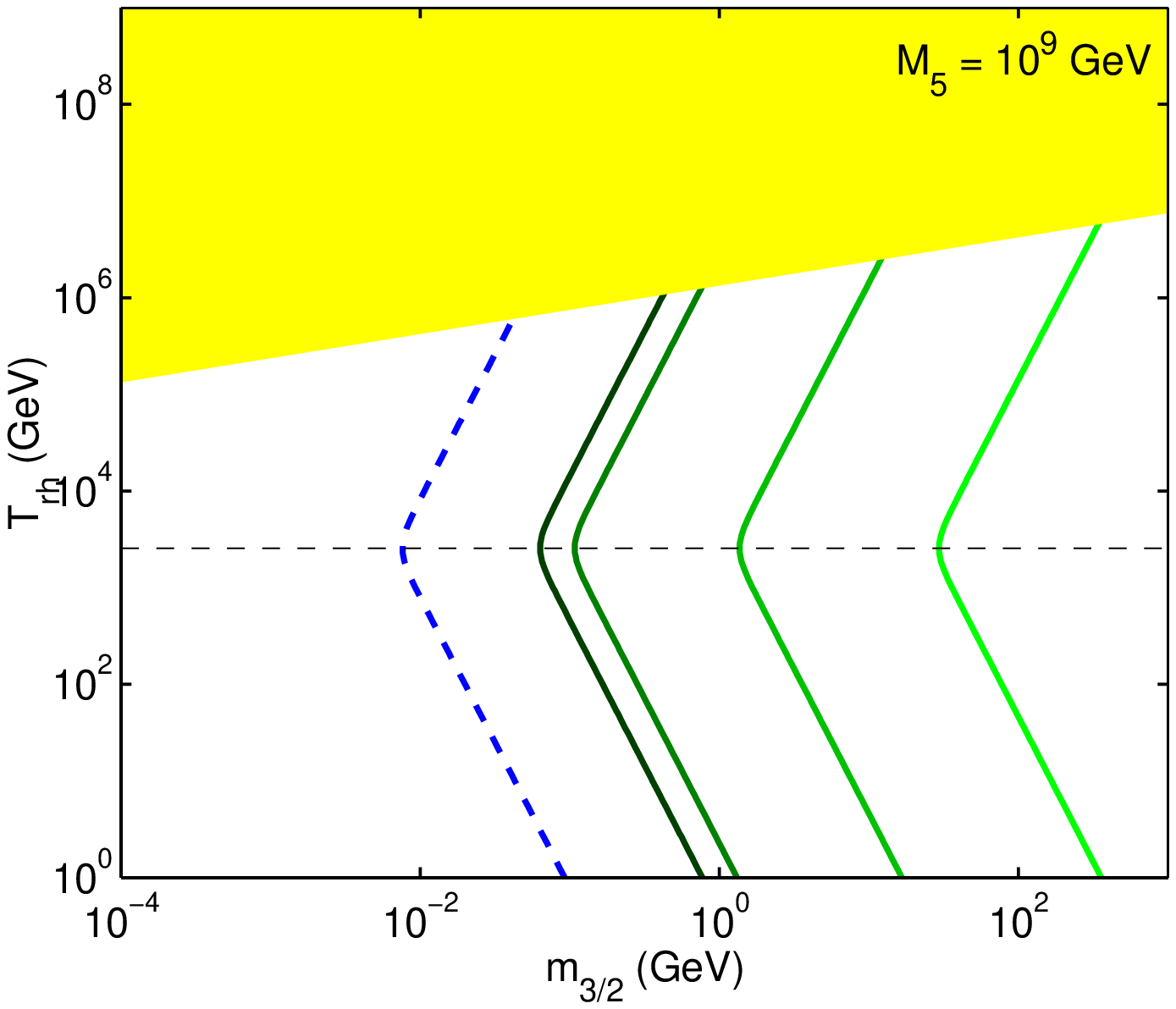}
\includegraphics[width=5.4cm]{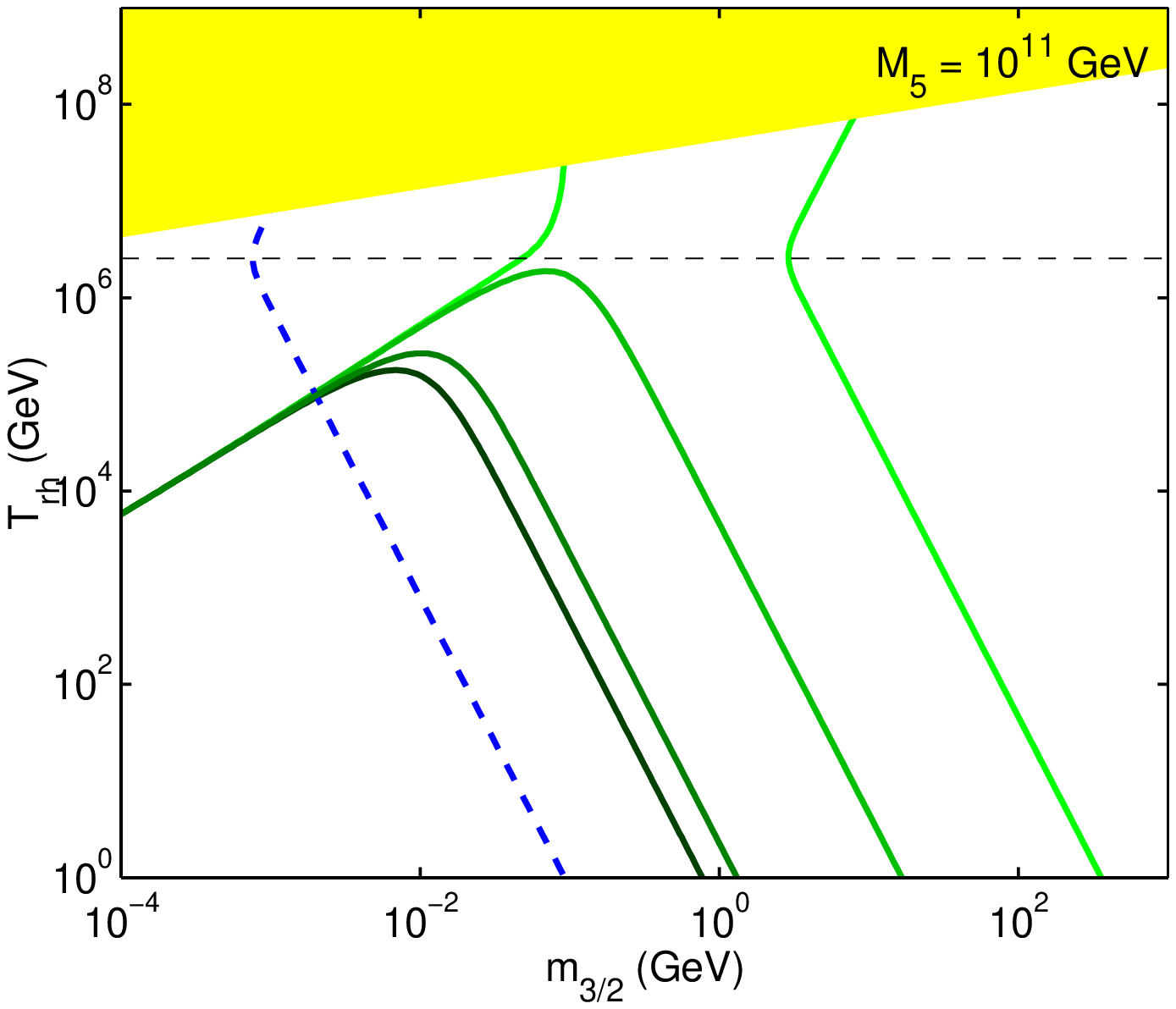}
\includegraphics[width=5.4cm]{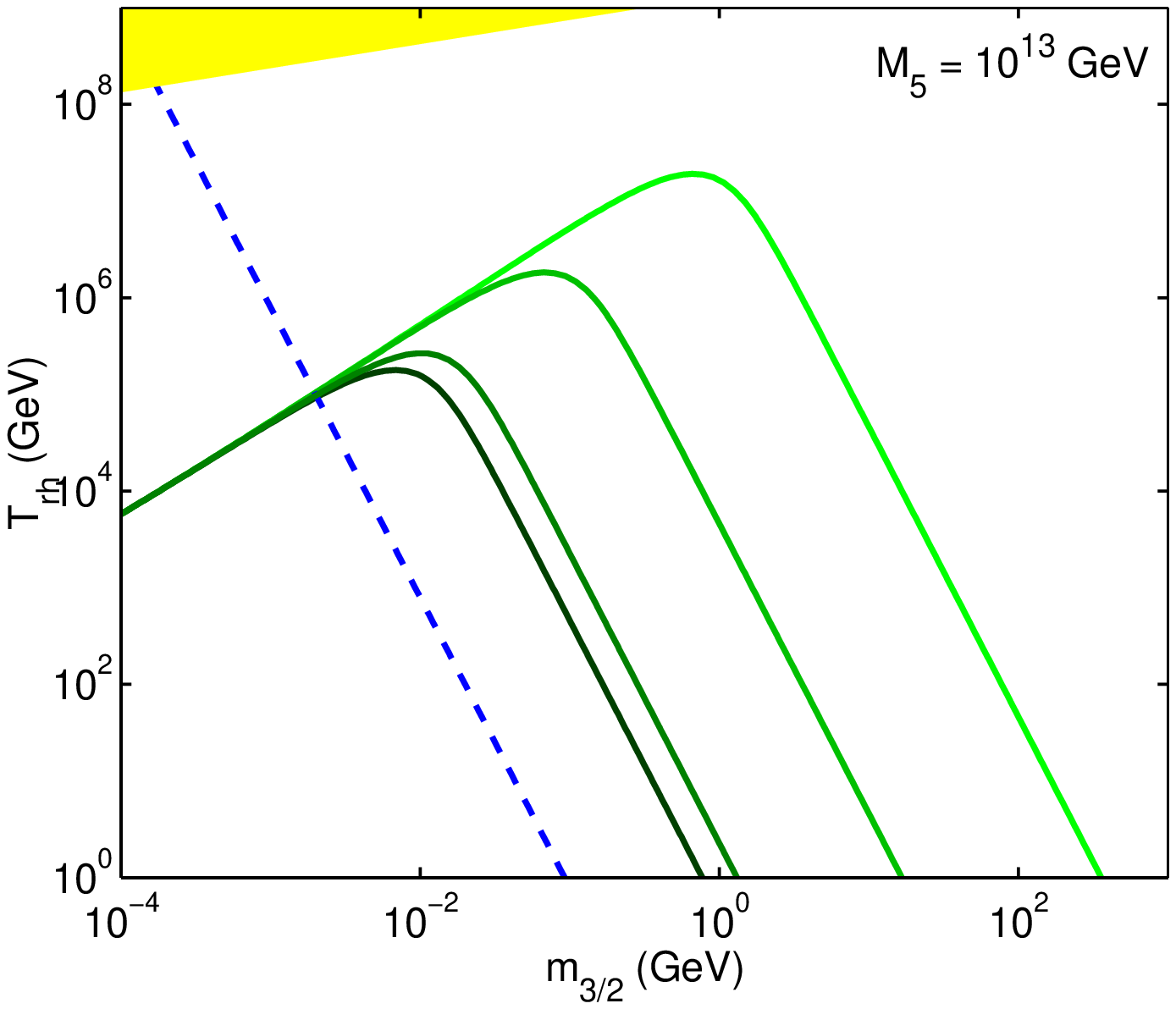}
\caption{(Color online) Same as in Fig.~\ref{fig:stable1}, but in
the $T_{rh}-m_{3/2}$ plane and for $\alpha_{3/2} B_{3/2}= 10^{-3},\,
4.5 \times 10^{-4},\, 10^{-5},\, 10^{-7}$ (darker to lighter, or
left to right, green full lines). \label{fig:stable2}}
\end{figure}
%---------------------------------------------------

If one now takes $T_{rh}$ as the free parameter one can derive the
contours of constant $\Omega_{3/2}h^2=\Omega_{DM}h^2$ in the
$T_{rh}-m_{3/2}$, shown in Fig.~\ref{fig:stable2} for different
values of $M_5$. Here the different (green) full lines correspond to
various values of $\alpha_{3/2} B_{3/2}$, and the gravitino dark
matter is only viable to the right/above the line which corresponds
to the upper bound on the branching ratio $B_{3/2} \lesssim 4.5
\times 10^{-4}$ [since $\alpha_{3/2} \lesssim {\cal O}(1)$]. Once
again one can distinguish the two behaviors for SC and high-energy
BC regimes. The contributions to the gravitino dark matter read as
\begin{align}
\Omega_{3/2}^X h^2 &= 1.9 \times 10^{-2}\, \left(
\dfrac{\alpha_{3/2}
B_{3/2}}{10^{-6}}\right)^{2/3}\left(\dfrac{m_{3/2}}{10^{-1}\,{\rm
GeV}}\right)\,\left(\dfrac{T_{rh}}{10^{6}\,{\rm
GeV}}\right)^{1/3} \,,\\
\Omega_{3/2}^{th} h^2 &=2.8 \times 10^{-1}\, \left[\dfrac{{\cal
A}(T_{t};m_{3/2} ,m_{{\tilde g}})}{10^{-14}~{\rm
GeV}^{-1}}\right]\, \left(\dfrac{m_{3/2}}{10^{-1}\,{\rm
GeV}}\right)\, \left(\dfrac{T_{rh}}{10^{6}\,{\rm GeV}}\right) \,,
\end{align}
in SC limit. In the BC regime we obtain
\begin{align}
\Omega_{3/2}^X h^2 &= 3.5 \times 10^{-3}\,
\left(\dfrac{\alpha_{3/2}\,B_{3/2}}{10^{-6}}\right)^{2/3}
\left(\dfrac{m_{3/2}}{10^{-1}\,{\rm GeV}}\right)
\left(\dfrac{T_{rh}}{10^{6}\,{\rm GeV}}\right)^{-1/3}
\left(\dfrac{M_{5}}{10^{10}\,{\rm GeV}}\right) \,,\\
\Omega_{3/2}^{th} h^2 &=5.5 \times 10^{-1}\, \left[\dfrac{{\cal
A}(T_{t};m_{3/2} ,m_{{\tilde g}})}{10^{-14}~{\rm
GeV}^{-1}}\right]\, \left(\dfrac{m_{3/2}}{10^{-1}\,{\rm
GeV}}\right)\, \left(\dfrac{T_{t}}{10^{6}\,{\rm GeV}}\right) \,.
\end{align}
As expected, the bound
\begin{align}
T_{rh} \lesssim 1.8 \times 10^6 \,{\rm GeV} \,
\left(\dfrac{\alpha_{3/2}\,B_{3/2}}{10^{-5}}\right)^{1/2} \,,
\end{align}
derived for the SC case in order to avoid the overclosure of the
universe by dark matter disappears when we consider sufficiently low
5D Planck scales $M_5$.

If $B_{3/2} > 4.5 \times 10^{-4}$, then the gravitinos can only
contribute to a fraction of the total amount of dark matter. In
Figs.~\ref{fig:stable1} and \ref{fig:stable2} the blue-dashed line
represents the contour for $\Omega_{3/2}h^2=0.12\Omega_{DM}h^2$,
the maximum contribution of gravitinos (warm) dark matter that
does not spoil the structure formation, for $B_{3/2}= 10^{-2}$. We
can see that in this case the bound becomes more stringent.

%---------------------------------------------------
\begin{figure}[t]
\includegraphics[width=5.4cm]{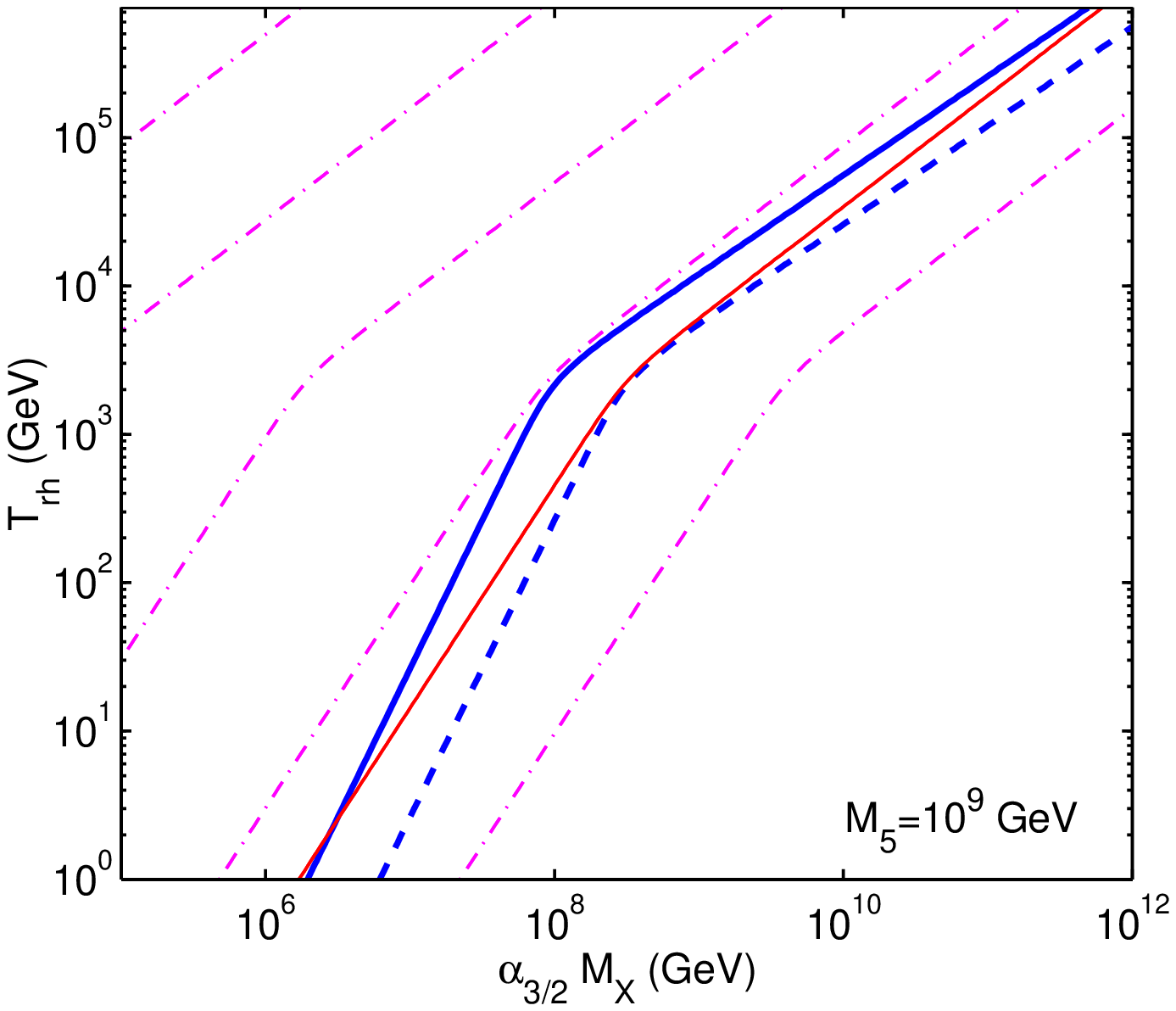}
\includegraphics[width=5.4cm]{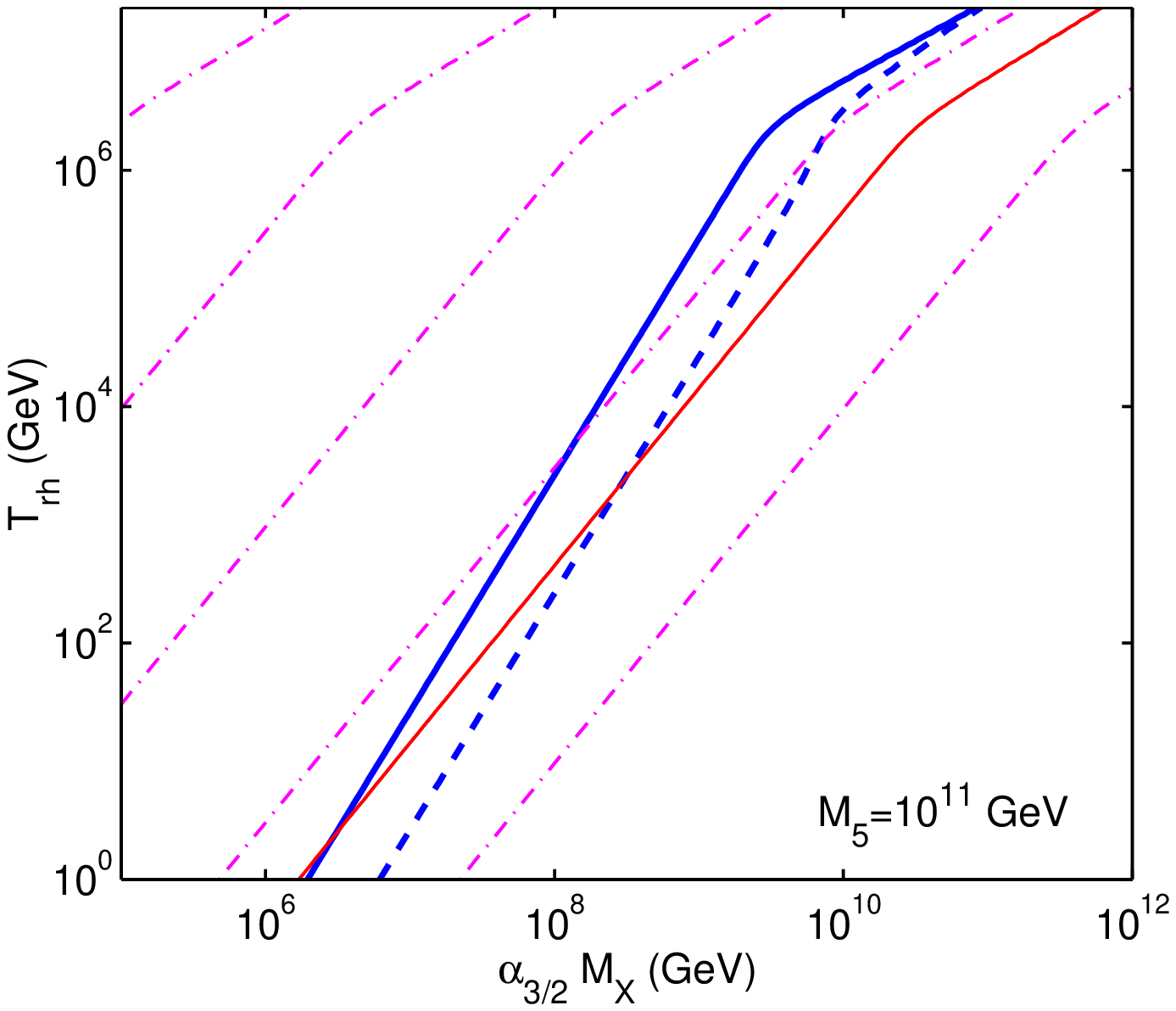}
\includegraphics[width=5.4cm]{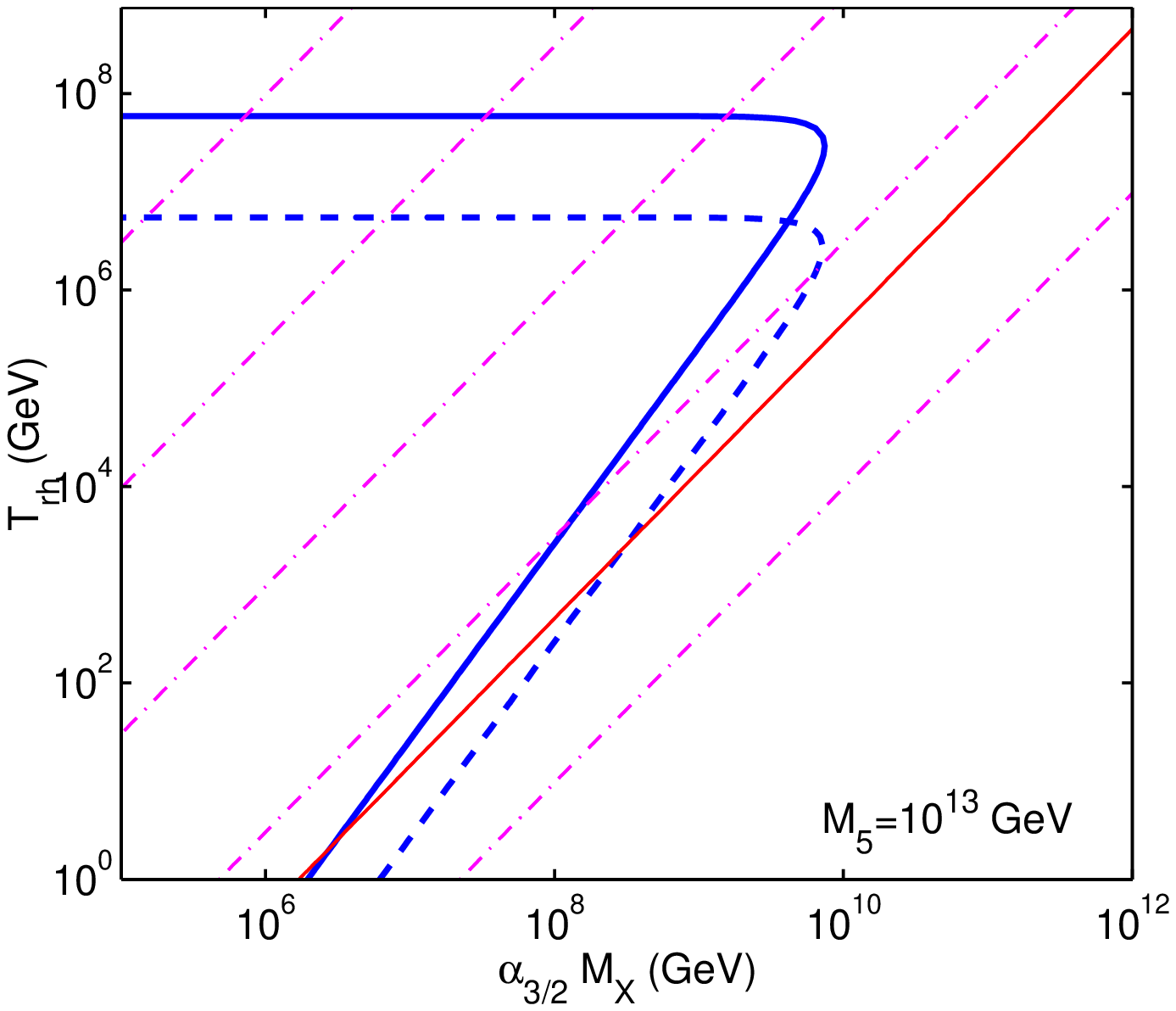}
\caption{(Color online) Contours of constant
$\Omega_{3/2}h^2=\Omega_{DM}h^2=0.143$ for $m_{3/2}=1$~GeV (full
bold line) and $m_{3/2}=0.1$~GeV (dotted bold line) in the
$T_{rh}-\alpha_{3/2}M_X$ plane. The dash-dotted lines represent the
contours of constant $B_{3/2}=
1,\,10^{-5},\,10^{-10},\,10^{-15},\,10^{-20}$ (from below to top),
and the full line is the contour for $B_{3/2}= 4.5 \times 10^{-4}$.
In the computation of the branching ratio we fixed $\alpha_{3/2}=1$.
The different panels correspond to various values for $M_5$.
\label{fig:stable3}}
\end{figure}
%---------------------------------------------------

Finally, in Fig.~\ref{fig:stable3} we show the contours of constant
$\Omega_{3/2}h^2=\Omega_{DM}h^2$, in the $T_{rh}-\alpha_{3/2}M_X$
plane, for two different gravitino masses: $m_{3/2}=1,~0.1$~GeV
(full and dotted bold lines, respectively) and $M_5= 10^9, 10^{11},
10^{13}$~GeV. Also shown are the contours of constant $B_{3/2}$
(dash-dotted lines) for a fixed $\alpha_{3/2}=1$. It is clearly seen
from the figure that, in the BC regime and for sufficiently low
values of $M_5$, higher reheating temperatures are allowed for large
$M_X$.

\section{Discussion and conclusion}
\label{sec:conclusion}

We have studied the gravitino production in the  braneworld
cosmology context. Our framework is the RSII construction in which
the Friedmann expansion law is modified with a quadratic term in the
energy density. As an approximation, we considered the gravitino as
a field localized on the UV brane and used 4D supersymmetry and the
usual Boltzmann equation to describe its production.

\begin{figure}[t]
\includegraphics[width=5.4cm]{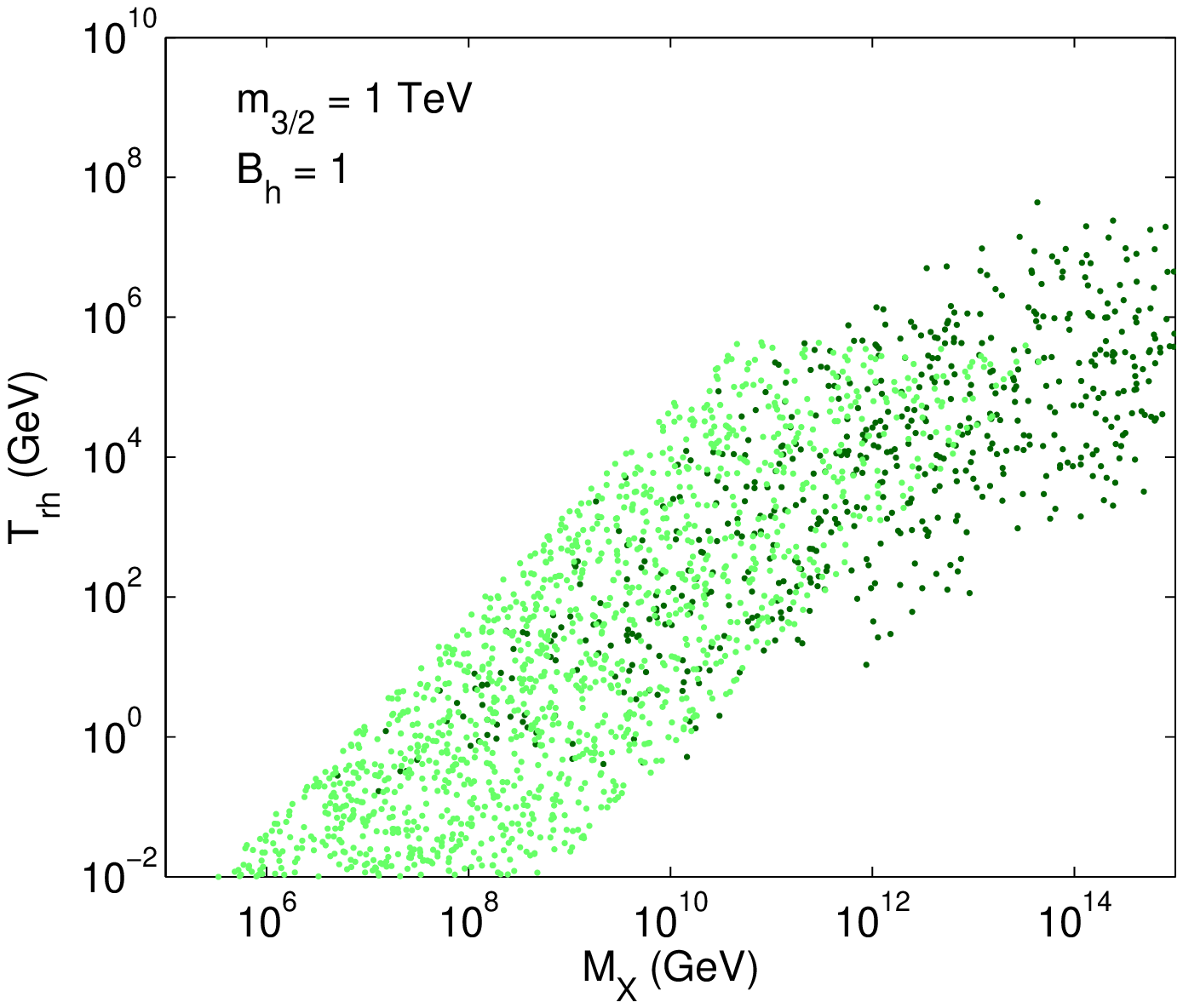}
\includegraphics[width=5.4cm]{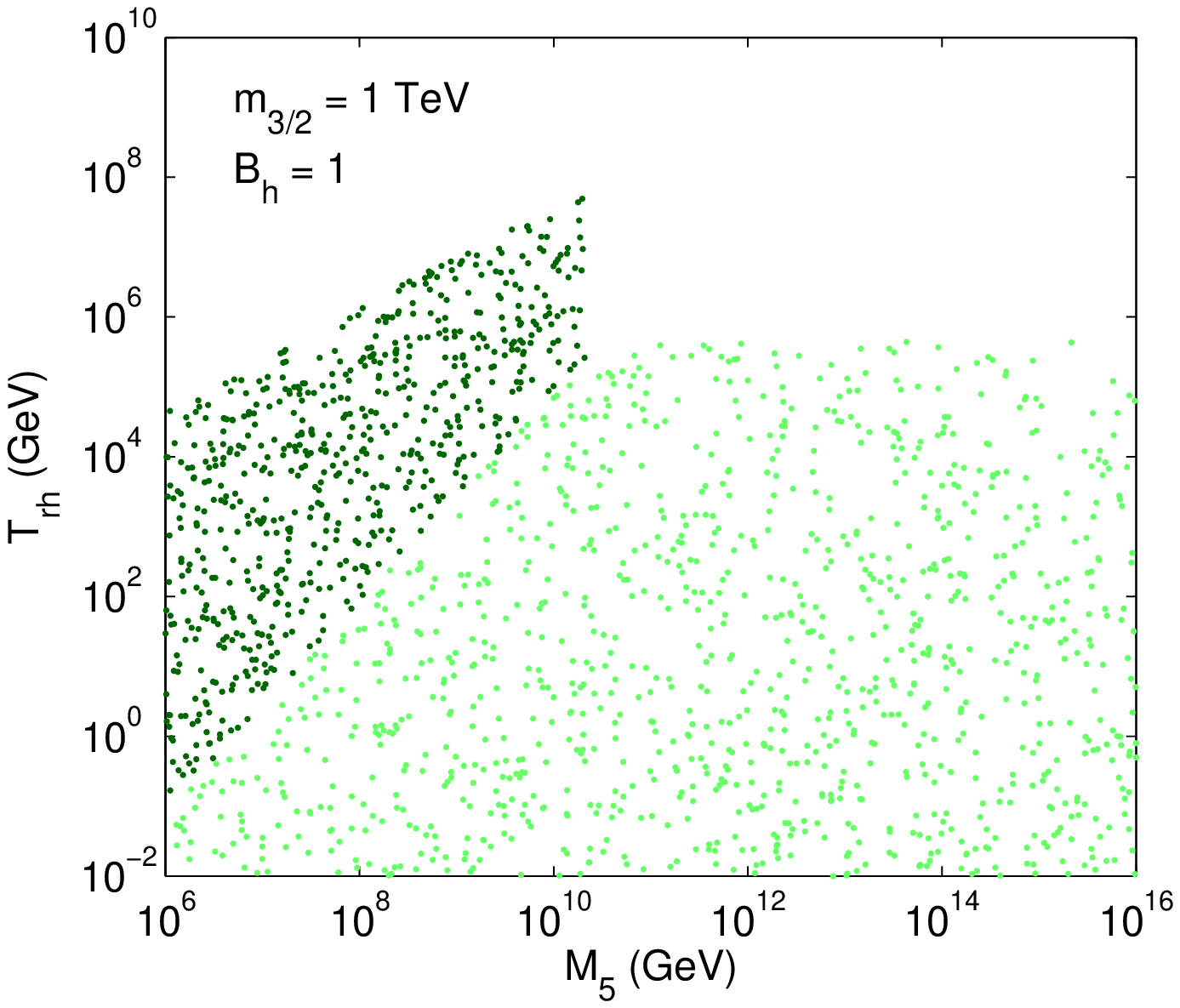}
\includegraphics[width=5.4cm]{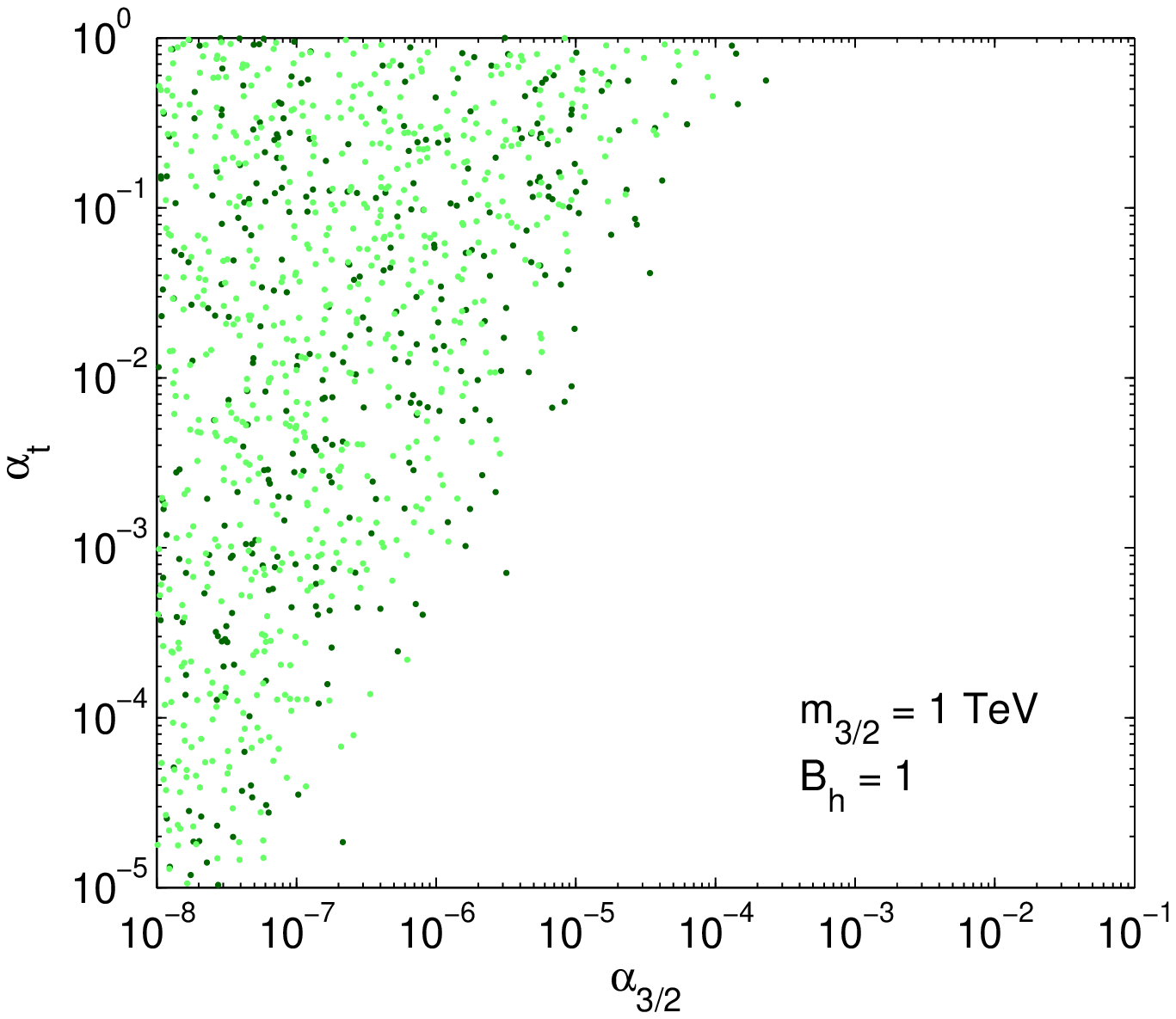}
\includegraphics[width=5.4cm]{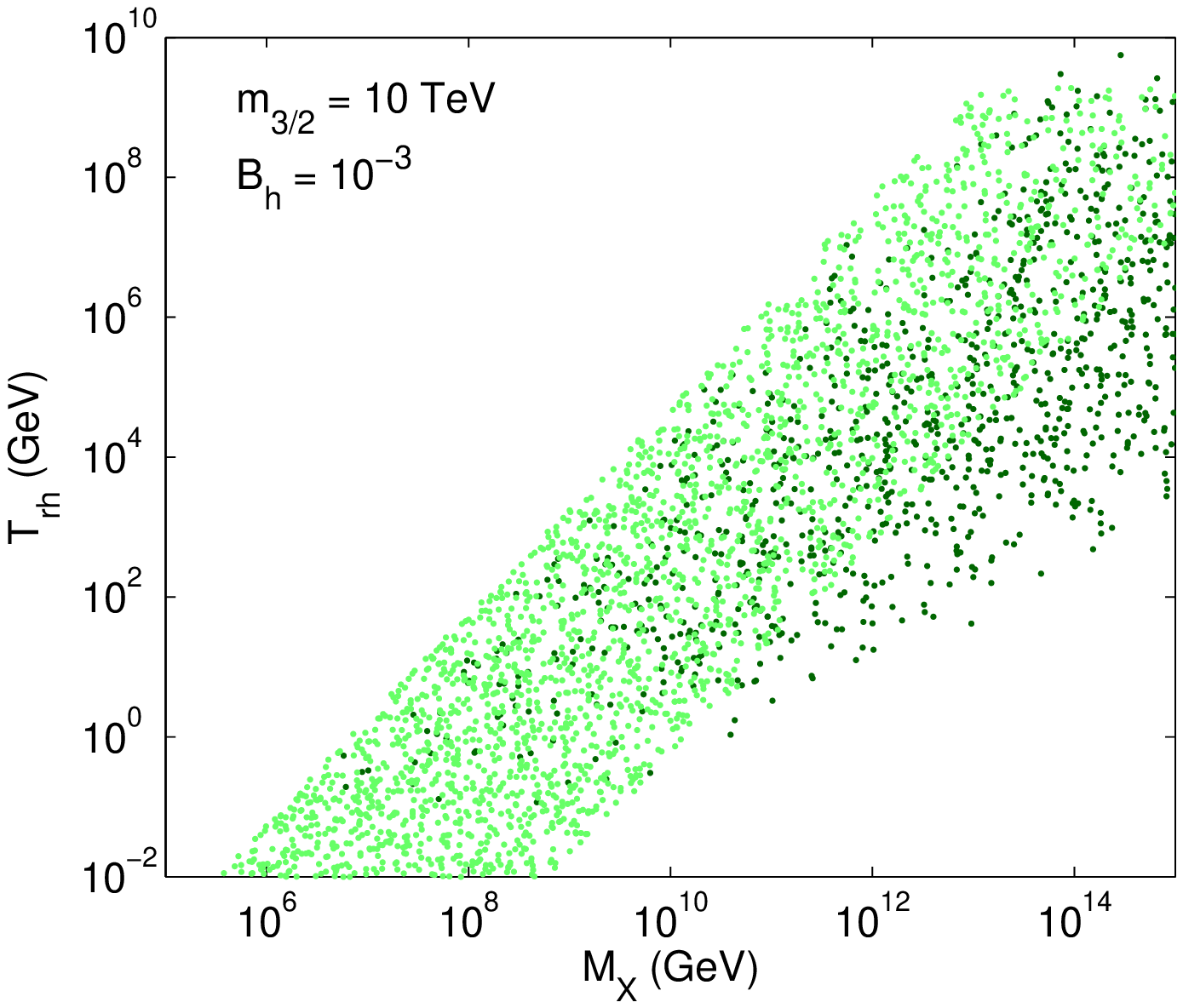}
\includegraphics[width=5.4cm]{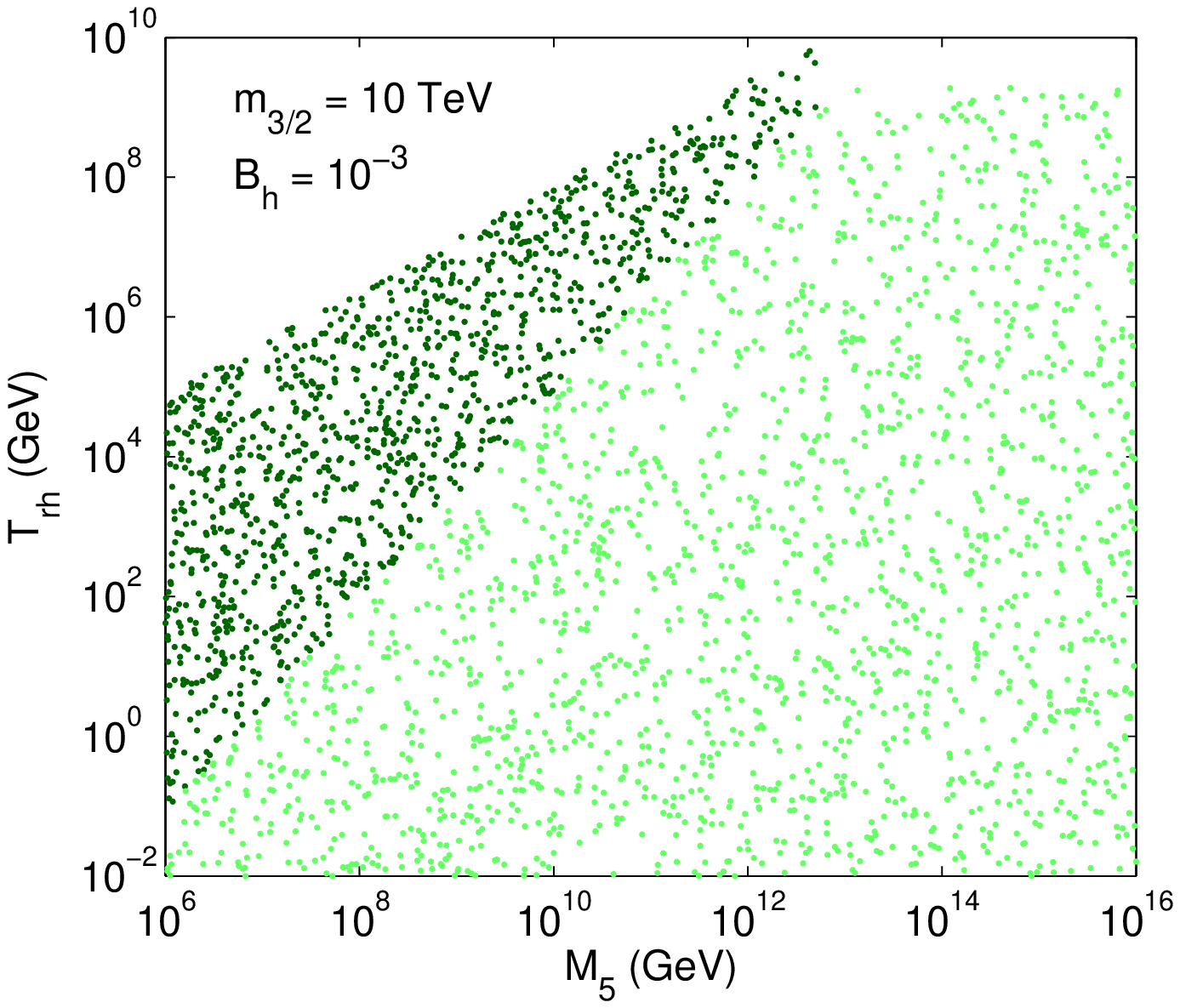}
\includegraphics[width=5.4cm]{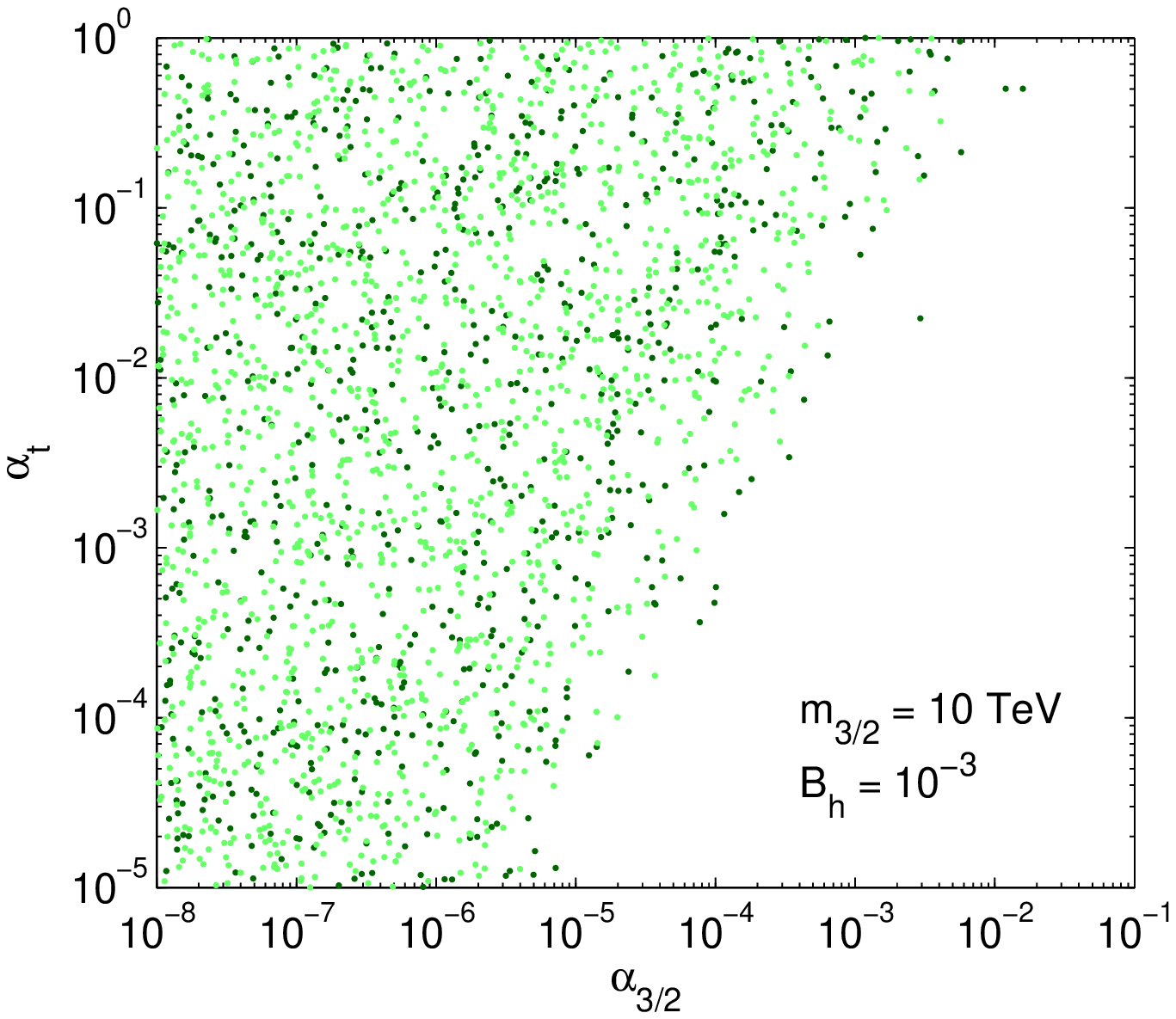}
\caption{(Color online) Parameter space allowed by the gravitino
abundance BBN bounds, in the $T_{rh}-M_X$, $T_{rh}-M_5$ and
$\alpha_t-\alpha_{3/2}$ planes. In the top panels we have
$Y_{3/2}^{BBN}=4 \times 10^{-17}$ ($m_{3/2}=1$~TeV and $B_h=1$), and
in the bottom one $Y_{3/2}^{BBN}=2 \times 10^{-13}$
($m_{3/2}=10$~TeV and $B_h=10^{-3}$). The light and dark green
points correspond to the SC and BC high energy regimes,
respectively. \label{fig:randomunstable}}
\end{figure}

\begin{figure}[t]
\includegraphics[width=5.4cm]{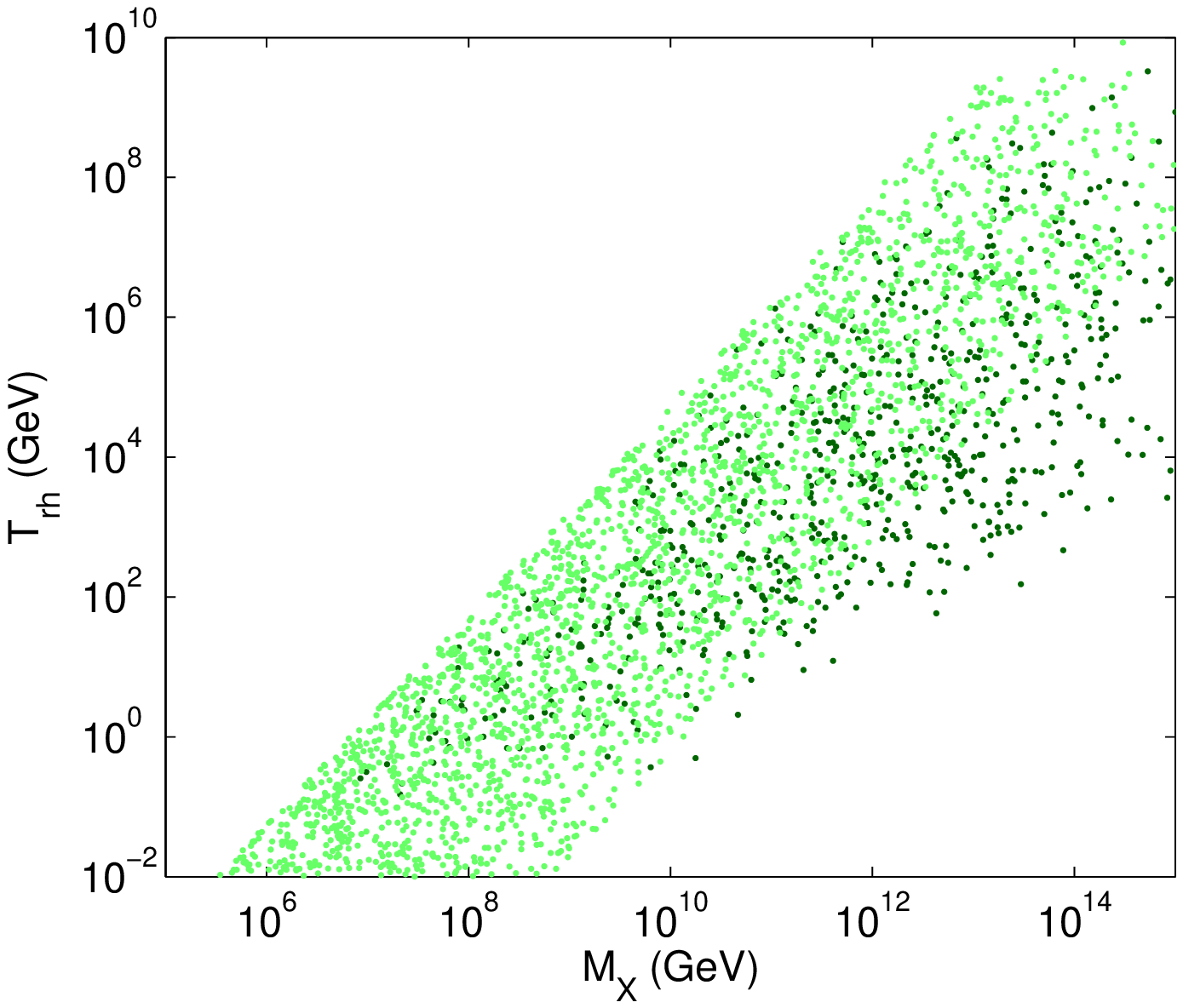}
\includegraphics[width=5.4cm]{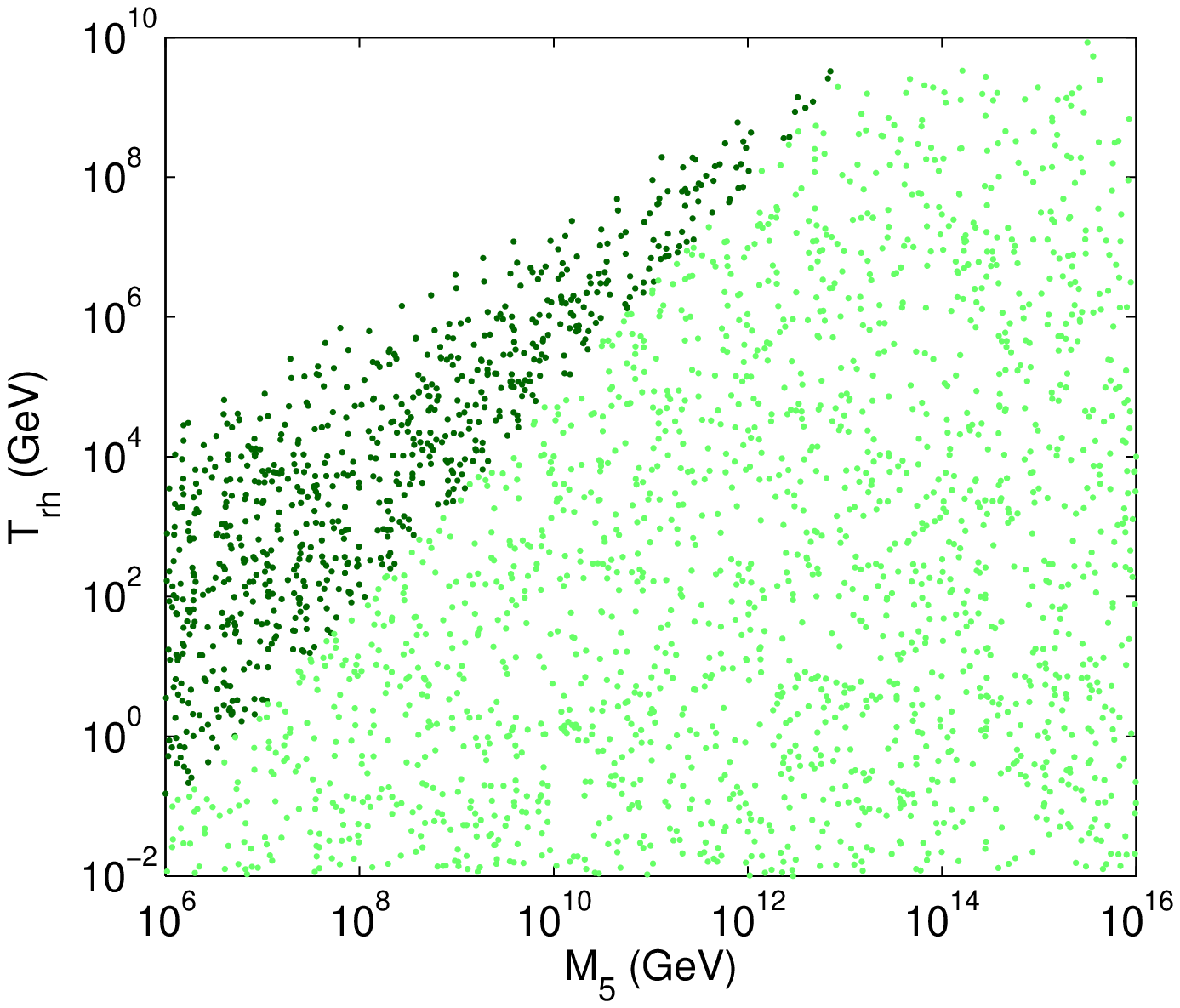}
\includegraphics[width=5.4cm]{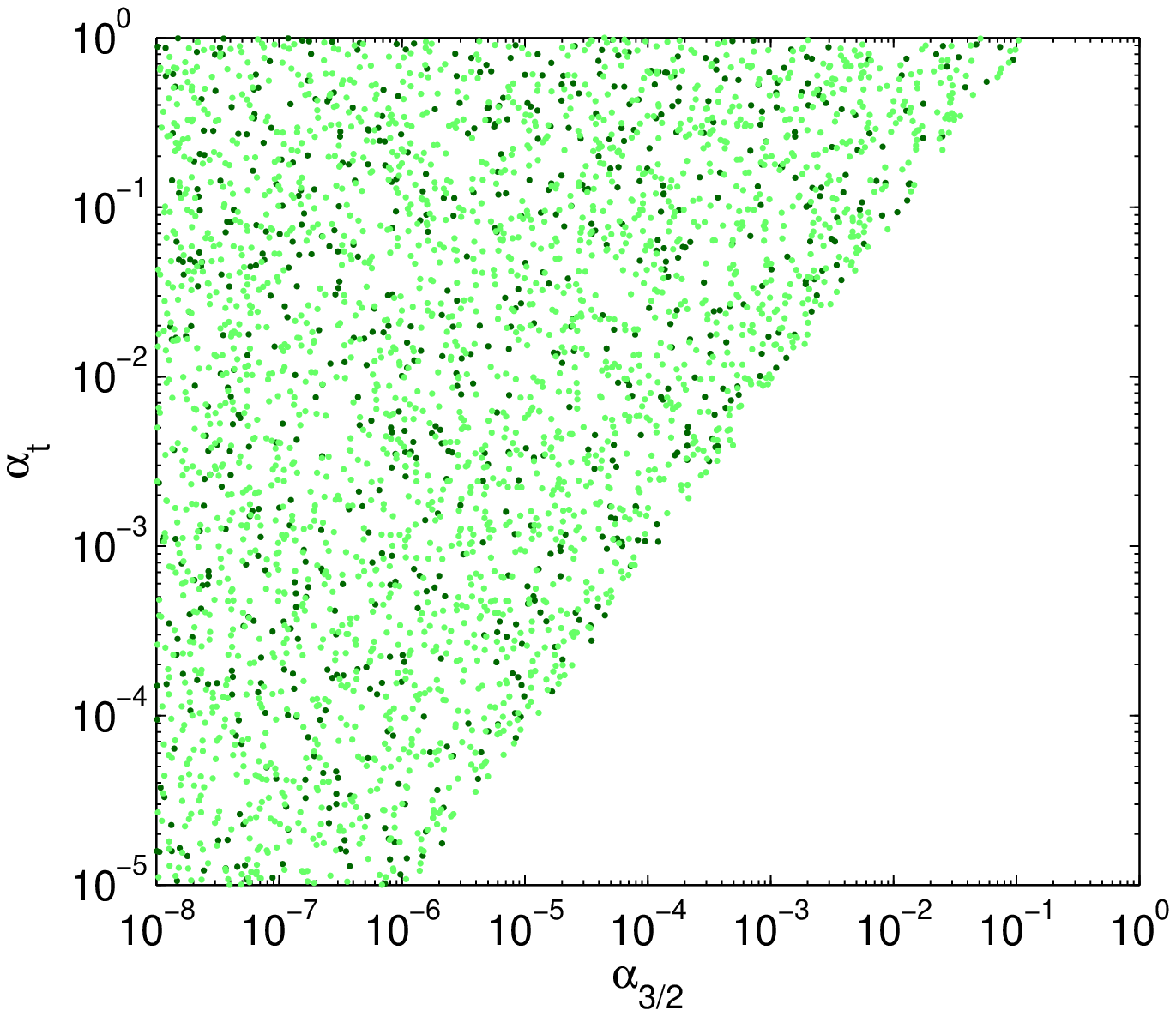}
\caption{(Color online) Parameter space allowed by the dark matter
bounds, $\Omega_{3/2}h^2 \leq \Omega_{DM}h^2=0.143$, and $B_{3/2}
\lesssim 4.5 \times 10^{-4}$ , in the $T_{rh}-M_X$, $T_{rh}-M_5$ and
$\alpha_t-\alpha_{3/2}$planes. The gravitino mass runs in the
interval $10^{-4} - 10^3$~GeV. The light and dark green points
correspond to the SC and BC high energy regimes, respectively.
\label{fig:randomstable}}
\end{figure}

In Figs.~\ref{fig:randomunstable} and \ref{fig:randomstable} we show
the points, in the $T_{rh}-M_X$, $T_{rh}-M_5$ and
$\alpha_t-\alpha_{3/2}$ planes, that satisfy the constraints
discussed in detail in the other sections. We have allowed a
variation of the couplings and masses in the ranges $10^{-8} \leq
\alpha_{3/2} \leq 1$, $\,10^{-5} \leq \alpha_t \leq 1$,
$\,10^5\,{\rm GeV}\leq M_X \leq 10^{15}\,{\rm GeV}$ and $10^6\,{\rm
GeV}\leq M_5 \leq 10^{16}\,{\rm GeV}$. For the stable gravitino case
one has $10^{-4}\,{\rm GeV} \lesssim m_{3/2} \lesssim10^3$~GeV. The
light and dark green points correspond to the SC and BC high energy
regimes, respectively.

%--------------------------------------------------
%-----------------------------------------------------

For the unstable gravitino case, one can derive the range of $M_5$
for which the brane correction to the standard expansion rate is
relevant. The upper bound for this range is dependent of the
gravitino mass and we have obtained $M_5 \lesssim 3.3 \times
10^{10}$~GeV and $M_5 \lesssim 9.7 \times 10^{12}$~GeV, for
$m_{3/2}= 1,\,10$~TeV, respectively.

In the BC regime the upper bounds on the mass of the heavy scalar
and the reheating temperature, coming from the gravitino abundance,
disappear. Thus, with the inclusion of the brane correction, the
parameter space with the correct gravitino abundance is enlarged.
Both, larger reheating temperatures and larger masses for the heavy
scalar are allowed. This is more evident for a smaller gravitino
mass and a larger hadronic branching ratio. Although $T_{rh}$ and
$M_X$ are not constrained by the gravitino abundance limits, they
are bounded by the fact that $H>m_{3/2}$ for the gravitino decay to
be possible.

On the other hand, in BC, like in SC, one also needs a small
coupling $\alpha_{3/2}$ and/or a small branching ratio $B_{3/2}$.
However, for a fixed $\alpha_t$, it is possible to have a slightly
larger $\alpha_{3/2}$ in BC. The required lower value for
$\alpha_{3/2}$ does not allow to solve the moduli problem  by simply
increasing the moduli mass, since, for the range of gravitino masses
considered, $\alpha_{3/2}= {\cal O} (1)$ is excluded. It also poses
severe constraints to inflationary models. In order to achieve a
small $\alpha_{3/2}$ one needs a VEV of $X$ (the inflaton in this
case), much smaller than $M_P$; on the other hand, one can decrease
$B_{3/2}$ if one increases $\alpha_t$, by demanding interactions of
$X$ with the other MSSM particles with strength much larger than
$M_P^{-1}$. A lower value for $\alpha_{3/2}$ can be achieved in some
new and hybrid inflationary models, and in the simplest chaotic
inflationary models, for which $G_X$ vanishes in the vacuum (see
Refs.~\cite{Asaka:2006bv,Kawasaki:2006hm,Endo:2006qk}) one obtains
$\alpha_{3/2}=0$ and then no gravitinos are produced in the decay of
$X$.

%--------------------------------------------------
%-----------------------------------------------------

In the stable gravitino case, similar bounds can be derived. Notice
however that the overclosure constraint alone allows to have
$B_{3/2} \sim 10^{-2}$. Therefore, it would be possible to solve the
moduli problem. But, if one takes into account the warm dark matter
constraint, then one needs $B_{3/2} \lesssim 10^{-4}$, which means
that $\alpha_{3/2} \lesssim 0.1\, \alpha_t$ (cf.
Fig.~\ref{fig:randomstable}). In Ref.~\cite{Asaka:2006bv} its was
suggested that such a suppression in the coupling might be possible
in some moduli stabilization mechanism. In that case it would be
possible to have gravitino warm dark matter from the decay of a
moduli field with mass as large as ${\cal O}(10^{11})-{\cal
O}(10^{12})$~GeV, for a gravitino with $m_{3/2}\lesssim 3$~GeV and
$M_5 \sim {\cal O}(10^{9})$ (cf. Fig.~\ref{fig:stable1}). In SC, the
upper bound for the mass of the moduli would be one order of
magnitude bellow the BC one. Moreover, we should point out that the
warm dark matter constraint disappears if the gravitino contributes
with less than $12\%$ to the total dark matter density.

Finally, it would be interesting to further investigate the
gravitino production in specific brane inflationary models. The
gravitino production analysis presented here could also be extended
to include the KK gravitino production process.

\begin{acknowledgments}
The author wishes to thank R. Gonz\'{a}lez Felipe for the useful
discussions and the careful reading of the manuscript.
\end{acknowledgments}

%\bibliography{D:/Work/Projects/References2papersnew}

\begin{thebibliography}{61}
\expandafter\ifx\csname
natexlab\endcsname\relax\def\natexlab#1{#1}\fi
\expandafter\ifx\csname bibnamefont\endcsname\relax
  \def\bibnamefont#1{#1}\fi
\expandafter\ifx\csname bibfnamefont\endcsname\relax
  \def\bibfnamefont#1{#1}\fi
\expandafter\ifx\csname citenamefont\endcsname\relax
  \def\citenamefont#1{#1}\fi
\expandafter\ifx\csname url\endcsname\relax
  \def\url#1{\texttt{#1}}\fi
\expandafter\ifx\csname urlprefix\endcsname\relax\def\urlprefix{URL
}\fi \providecommand{\bibinfo}[2]{#2}
\providecommand{\eprint}[2][]{\url{#2}}

\bibitem[{\citenamefont{Khlopov and Linde}(1984)}]{Khlopov:1984pf}
\bibinfo{author}{\bibfnamefont{M.~Y.} \bibnamefont{Khlopov}} \bibnamefont{and}
  \bibinfo{author}{\bibfnamefont{A.~D.} \bibnamefont{Linde}},
  \bibinfo{journal}{Phys. Lett.} \textbf{\bibinfo{volume}{B138}},
  \bibinfo{pages}{265} (\bibinfo{year}{1984});
%\bibitem[{\citenamefont{Falomkin et~al.}(1984)}]{Falomkin:1984eu}
\bibinfo{author}{\bibfnamefont{I.~V.} \bibnamefont{Falomkin}}
  \bibnamefont{et~al.}, \bibinfo{journal}{Nuovo Cim.}
  \textbf{\bibinfo{volume}{A79}}, \bibinfo{pages}{193}
  (\bibinfo{year}{1984});
%\bibitem[{\citenamefont{Ellis et~al.}(1984)\citenamefont{Ellis, Kim, and
%  Nanopoulos}}]{Ellis:1984eq}
\bibinfo{author}{\bibfnamefont{J.~R.} \bibnamefont{Ellis}},
  \bibinfo{author}{\bibfnamefont{J.~E.} \bibnamefont{Kim}}, \bibnamefont{and}
  \bibinfo{author}{\bibfnamefont{D.~V.} \bibnamefont{Nanopoulos}},
  \bibinfo{journal}{Phys. Lett.} \textbf{\bibinfo{volume}{B145}},
  \bibinfo{pages}{181} (\bibinfo{year}{1984});
%\bibitem[{\citenamefont{Cyburt et~al.}(2003)\citenamefont{Cyburt, Ellis,
%  Fields, and Olive}}]{Cyburt:2002uv}
\bibinfo{author}{\bibfnamefont{R.~H.} \bibnamefont{Cyburt}},
  \bibinfo{author}{\bibfnamefont{J.~R.} \bibnamefont{Ellis}},
  \bibinfo{author}{\bibfnamefont{B.~D.} \bibnamefont{Fields}},
  \bibnamefont{and} \bibinfo{author}{\bibfnamefont{K.~A.} \bibnamefont{Olive}},
  \bibinfo{journal}{Phys. Rev.} \textbf{\bibinfo{volume}{D67}},
  \bibinfo{pages}{103521} (\bibinfo{year}{2003}), \eprint{astro-ph/0211258}.

\bibitem[{\citenamefont{Endo et~al.}(2006{\natexlab{a}})\citenamefont{Endo,
  Hamaguchi, and Takahashi}}]{Endo:2006zj}
\bibinfo{author}{\bibfnamefont{M.}~\bibnamefont{Endo}},
  \bibinfo{author}{\bibfnamefont{K.}~\bibnamefont{Hamaguchi}},
  \bibnamefont{and}
  \bibinfo{author}{\bibfnamefont{F.}~\bibnamefont{Takahashi}},
  \bibinfo{journal}{Phys. Rev. Lett.} \textbf{\bibinfo{volume}{96}},
  \bibinfo{pages}{211301} (\bibinfo{year}{2006}{\natexlab{a}}),
  \eprint{hep-ph/0602061}.

\bibitem[{\citenamefont{Nakamura and Yamaguchi}(2006)}]{Nakamura:2006uc}
\bibinfo{author}{\bibfnamefont{S.}~\bibnamefont{Nakamura}} \bibnamefont{and}
  \bibinfo{author}{\bibfnamefont{M.}~\bibnamefont{Yamaguchi}},
  \bibinfo{journal}{Phys. Lett.} \textbf{\bibinfo{volume}{B638}},
  \bibinfo{pages}{389} (\bibinfo{year}{2006}), \eprint{hep-ph/0602081}.

\bibitem[{\citenamefont{Coughlan et~al.}(1983)\citenamefont{Coughlan, Fischler,
  Kolb, Raby, and Ross}}]{Coughlan:1983ci}
\bibinfo{author}{\bibfnamefont{G.~D.} \bibnamefont{Coughlan}},
  \bibinfo{author}{\bibfnamefont{W.}~\bibnamefont{Fischler}},
  \bibinfo{author}{\bibfnamefont{E.~W.} \bibnamefont{Kolb}},
  \bibinfo{author}{\bibfnamefont{S.}~\bibnamefont{Raby}}, \bibnamefont{and}
  \bibinfo{author}{\bibfnamefont{G.~G.} \bibnamefont{Ross}},
  \bibinfo{journal}{Phys. Lett.} \textbf{\bibinfo{volume}{B131}},
  \bibinfo{pages}{59} (\bibinfo{year}{1983}).

\bibitem[{\citenamefont{Hashimoto et~al.}(1998)\citenamefont{Hashimoto, Izawa,
  Yamaguchi, and Yanagida}}]{Hashimoto:1998mu}
\bibinfo{author}{\bibfnamefont{M.}~\bibnamefont{Hashimoto}},
  \bibinfo{author}{\bibfnamefont{K.~I.} \bibnamefont{Izawa}},
  \bibinfo{author}{\bibfnamefont{M.}~\bibnamefont{Yamaguchi}},
  \bibnamefont{and} \bibinfo{author}{\bibfnamefont{T.}~\bibnamefont{Yanagida}},
  \bibinfo{journal}{Prog. Theor. Phys.} \textbf{\bibinfo{volume}{100}},
  \bibinfo{pages}{395} (\bibinfo{year}{1998}), \eprint{hep-ph/9804411}.

\bibitem[{\citenamefont{Lyth and Riotto}(1999)}]{Lyth:1998xn}
\bibinfo{author}{\bibfnamefont{D.~H.} \bibnamefont{Lyth}} \bibnamefont{and}
  \bibinfo{author}{\bibfnamefont{A.}~\bibnamefont{Riotto}},
  \bibinfo{journal}{Phys. Rept.} \textbf{\bibinfo{volume}{314}},
  \bibinfo{pages}{1} (\bibinfo{year}{1999}), \eprint{hep-ph/9807278}.

\bibitem[{\citenamefont{Kawasaki
  et~al.}(2006{\natexlab{a}})\citenamefont{Kawasaki, Takahashi, and
  Yanagida}}]{Kawasaki:2006gs}
\bibinfo{author}{\bibfnamefont{M.}~\bibnamefont{Kawasaki}},
  \bibinfo{author}{\bibfnamefont{F.}~\bibnamefont{Takahashi}},
  \bibnamefont{and} \bibinfo{author}{\bibfnamefont{T.~T.}
  \bibnamefont{Yanagida}}, \bibinfo{journal}{Phys. Lett.}
  \textbf{\bibinfo{volume}{B638}}, \bibinfo{pages}{8}
  (\bibinfo{year}{2006}{\natexlab{a}}), \eprint{hep-ph/0603265}.

\bibitem[{\citenamefont{Kawasaki
  et~al.}(2006{\natexlab{b}})\citenamefont{Kawasaki, Takahashi, and
  Yanagida}}]{Kawasaki:2006hm}
\bibinfo{author}{\bibfnamefont{M.}~\bibnamefont{Kawasaki}},
  \bibinfo{author}{\bibfnamefont{F.}~\bibnamefont{Takahashi}},
  \bibnamefont{and} \bibinfo{author}{\bibfnamefont{T.~T.}
  \bibnamefont{Yanagida}}, \bibinfo{journal}{Phys. Rev.}
  \textbf{\bibinfo{volume}{D74}}, \bibinfo{pages}{043519}
  (\bibinfo{year}{2006}{\natexlab{b}}), \eprint{hep-ph/0605297}.

\bibitem[{\citenamefont{Asaka et~al.}(2006)\citenamefont{Asaka, Nakamura, and
  Yamaguchi}}]{Asaka:2006bv}
\bibinfo{author}{\bibfnamefont{T.}~\bibnamefont{Asaka}},
  \bibinfo{author}{\bibfnamefont{S.}~\bibnamefont{Nakamura}}, \bibnamefont{and}
  \bibinfo{author}{\bibfnamefont{M.}~\bibnamefont{Yamaguchi}},
  \bibinfo{journal}{Phys. Rev.} \textbf{\bibinfo{volume}{D74}},
  \bibinfo{pages}{023520} (\bibinfo{year}{2006}), \eprint{hep-ph/0604132}.

\bibitem[{\citenamefont{Endo et~al.}(2006{\natexlab{b}})\citenamefont{Endo,
  Kawasaki, Takahashi, and Yanagida}}]{Endo:2006qk}
\bibinfo{author}{\bibfnamefont{M.}~\bibnamefont{Endo}},
  \bibinfo{author}{\bibfnamefont{M.}~\bibnamefont{Kawasaki}},
  \bibinfo{author}{\bibfnamefont{F.}~\bibnamefont{Takahashi}},
  \bibnamefont{and} \bibinfo{author}{\bibfnamefont{T.~T.}
  \bibnamefont{Yanagida}}, \bibinfo{journal}{Phys. Lett.}
  \textbf{\bibinfo{volume}{B642}}, \bibinfo{pages}{518}
  (\bibinfo{year}{2006}{\natexlab{b}}), \eprint{hep-ph/0607170}.

\bibitem[{\citenamefont{Endo et~al.}(2006{\natexlab{c}})\citenamefont{Endo,
  Hamaguchi, and Takahashi}}]{Endo:2006tf}
\bibinfo{author}{\bibfnamefont{M.}~\bibnamefont{Endo}},
  \bibinfo{author}{\bibfnamefont{K.}~\bibnamefont{Hamaguchi}},
  \bibnamefont{and}
  \bibinfo{author}{\bibfnamefont{F.}~\bibnamefont{Takahashi}},
  \bibinfo{journal}{Phys. Rev.} \textbf{\bibinfo{volume}{D74}},
  \bibinfo{pages}{023531} (\bibinfo{year}{2006}{\natexlab{c}}),
  \eprint{hep-ph/0605091}.

\bibitem[{\citenamefont{Dine et~al.}(2006)\citenamefont{Dine, Kitano, Morisse,
  and Shirman}}]{Dine:2006ii}
\bibinfo{author}{\bibfnamefont{M.}~\bibnamefont{Dine}},
  \bibinfo{author}{\bibfnamefont{R.}~\bibnamefont{Kitano}},
  \bibinfo{author}{\bibfnamefont{A.}~\bibnamefont{Morisse}}, \bibnamefont{and}
  \bibinfo{author}{\bibfnamefont{Y.}~\bibnamefont{Shirman}},
  \bibinfo{journal}{Phys. Rev.} \textbf{\bibinfo{volume}{D73}},
  \bibinfo{pages}{123518} (\bibinfo{year}{2006}), \eprint{hep-ph/0604140}.

\bibitem[{\citenamefont{Maartens}(2004)}]{Maartens:2003tw}
\bibinfo{author}{\bibfnamefont{R.}~\bibnamefont{Maartens}},
  \bibinfo{journal}{Living Rev. Rel.} \textbf{\bibinfo{volume}{7}},
  \bibinfo{pages}{7} (\bibinfo{year}{2004}), \eprint{gr-qc/0312059}.

\bibitem[{\citenamefont{Dvali et~al.}(2000)\citenamefont{Dvali, Gabadadze, and
  Porrati}}]{Dvali:2000hr}
\bibinfo{author}{\bibfnamefont{G.~R.} \bibnamefont{Dvali}},
  \bibinfo{author}{\bibfnamefont{G.}~\bibnamefont{Gabadadze}},
  \bibnamefont{and} \bibinfo{author}{\bibfnamefont{M.}~\bibnamefont{Porrati}},
  \bibinfo{journal}{Phys. Lett.} \textbf{\bibinfo{volume}{B485}},
  \bibinfo{pages}{208} (\bibinfo{year}{2000}),
  \eprint{hep-th/0005016};
%\bibitem[{\citenamefont{Deffayet}(2001)}]{Deffayet:2000uy}
\bibinfo{author}{\bibfnamefont{C.}~\bibnamefont{Deffayet}},
  \bibinfo{journal}{Phys. Lett.} \textbf{\bibinfo{volume}{B502}},
  \bibinfo{pages}{199} (\bibinfo{year}{2001}), \eprint{hep-th/0010186}.

\bibitem[{\citenamefont{Randall and Sundrum}(1999)}]{Randall:1999vf}
\bibinfo{author}{\bibfnamefont{L.}~\bibnamefont{Randall}} \bibnamefont{and}
  \bibinfo{author}{\bibfnamefont{R.}~\bibnamefont{Sundrum}},
  \bibinfo{journal}{Phys. Rev. Lett.} \textbf{\bibinfo{volume}{83}},
  \bibinfo{pages}{4690} (\bibinfo{year}{1999}), \eprint{hep-th/9906064}.

\bibitem[{\citenamefont{Binetruy
  et~al.}(2000{\natexlab{a}})\citenamefont{Binetruy, Deffayet, and
  Langlois}}]{Binetruy:1999ut}
\bibinfo{author}{\bibfnamefont{P.}~\bibnamefont{Binetruy}},
  \bibinfo{author}{\bibfnamefont{C.}~\bibnamefont{Deffayet}}, \bibnamefont{and}
  \bibinfo{author}{\bibfnamefont{D.}~\bibnamefont{Langlois}},
  \bibinfo{journal}{Nucl. Phys.} \textbf{\bibinfo{volume}{B565}},
  \bibinfo{pages}{269} (\bibinfo{year}{2000}{\natexlab{a}}),
  \eprint{hep-th/9905012}.

\bibitem[{\citenamefont{Csaki et~al.}(1999)\citenamefont{Csaki, Graesser,
  Kolda, and Terning}}]{Csaki:1999jh}
\bibinfo{author}{\bibfnamefont{C.}~\bibnamefont{Csaki}},
  \bibinfo{author}{\bibfnamefont{M.}~\bibnamefont{Graesser}},
  \bibinfo{author}{\bibfnamefont{C.~F.} \bibnamefont{Kolda}}, \bibnamefont{and}
  \bibinfo{author}{\bibfnamefont{J.}~\bibnamefont{Terning}},
  \bibinfo{journal}{Phys. Lett.} \textbf{\bibinfo{volume}{B462}},
  \bibinfo{pages}{34} (\bibinfo{year}{1999}), \eprint{hep-ph/9906513}.

\bibitem[{\citenamefont{Cline et~al.}(1999)\citenamefont{Cline, Grojean, and
  Servant}}]{Cline:1999ts}
\bibinfo{author}{\bibfnamefont{J.~M.} \bibnamefont{Cline}},
  \bibinfo{author}{\bibfnamefont{C.}~\bibnamefont{Grojean}}, \bibnamefont{and}
  \bibinfo{author}{\bibfnamefont{G.}~\bibnamefont{Servant}},
  \bibinfo{journal}{Phys. Rev. Lett.} \textbf{\bibinfo{volume}{83}},
  \bibinfo{pages}{4245} (\bibinfo{year}{1999}), \eprint{hep-ph/9906523}.

\bibitem[{\citenamefont{Shiromizu et~al.}(2000)\citenamefont{Shiromizu, Maeda,
  and Sasaki}}]{Shiromizu:1999wj}
\bibinfo{author}{\bibfnamefont{T.}~\bibnamefont{Shiromizu}},
  \bibinfo{author}{\bibfnamefont{K.-i.} \bibnamefont{Maeda}}, \bibnamefont{and}
  \bibinfo{author}{\bibfnamefont{M.}~\bibnamefont{Sasaki}},
  \bibinfo{journal}{Phys. Rev.} \textbf{\bibinfo{volume}{D62}},
  \bibinfo{pages}{024012} (\bibinfo{year}{2000}), \eprint{gr-qc/9910076}.

\bibitem[{\citenamefont{Binetruy
  et~al.}(2000{\natexlab{b}})\citenamefont{Binetruy, Deffayet, Ellwanger, and
  Langlois}}]{Binetruy:1999hy}
\bibinfo{author}{\bibfnamefont{P.}~\bibnamefont{Binetruy}},
  \bibinfo{author}{\bibfnamefont{C.}~\bibnamefont{Deffayet}},
  \bibinfo{author}{\bibfnamefont{U.}~\bibnamefont{Ellwanger}},
  \bibnamefont{and} \bibinfo{author}{\bibfnamefont{D.}~\bibnamefont{Langlois}},
  \bibinfo{journal}{Phys. Lett.} \textbf{\bibinfo{volume}{B477}},
  \bibinfo{pages}{285} (\bibinfo{year}{2000}{\natexlab{b}}),
  \eprint{hep-th/9910219}.

\bibitem[{\citenamefont{Kraus}(1999)}]{Kraus:1999it}
\bibinfo{author}{\bibfnamefont{P.}~\bibnamefont{Kraus}},
  \bibinfo{journal}{JHEP} \textbf{\bibinfo{volume}{12}}, \bibinfo{pages}{011}
  (\bibinfo{year}{1999}), \eprint{hep-th/9910149}.

\bibitem[{\citenamefont{Ida}(2000)}]{Ida:1999ui}
\bibinfo{author}{\bibfnamefont{D.}~\bibnamefont{Ida}}, \bibinfo{journal}{JHEP}
  \textbf{\bibinfo{volume}{09}}, \bibinfo{pages}{014} (\bibinfo{year}{2000}),
  \eprint{gr-qc/9912002}.

\bibitem[{\citenamefont{Mukohyama et~al.}(2000)\citenamefont{Mukohyama,
  Shiromizu, and Maeda}}]{Mukohyama:1999wi}
\bibinfo{author}{\bibfnamefont{S.}~\bibnamefont{Mukohyama}},
  \bibinfo{author}{\bibfnamefont{T.}~\bibnamefont{Shiromizu}},
  \bibnamefont{and} \bibinfo{author}{\bibfnamefont{K.-i.} \bibnamefont{Maeda}},
  \bibinfo{journal}{Phys. Rev.} \textbf{\bibinfo{volume}{D62}},
  \bibinfo{pages}{024028} (\bibinfo{year}{2000}), \eprint{hep-th/9912287}.

\bibitem[{\citenamefont{Maartens et~al.}(2000)\citenamefont{Maartens, Wands,
  Bassett, and Heard}}]{Maartens:1999hf}
\bibinfo{author}{\bibfnamefont{R.}~\bibnamefont{Maartens}},
  \bibinfo{author}{\bibfnamefont{D.}~\bibnamefont{Wands}},
  \bibinfo{author}{\bibfnamefont{B.~A.} \bibnamefont{Bassett}},
  \bibnamefont{and} \bibinfo{author}{\bibfnamefont{I.}~\bibnamefont{Heard}},
  \bibinfo{journal}{Phys. Rev.} \textbf{\bibinfo{volume}{D62}},
  \bibinfo{pages}{041301} (\bibinfo{year}{2000}),
  \eprint{hep-ph/9912464};
%\bibitem[{\citenamefont{Langlois et~al.}(2000)\citenamefont{Langlois, Maartens,
%  and Wands}}]{Langlois:2000ns}
\bibinfo{author}{\bibfnamefont{D.}~\bibnamefont{Langlois}},
  \bibinfo{author}{\bibfnamefont{R.}~\bibnamefont{Maartens}}, \bibnamefont{and}
  \bibinfo{author}{\bibfnamefont{D.}~\bibnamefont{Wands}},
  \bibinfo{journal}{Phys. Lett.} \textbf{\bibinfo{volume}{B489}},
  \bibinfo{pages}{259} (\bibinfo{year}{2000}),
  \eprint{hep-th/0006007};
%\bibitem[{\citenamefont{Bento et~al.}(2003{\natexlab{a}})\citenamefont{Bento,
%  Bertolami, and Sen}}]{Bento:2002kp}
\bibinfo{author}{\bibfnamefont{M.~C.} \bibnamefont{Bento}},
  \bibinfo{author}{\bibfnamefont{O.}~\bibnamefont{Bertolami}},
  \bibnamefont{and} \bibinfo{author}{\bibfnamefont{A.~A.} \bibnamefont{Sen}},
  \bibinfo{journal}{Phys. Rev.} \textbf{\bibinfo{volume}{D67}},
  \bibinfo{pages}{023504} (\bibinfo{year}{2003}{\natexlab{a}}),
  \eprint{gr-qc/0204046};
%\bibitem[{\citenamefont{Bento et~al.}(2003{\natexlab{b}})\citenamefont{Bento,
%  Bertolami, and Sen}}]{Bento:2002np}
\bibinfo{author}{\bibfnamefont{M.~C.} \bibnamefont{Bento}},
  \bibinfo{author}{\bibfnamefont{O.}~\bibnamefont{Bertolami}},
  \bibnamefont{and} \bibinfo{author}{\bibfnamefont{A.~A.} \bibnamefont{Sen}},
  \bibinfo{journal}{Phys. Rev.} \textbf{\bibinfo{volume}{D67}},
  \bibinfo{pages}{063511} (\bibinfo{year}{2003}{\natexlab{b}}),
  \eprint{hep-th/0208124};
%\bibitem[{\citenamefont{Bento et~al.}(2004{\natexlab{a}})\citenamefont{Bento,
%  Santos, and Sen}}]{Bento:2003qe}
\bibinfo{author}{\bibfnamefont{M.~C.} \bibnamefont{Bento}},
  \bibinfo{author}{\bibfnamefont{N.~M.~C.} \bibnamefont{Santos}},
  \bibnamefont{and} \bibinfo{author}{\bibfnamefont{A.~A.} \bibnamefont{Sen}},
  \bibinfo{journal}{Int. J. Mod. Phys.} \textbf{\bibinfo{volume}{D13}},
  \bibinfo{pages}{1927} (\bibinfo{year}{2004}{\natexlab{a}}),
  \eprint{astro-ph/0307292};
%\bibitem[{\citenamefont{Bento et~al.}(2004{\natexlab{b}})\citenamefont{Bento,
%  Santos, and Sen}}]{Bento:2003eu}
\bibinfo{author}{\bibfnamefont{M.~C.} \bibnamefont{Bento}},
  \bibinfo{author}{\bibfnamefont{N.~M.~C.} \bibnamefont{Santos}},
  \bibnamefont{and} \bibinfo{author}{\bibfnamefont{A.~A.} \bibnamefont{Sen}},
  \bibinfo{journal}{Phys. Rev.} \textbf{\bibinfo{volume}{D69}},
  \bibinfo{pages}{023508} (\bibinfo{year}{2004}{\natexlab{b}}),
  \eprint{astro-ph/0307093};
%\bibitem[{\citenamefont{Felipe}(2005)}]{Felipe:2004wh}
\bibinfo{author}{\bibfnamefont{R.~G.} \bibnamefont{Felipe}},
  \bibinfo{journal}{Phys. Lett.} \textbf{\bibinfo{volume}{B618}},
  \bibinfo{pages}{7} (\bibinfo{year}{2005}),
  \eprint{hep-ph/0411349};
%\bibitem[{\citenamefont{Bento et~al.}(2006{\natexlab{a}})\citenamefont{Bento,
%  Gonz\'alez~Felipe, and Santos}}]{Bento:2006sr}
\bibinfo{author}{\bibfnamefont{M.~C.} \bibnamefont{Bento}},
  \bibinfo{author}{\bibfnamefont{R.}~\bibnamefont{Gonz\'alez~Felipe}},
  \bibnamefont{and} \bibinfo{author}{\bibfnamefont{N.~M.~C.}
  \bibnamefont{Santos}}, \bibinfo{journal}{Phys. Rev.}
  \textbf{\bibinfo{volume}{D74}}, \bibinfo{pages}{083503}
  (\bibinfo{year}{2006}{\natexlab{a}}), \eprint{astro-ph/0606047}.

\bibitem[{\citenamefont{Mazumdar}(2001)}]{Mazumdar:2000gj}
\bibinfo{author}{\bibfnamefont{A.}~\bibnamefont{Mazumdar}},
  \bibinfo{journal}{Phys. Rev.} \textbf{\bibinfo{volume}{D64}},
  \bibinfo{pages}{027304} (\bibinfo{year}{2001}),
  \eprint{hep-ph/0007269};
%\bibitem[{\citenamefont{Allahverdi et~al.}(2001)\citenamefont{Allahverdi,
%  Enqvist, Mazumdar, and Perez-Lorenzana}}]{Allahverdi:2001dm}
\bibinfo{author}{\bibfnamefont{R.}~\bibnamefont{Allahverdi}},
  \bibinfo{author}{\bibfnamefont{K.}~\bibnamefont{Enqvist}},
  \bibinfo{author}{\bibfnamefont{A.}~\bibnamefont{Mazumdar}}, \bibnamefont{and}
  \bibinfo{author}{\bibfnamefont{A.}~\bibnamefont{Perez-Lorenzana}},
  \bibinfo{journal}{Nucl. Phys.} \textbf{\bibinfo{volume}{B618}},
  \bibinfo{pages}{277} (\bibinfo{year}{2001}),
  \eprint{hep-ph/0108225};
%\bibitem[{\citenamefont{Mazumdar and Perez-Lorenzana}(2002)}]{Mazumdar:2001nw}
\bibinfo{author}{\bibfnamefont{A.}~\bibnamefont{Mazumdar}} \bibnamefont{and}
  \bibinfo{author}{\bibfnamefont{A.}~\bibnamefont{Perez-Lorenzana}},
  \bibinfo{journal}{Phys. Rev.} \textbf{\bibinfo{volume}{D65}},
  \bibinfo{pages}{107301} (\bibinfo{year}{2002}),
  \eprint{hep-ph/0103215};
%\bibitem[{\citenamefont{Matsuda}(2002)}]{Matsuda:2002jv}
\bibinfo{author}{\bibfnamefont{T.}~\bibnamefont{Matsuda}},
  \bibinfo{journal}{Phys. Rev.} \textbf{\bibinfo{volume}{D65}},
  \bibinfo{pages}{103501} (\bibinfo{year}{2002}),
  \eprint{hep-ph/0202209};
%\bibitem[{\citenamefont{Shiromizu and Koyama}(2004)}]{Shiromizu:2004cb}
\bibinfo{author}{\bibfnamefont{T.}~\bibnamefont{Shiromizu}} \bibnamefont{and}
  \bibinfo{author}{\bibfnamefont{K.}~\bibnamefont{Koyama}},
  \bibinfo{journal}{JCAP} \textbf{\bibinfo{volume}{0407}}, \bibinfo{pages}{011}
  (\bibinfo{year}{2004}), \eprint{hep-ph/0403231};
%\bibitem[{\citenamefont{Bento et~al.}(2005)\citenamefont{Bento,
%  Gonz\'alez~Felipe, and Santos}}]{Bento:2005xk}
\bibinfo{author}{\bibfnamefont{M.~C.} \bibnamefont{Bento}},
  \bibinfo{author}{\bibfnamefont{R.}~\bibnamefont{Gonz\'alez~Felipe}},
  \bibnamefont{and} \bibinfo{author}{\bibfnamefont{N.~M.~C.}
  \bibnamefont{Santos}}, \bibinfo{journal}{Phys. Rev.}
  \textbf{\bibinfo{volume}{D71}}, \bibinfo{pages}{123517}
  (\bibinfo{year}{2005}), \eprint{hep-ph/0504113};
%\bibitem[{\citenamefont{Bento et~al.}(2006{\natexlab{b}})\citenamefont{Bento,
%  Gonz\'alez~Felipe, and Santos}}]{Bento:2005je}
\bibinfo{author}{\bibfnamefont{M.~C.} \bibnamefont{Bento}},
  \bibinfo{author}{\bibfnamefont{R.}~\bibnamefont{Gonz\'alez~Felipe}},
  \bibnamefont{and} \bibinfo{author}{\bibfnamefont{N.~M.~C.}
  \bibnamefont{Santos}}, \bibinfo{journal}{Phys. Rev.}
  \textbf{\bibinfo{volume}{D73}}, \bibinfo{pages}{023506}
  (\bibinfo{year}{2006}{\natexlab{b}}), \eprint{hep-ph/0508213};
%\bibitem[{\citenamefont{Okada and Seto}(2006)}]{Okada:2005kv}
\bibinfo{author}{\bibfnamefont{N.}~\bibnamefont{Okada}} \bibnamefont{and}
  \bibinfo{author}{\bibfnamefont{O.}~\bibnamefont{Seto}},
  \bibinfo{journal}{Phys. Rev.} \textbf{\bibinfo{volume}{D73}},
  \bibinfo{pages}{063505} (\bibinfo{year}{2006}), \eprint{hep-ph/0507279}.

\bibitem[{\citenamefont{Bento et~al.}(2004{\natexlab{c}})\citenamefont{Bento,
  Gonz\'alez~Felipe, and Santos}}]{Bento:2004pz}
\bibinfo{author}{\bibfnamefont{M.~C.} \bibnamefont{Bento}},
  \bibinfo{author}{\bibfnamefont{R.}~\bibnamefont{Gonz\'alez~Felipe}},
  \bibnamefont{and} \bibinfo{author}{\bibfnamefont{N.~M.~C.}
  \bibnamefont{Santos}}, \bibinfo{journal}{Phys. Rev.}
  \textbf{\bibinfo{volume}{D69}}, \bibinfo{pages}{123513}
  (\bibinfo{year}{2004}{\natexlab{c}}), \eprint{hep-ph/0402276}.

\bibitem[{\citenamefont{Okada and Seto}(2004)}]{Okada:2004nc}
\bibinfo{author}{\bibfnamefont{N.}~\bibnamefont{Okada}} \bibnamefont{and}
  \bibinfo{author}{\bibfnamefont{O.}~\bibnamefont{Seto}},
  \bibinfo{journal}{Phys. Rev.} \textbf{\bibinfo{volume}{D70}},
  \bibinfo{pages}{083531} (\bibinfo{year}{2004}),
  \eprint{hep-ph/0407092};
%\bibitem[{\citenamefont{Nihei et~al.}(2005)\citenamefont{Nihei, Okada, and
%  Seto}}]{Nihei:2004xv}
\bibinfo{author}{\bibfnamefont{T.}~\bibnamefont{Nihei}},
  \bibinfo{author}{\bibfnamefont{N.}~\bibnamefont{Okada}}, \bibnamefont{and}
  \bibinfo{author}{\bibfnamefont{O.}~\bibnamefont{Seto}},
  \bibinfo{journal}{Phys. Rev.} \textbf{\bibinfo{volume}{D71}},
  \bibinfo{pages}{063535} (\bibinfo{year}{2005}),
  \eprint{hep-ph/0409219};
%\bibitem[{\citenamefont{Panotopoulos}(2007)}]{Panotopoulos:2007fg}
\bibinfo{author}{\bibfnamefont{G.}~\bibnamefont{Panotopoulos}}
  (\bibinfo{year}{2007}), \eprint{hep-ph/0701233}.

\bibitem[{\citenamefont{Okada and Seto}(2005)}]{Okada:2004mh}
\bibinfo{author}{\bibfnamefont{N.}~\bibnamefont{Okada}} \bibnamefont{and}
  \bibinfo{author}{\bibfnamefont{O.}~\bibnamefont{Seto}},
  \bibinfo{journal}{Phys. Rev.} \textbf{\bibinfo{volume}{D71}},
  \bibinfo{pages}{023517} (\bibinfo{year}{2005}), \eprint{hep-ph/0407235}.

\bibitem[{\citenamefont{Gherghetta and Pomarol}(2000)}]{Gherghetta:2000qt}
\bibinfo{author}{\bibfnamefont{T.}~\bibnamefont{Gherghetta}} \bibnamefont{and}
  \bibinfo{author}{\bibfnamefont{A.}~\bibnamefont{Pomarol}},
  \bibinfo{journal}{Nucl. Phys.} \textbf{\bibinfo{volume}{B586}},
  \bibinfo{pages}{141} (\bibinfo{year}{2000}),
  \eprint{hep-ph/0003129};
%\bibitem[{\citenamefont{Altendorfer et~al.}(2001)\citenamefont{Altendorfer,
%  Bagger, and Nemeschansky}}]{Altendorfer:2000rr}
\bibinfo{author}{\bibfnamefont{R.}~\bibnamefont{Altendorfer}},
  \bibinfo{author}{\bibfnamefont{J.}~\bibnamefont{Bagger}}, \bibnamefont{and}
  \bibinfo{author}{\bibfnamefont{D.}~\bibnamefont{Nemeschansky}},
  \bibinfo{journal}{Phys. Rev.} \textbf{\bibinfo{volume}{D63}},
  \bibinfo{pages}{125025} (\bibinfo{year}{2001}), \eprint{hep-th/0003117}.

\bibitem[{\citenamefont{Randall and Thomas}(1995)}]{Randall:1994fr}
\bibinfo{author}{\bibfnamefont{L.}~\bibnamefont{Randall}} \bibnamefont{and}
  \bibinfo{author}{\bibfnamefont{S.~D.} \bibnamefont{Thomas}},
  \bibinfo{journal}{Nucl. Phys.} \textbf{\bibinfo{volume}{B449}},
  \bibinfo{pages}{229} (\bibinfo{year}{1995}),
  \eprint{hep-ph/9407248};
%\bibitem[{\citenamefont{Moroi et~al.}(1995)\citenamefont{Moroi, Yamaguchi, and
%  Yanagida}}]{Moroi:1994rs}
\bibinfo{author}{\bibfnamefont{T.}~\bibnamefont{Moroi}},
  \bibinfo{author}{\bibfnamefont{M.}~\bibnamefont{Yamaguchi}},
  \bibnamefont{and} \bibinfo{author}{\bibfnamefont{T.}~\bibnamefont{Yanagida}},
  \bibinfo{journal}{Phys. Lett.} \textbf{\bibinfo{volume}{B342}},
  \bibinfo{pages}{105} (\bibinfo{year}{1995}), \eprint{hep-ph/9409367}.

\bibitem[{\citenamefont{Kohri et~al.}(2004)\citenamefont{Kohri, Yamaguchi, and
  Yokoyama}}]{Kohri:2004qu}
\bibinfo{author}{\bibfnamefont{K.}~\bibnamefont{Kohri}},
  \bibinfo{author}{\bibfnamefont{M.}~\bibnamefont{Yamaguchi}},
  \bibnamefont{and} \bibinfo{author}{\bibfnamefont{J.}~\bibnamefont{Yokoyama}},
  \bibinfo{journal}{Phys. Rev.} \textbf{\bibinfo{volume}{D70}},
  \bibinfo{pages}{043522} (\bibinfo{year}{2004}), \eprint{hep-ph/0403043}.

\bibitem[{\citenamefont{Bolz et~al.}(2001)\citenamefont{Bolz, Brandenburg, and
  Buchmuller}}]{Bolz:2000fu}
\bibinfo{author}{\bibfnamefont{M.}~\bibnamefont{Bolz}},
  \bibinfo{author}{\bibfnamefont{A.}~\bibnamefont{Brandenburg}},
  \bibnamefont{and}
  \bibinfo{author}{\bibfnamefont{W.}~\bibnamefont{Buchmuller}},
  \bibinfo{journal}{Nucl. Phys.} \textbf{\bibinfo{volume}{B606}},
  \bibinfo{pages}{518} (\bibinfo{year}{2001}), \eprint{hep-ph/0012052}.

\bibitem[{\citenamefont{Kohri et~al.}(2006)\citenamefont{Kohri, Moroi, and
  Yotsuyanagi}}]{Kohri:2005wn}
\bibinfo{author}{\bibfnamefont{K.}~\bibnamefont{Kohri}},
  \bibinfo{author}{\bibfnamefont{T.}~\bibnamefont{Moroi}}, \bibnamefont{and}
  \bibinfo{author}{\bibfnamefont{A.}~\bibnamefont{Yotsuyanagi}},
  \bibinfo{journal}{Phys. Rev.} \textbf{\bibinfo{volume}{D73}},
  \bibinfo{pages}{123511} (\bibinfo{year}{2006}), \eprint{hep-ph/0507245}.

\bibitem[{\citenamefont{Bennett et~al.}(2003)}]{Bennett:2003bz}
\bibinfo{author}{\bibfnamefont{C.~L.} \bibnamefont{Bennett}}
  \bibnamefont{et~al.}, \bibinfo{journal}{Astrophys. J. Suppl.}
  \textbf{\bibinfo{volume}{148}}, \bibinfo{pages}{1} (\bibinfo{year}{2003}),
  \eprint{astro-ph/0302207}.

\bibitem[{\citenamefont{Jedamzik et~al.}(2006)\citenamefont{Jedamzik, Lemoine,
  and Moultaka}}]{Jedamzik:2005sx}
\bibinfo{author}{\bibfnamefont{K.}~\bibnamefont{Jedamzik}},
  \bibinfo{author}{\bibfnamefont{M.}~\bibnamefont{Lemoine}}, \bibnamefont{and}
  \bibinfo{author}{\bibfnamefont{G.}~\bibnamefont{Moultaka}},
  \bibinfo{journal}{JCAP} \textbf{\bibinfo{volume}{0607}}, \bibinfo{pages}{010}
  (\bibinfo{year}{2006}), \eprint{astro-ph/0508141}.

\bibitem[{\citenamefont{Borgani et~al.}(1996)\citenamefont{Borgani, Masiero,
  and Yamaguchi}}]{Borgani:1996ag}
\bibinfo{author}{\bibfnamefont{S.}~\bibnamefont{Borgani}},
  \bibinfo{author}{\bibfnamefont{A.}~\bibnamefont{Masiero}}, \bibnamefont{and}
  \bibinfo{author}{\bibfnamefont{M.}~\bibnamefont{Yamaguchi}},
  \bibinfo{journal}{Phys. Lett.} \textbf{\bibinfo{volume}{B386}},
  \bibinfo{pages}{189} (\bibinfo{year}{1996}), \eprint{hep-ph/9605222}.

\bibitem[{\citenamefont{Steffen}(2006)}]{Steffen:2006hw}
\bibinfo{author}{\bibfnamefont{F.~D.} \bibnamefont{Steffen}},
  \bibinfo{journal}{JCAP} \textbf{\bibinfo{volume}{0609}}, \bibinfo{pages}{001}
  (\bibinfo{year}{2006}), \eprint{hep-ph/0605306}.

\bibitem[{\citenamefont{Viel et~al.}(2005)\citenamefont{Viel, Lesgourgues,
  Haehnelt, Matarrese, and Riotto}}]{Viel:2005qj}
\bibinfo{author}{\bibfnamefont{M.}~\bibnamefont{Viel}},
  \bibinfo{author}{\bibfnamefont{J.}~\bibnamefont{Lesgourgues}},
  \bibinfo{author}{\bibfnamefont{M.~G.} \bibnamefont{Haehnelt}},
  \bibinfo{author}{\bibfnamefont{S.}~\bibnamefont{Matarrese}},
  \bibnamefont{and} \bibinfo{author}{\bibfnamefont{A.}~\bibnamefont{Riotto}},
  \bibinfo{journal}{Phys. Rev.} \textbf{\bibinfo{volume}{D71}},
  \bibinfo{pages}{063534} (\bibinfo{year}{2005}), \eprint{astro-ph/0501562}.

\bibitem[{\citenamefont{Seljak et~al.}(2006)\citenamefont{Seljak, Makarov,
  McDonald, and Trac}}]{Seljak:2006qw}
\bibinfo{author}{\bibfnamefont{U.}~\bibnamefont{Seljak}},
  \bibinfo{author}{\bibfnamefont{A.}~\bibnamefont{Makarov}},
  \bibinfo{author}{\bibfnamefont{P.}~\bibnamefont{McDonald}}, \bibnamefont{and}
  \bibinfo{author}{\bibfnamefont{H.}~\bibnamefont{Trac}},
  \bibinfo{journal}{Phys. Rev. Lett.} \textbf{\bibinfo{volume}{97}},
  \bibinfo{pages}{191303} (\bibinfo{year}{2006}), \eprint{astro-ph/0602430}.

\end{thebibliography}

\end{document}